\documentclass[]{raa}           

\usepackage{graphicx,times}
\usepackage{natbib}
\usepackage{amssymb,amsmath}
\bibpunct{(}{)}{;}{a}{}{,}
\usepackage{multirow}
\usepackage[section]{placeins}
\usepackage{float}
\usepackage{threeparttable}
\usepackage{subcaption}

\usepackage[pagebackref=true]{hyperref}
\usepackage{setspace} 

\begin{document}

%   \title{The \ww{True Mass of a Sample of} Radial-Velocity Exoplanets \ww{with Astrometric Measurements} }
   \title{The Masses of a Sample of Radial-Velocity Exoplanets with Astrometric Measurements }

 \volnopage{ {\bf 20XX} Vol.\ {\bf X} No. {\bf XX}, 000--000}
   \setcounter{page}{1}

   \author{Guang-Yao Xiao
   \inst{1,2,3,4}, Yu-Juan Liu\inst{3}\thanks{E-mail: lyj@bao.ac.cn}, Huan-Yu Teng\inst{5}, 
   Wei Wang\inst{3}, Timothy D. Brandt\inst{6}, 
   Gang Zhao\inst{3},  Fei Zhao\inst{3}, Meng Zhai\inst{3}, 
   Qi Gao\inst{1,2}\thanks{E-mail: fbc1980@163.com}
   }

   \institute{ Department of Physics, College of Science, Tibet University, Lhasa 850000, 
P. R. China\\
\and
Key Laboratory of Cosmic Rays (Tibet University), Ministry of Education, Lhasa 850000, P. R. China\\
\and
CAS Key Laboratory of Optical Astronomy, National Astronomical Observatories, Chinese
Academy of Sciences, Beijing 100101, China\\
\and School of Astronomy and Space Science, University of Chinese Academy of Science, Beijing 100049, China
\\
\and 
Department of Earth and Planetary Sciences, School of Science, Tokyo Institute of
Technology, 2-12-1 Ookayama, Meguro-ku, Tokyo 152-8551, Japan\\
\and
Department of Physics, University of California, Santa Barbara, Santa Barbara, CA 93106, USA
\vs \no
   {\small Received 20XX Month Day; accepted 20XX Month Day}
}

\abstract{Being one of the most fundamental physical parameter of  astronomical objects, mass plays a vital role in the study of exoplanets, including their temperature structure, chemical composition, formation, and evolution. However, nearly a quarter of the known confirmed exoplanets lack measurements of their masses. This is particularly severe for those discovered via the radial-velocity (RV) technique, which alone could only yield the minimum mass of planets. In this study, we use published RV data combined with astrometric data from a cross-calibrated Hipparcos-Gaia Catalog of Accelerations (HGCA) to jointly constrain the masses of 115 RV-detected substellar companions, by conducting full orbital fits using the public tool \texttt{orvara}. Among them, 9 exoplanets with $M_{\rm p}\,{\rm sin}\,i<13.5\ M_{\rm Jup}$ are reclassified to the brown dwarf (BD) regime, and 16 BD candidates ($13.5\leqslant M_{\rm p}\,{\rm sin}\,i<80\,M_{\rm Jup}$) turn out to be low-mass M dwarfs. 
%We also verify and refine the nature of two unconfirmed exoplanet candidates: HD\,167677 b and HD\,89839 b. In addition, we report the discovery of two low-mass BDs: HD\,165131 b ($M_{\rm p}={18.7}_{-1.0}^{+1.4}\ M_{\rm Jup}$) and HD\,62364 b ($M_{\rm p}={17.46}_{-0.59}^{+0.62}\ M_{\rm Jup}$), based on the published European Southern Observatory (ESO) archive data.
We point out the presence of a transition in the BD regime as seen in the distributions of host star metallicity and orbital eccentricity with respect to planet masses. We confirm the previous findings that companions with masses below $42.5\ M_{\rm Jup}$ might primarily form in the protoplanetary disc through core accretion or disc gravitational instability, while those with masses above $42.5\ M_{\rm Jup}$ formed through the gravitational instability of molecular cloud like stars. Selection effects and detection biases which may affect our analysis to some extent, are discussed. 
%\ww{\sout{Further homogeneous and better quality data are required to check our results.}}
}

   \authorrunning{G.-Y. Xiao et al. }            %author_head in even pages
   \titlerunning{The Masses of Radial-Velocity Exoplanets}  % title_head in odd pages
   \maketitle

%________________________________________________ sections below
% 
\section{Introduction}           %% first-level sections will be auto-capitalized
\label{sect:intro}

Since the first exoplanet orbiting a solar-type star, 51 Pegasi b, was detected by the RV technique in 1995 \citep{Mayor1995}, the field of exoplanet science has achieved flourishing development. Up to now, more than 5000 exoplanets \citep{Akeson2013} have been discovered and confirmed through various methods, such as RV, transit, direct imaging, astrometry, and microlensing \citep{Deeg2018}. As one of the earliest adopted one, the RV technique is still active and developing rapidly with measuring precision dramatically improved from $\sim10 {\rm\ m\ s^{-1}}$ to $\sim0.1\ {\rm m\ s^{-1}}$ level in the past three decades, 
allowing detection of Earth-like planets around low mass stars~ \citep{Fischer2016}.

For a star with a companion, when it moves in a Keplerian orbit around the system's barycenter, its RV will change periodically owing to the perturbations induced by the companion. As a result, the velocity variation will cause the Doppler shift in the stellar spectrum \citep{Fischer2014,Fischer2016,Deeg2018}, which can be used to indicate the presence and infer the properties of the companions. According to Kepler's law, the RV semi-amplitude is proportional to the mass of the companion, the sine of orbital inclination, the reciprocal of the square root of the semi-major axis, and the total mass of the system (or the mass of primary for simplicity in case of small companion-to-main mass ratio). Therefore, those companions with massive mass and close-in orbit are much easier to be detected by the RV method. Among the exoplanets found by RV method, more than half can be regarded as Jupiter analogs orbiting around solar-type stars with semi-major axis within the snow line ($\lessapprox 3$\,AU~\citealt{Wittenmyer2016}). Fortunately, as the temporal baseline grows, more and more long-period giant planets beyond the ice line have been found recently, which provides new insight into their orbital properties and formation scenarios \citep{Marmier2013,Rickman2019,Kiefer2019,Dalal2021}.
%\ww{\sout{However, it is still challenge to detect wide-orbit terrestrial and  Neptunian planets, due to their host stars' activity and insufficient orbital coverage.}}

The RV precision is susceptible to chromospheric activity \citep{Fischer2014},  which may cause a false positive signal in some cases (e.g., GJ\,1151\,b; \citealt{Mahadevan2021}). In order to improve the reliability of detection, many RV surveys focused on main-sequence stars with relatively quiet chromosphere and abundant absorption lines (e.g., \citealt{Udry2000, Howard2012}). 
However, the most significant limitation of RV method is the so-called $M_{\rm p}\,{\rm sin}\,i$ degeneracy, where $i$ is the orbital inclination. It means that RV method can only measure a minimum mass instead of true mass, simply because the observed quantities are radial velocities instead of true velocities. In other words, the planet candidates with $M_{\rm p}\,{\rm sin}\,i<13.5\ M_{\rm Jup}$ has a non-ignorable probability of being brown dwarfs (13.5 $\sim$ 80 $M_{\rm Jup}$; \citealt{Burrows1997, Spiegel2011}) and even low-mass M dwarfs ($>$ 80 $M_{\rm Jup}$) if they have nearly face-on orbital configurations.
%\ww{OR: if they have quite small $i$}.

So far, among the $\sim$1000 RV planets, only about 150 of them have mass estimations, according to the \href{http://exoplanet.eu/catalog/}{exoplanet.eu} database \citep{Schneider2011}, leaving the mass and thus real nature of most RV-detected ``planets'' to be uncertain. For those companion with nearly edge-on orbits that transit may occur, the mass of the companions could be determined by joint fits of the RV curves and light curves, because the latter data can put strong constrain on $i$ and thus can break the $M_{\rm p}\,{\rm sin}\,i$ degeneracy \citep{Fischer2014,Deeg2018}. However, for those RV-detected companions with small $i$ and thus low transit probability, high-accuracy astrometry is superior to other methods in breaking the $M_{\rm p}\,{\rm sin}\,i$ degeneracy. Astrometric measurements represent the transverse component of the host star's displacement (or proper motion or acceleration), which can be used to reveal the 3D stellar reflex motion perturbed by unseen companions when combining RV measurements~\citep{Lindegren2003}. The amplitude of astrometric signal (or angular semi-major axis) is proportional to the companion-to-primary mass ratio $q$, the semi-major axis of the companion's orbit, and inversely, the distance to us. 
Therefore, the astrometry technique is particularly powerful in the detection of long-period and massive companions and their mass assessment~\citep{Huang2017,Deeg2018,Xu2017}.

%\ww{\sout{The pioneer of applying astrometry to constrain the mass of exoplanet candidates is the Fine Guidance Sensor (FGS), equipped on the Hubble Space Telescope (HST). HST/FGS has determined the  masses of several companions with submilliarcsecond precision, which permits the detection of the reflex astrometric motion of planet hosts. In particular,} 

The first  exoplanet with mass determined by astrometry is GJ\,876\,b by \citet{Benedict2002}, based on a joint analysis of the astrometric measurements from HST Fine Guidance Sensor (FGS) and archive RV data. Subsequently, with mass revised by the inclusion of HST astrometric data,
some systems were limited to be planets (e.g., $\upsilon$\,And\,d: \citealt{McArthur2010}, $\gamma$\,Cep Ab: \citealt{Benedict2018}, $\mu$\,Arae\,b, d, e: \citealt{Benedict2022}), 
and a few planet candidates were found to be BDs, even M dwarfs \citep{Bean2007,Benedict2010,Benedict2017}. 
Those multiple systems  with available measurements of mutual inclination can even allow for rigorous dynamical analysis (e.g., $\gamma$\,Cep Ab: \citealt{Huang2022}).
In addition Hipparcos intermediate astrometry data (IAD) \citep{Perryman1997,vanLeeuwen2007} have been widely used to yield the masses of the known RV-detected planetary systems. However, due to poor precision, only systems with companion's masses in the BD or M-dwarf regimes could be reliably characterized thanks to their relatively large astrometry amplitude (e.g., \citealt{Zucker2001,Sahlmann2011,Diaz2012,Wilson2016,Kiefer2019}). More recently, the early Gaia astrometric data were rapidly used to measure the mass and inclination of exoplanet candidates. For example, \cite{Kiefer2021} employed a tool named Gaia Astrometric noise Simulation To derived Orbit iNclination (\texttt{GASTON}; \citealt{Kiefer2019}) to assess the nature of hundreds of RV-detected exoplanets. They replaced 9 of them into the BD or low mass star domain and confirmed the presence of a void of BD populations below $\sim100$ days~\citep{Kiefer2019, Ma2014}. Very recently, the Gaia data release 3 (Gaia DR3) delivered astrometric orbital solutions for non-single stars for the first time~\citep{Holl2022}. Nine RV-detected planets with periods shorter than the baseline of Gaia (34 months) have been directly verified by astrometry alone in planetary-mass regime~\citep{GaiaCollaboration2022}. 
In the future, it is expected that Gaia will obtain orbital solutions of thousands of exoplanets and provide the most precise astrometry to characterize the nature of longer-period companions.

Recently, a new method that utilizes RV data with proper-motion data from both Hipparcos and Gaia has been developed independently by several groups \citep{ Feng2019,Feng2021,Venner2021,Brandt2021b,Kervella2022}. The positional differences between the Hipparcos and Gaia measurements with a $\sim25$\,yr span can offer new and precise proper motions of individual stars, the variation of which may indicate the presence of unseen companions. We note that Hipparcos and Gaia only provide one-epoch proper motion measurement for the entire mission baseline, instead of instantaneous proper motions. Given the systematic error between Hipparcos and Gaia DR2, \citet{Brandt2018} made a cross-calibration to correct for the underestimate of the nominal uncertainties and the rotation of the reference frames. Then he merged the two catalogs in one common frame and provided a Hipparcos-Gaia catalog of Accelerations (HGCA) to allow the measurement of acceleration. Subsequently, the HGCA of 
Gaia EDR3 version was also published \citep{Brandt2021}. The absolute astrometry from the HGCA has been proved reliable and has been applied to combine RVs and/or relative astrometry to break the $M_{\rm p}\,{\rm sin}\,i$ degeneracy. For example, \citet{Li2021} used the Keplerian orbital code \texttt{orvara} \citep{Brandt2021b}\footnote{\url{https://github.com/t-brandt/orvara}} to fit the published RV and HGCA astrometry (Gaia EDR3 version) of nine single and massive RV-detected exoplanets and obtained accurate determination of their masses.

In this study, we target to derive orbital solutions and masses of 115 RV-detected companions (113 systems) via the joint analysis of published RVs and astrometry data from the HGCA or direct imaging. We also use \texttt{orvara}, an open source orbit-fitting package using the parallel-tempering Markov Chain Monte Carlo (MCMC) sampler \texttt{ptemcee} \citep{Vousden2016} to sample the posterior distributions, to perform two-body fit for 102 systems and three-body fit for 11 systems. \texttt{orvara} is designed to fit full orbital parameters to any combination of RVs, relative and absolute astrometry, and has the capability of predicting the positions of interesting targets for follow-up imaging. Because the epoch astrometry data of Gaia is currently not published, \texttt{orvara} uses the intermediate astrometry fitting package \texttt{htof} \footnote{\url{https://github.com/gmbrandt/HTOF}} \citep{GMBrandt2021} to fit astrometry parameters (e.g., positions, proper motions and parallax) to Hipparcos IAD and synthetic Gaia data (see section \ref{sect:fit} for detail). Following \citet{Li2021}, We mainly analyze the systems with long-period, massive, and single RV-detected companion. However, there are some exceptions. Two multi-planetary systems, HD\,74156, and GJ\,832, were found hosting two planets, \citep{Segransan2010,Naef2004,Wittenmyer2014} and the inner one has a negligible effect on the host star’s proper motions. We thus treat them as two-body systems like most other companions. 

This paper is organised as follows. In Section \ref{sect:Sample}, we present the definitions and characteristics of our sample. The RV and astrometry data are described in Section \ref{sect:data}. The principle
of joint analysis is briefly demonstrated in Section \ref{sect:fit}. We summarise the results and provide the updated orbital parameters in Section \ref{sect:result}. Our discussions and conclusions are presented in Section \ref{sect:discussion} and Section \ref{sect:conclusion}, respectively.
% Authors can give a citation as `\citealt{Michel+etal+1992}'.
% You may also use \cite, \citep and \citet for citation, and use Table$\sim$1
% or Figure$\sim$1 and so forth. Using \ref and \label for cross-references of
% Tables/Figures is a good way in adjusting/adding/removing text, tables or
% figures.
\section{Sample Characteristics}
\label{sect:Sample}

In order to derive reliable mass and inclination of the RV-detected companions, we first set some criteria for sample selection. The targets of exoplanets are mainly selected from two online catalogs: the \href{http://exoplanet.eu/catalog/}{exoplanet.eu} \citep{Schneider2011} and NASA exoplanet archive \footnote{\url{https://exoplanetarchive.ipac.caltech.edu/index.html}} \citep{Akeson2013}. While for BDs and M dwarfs, we compile a sample from several high precision RV surveys, such as the CORALIE survey \citep{Queloz2000}, the High Accuracy Radial-velocity Planet Searcher (HARPS) survey \citep{Pepe2000}, the Anglo-Australian Planet Search (AAPS; \citealt{Tinney2001}), the California Planets Search (CPS; \citealt{Howard2010}) and so on. The basic selection criteria are as follows: 

1. The companions should be discovered by RV method, and the time coverage of corresponding RV monitoring should be more than 1000 days;

2. The systems should be or can be regarded as a single-planet system, or accompanied a wide stellar companion whose relative astrometry can be derived from Gaia or direct imaging;

3. The orbital period $P$ should be $>1000$ days, and the minimum mass $M_{\rm p}\,{\rm sin}\,i$ of companion should be $>1\ M_{\rm Jup}$ and $<500\ M_{\rm Jup}$;

4. The RV trend should be negligible unless additional astrometry for the third object is available;

5. The significance $\chi^2$ of accelerations from HGCA should be $>6$ as far as possible (see below).

Criteria 2 and 4 are set to ensure the acceleration variations of the host star can only attribute to the known companions. 
Three planetary systems, HD\,111232, HD\,204313 and HD\,73267, are recently refined as multi-planetary systems \citep{Feng2022, Diaz2016}. Therefore, even without additional astrometry data, we perform a 3-body fit for them. 
Since the time baseline of Gaia EDR3 is comparable to 1000 days, and the HGCA astrometry is sensitive to stars hosting massive and long-period companions, we utilize criteria 3  to filter short-period companions. However, we add 8 additionally companions, with orbital period slightly smaller than the critical value, but with relatively large $M_{\rm p}\,{\rm sin}\,i$ or acceleration $\chi^2$, to assess the quality of the inclinations derived by \texttt{orvara}. Some of them have the astrometrically determined mass and inclination in Gaia DR3 \citep{GaiaCollaboration2022}. Although the last criterion is not essential, \citet{Li2021} found that stars with significant astrometric accelerations would yield a much higher mass or a face-on inclination. We therefore set a relatively small threshold value to guarantee a reasonably complete sample. In addition, a small part of stars with $\chi^2 < 6$ that are randomly selected from literatures are included to test feasibility. Actually, we find that stars with relatively large $\chi^2$ can usually succeed in yielding an acceptable solution, but those stars with small $\chi^2$ can not. 

After applying the above criteria, 263 companions are initially obtained. The stellar spectral type and apparent $V$ magnitude are selected from Hipparcos photometry \citep{Perryman1997} or the SIMBAD database \citep{Wenger2000}, and the parallax can be found in Gaia EDR3 \citep{GaiaCollaboration2021}. Due to the high reliability of spectroscopic data, the stellar atmospheric parameters, including the effective temperature $T_{\rm eff}$, surface gravity ${\rm log}g$ and metallicity [Fe/H], are directly obtained from the corresponding RV surveys. For the stars with no given mass uncertainty, we use \texttt{isochrones} \citep{Morton2015} to globally fit stellar parameters and obtain the posterior distribution of stellar masses. The mass of stars will be used as a Gaussian prior in our joint orbital analysis. 

\begin{figure}
    \centering
    \subcaptionbox{$H\text{-}R$ diagram\label{subfig:HRD}}[.495\linewidth]{
        \includegraphics[width=72mm]{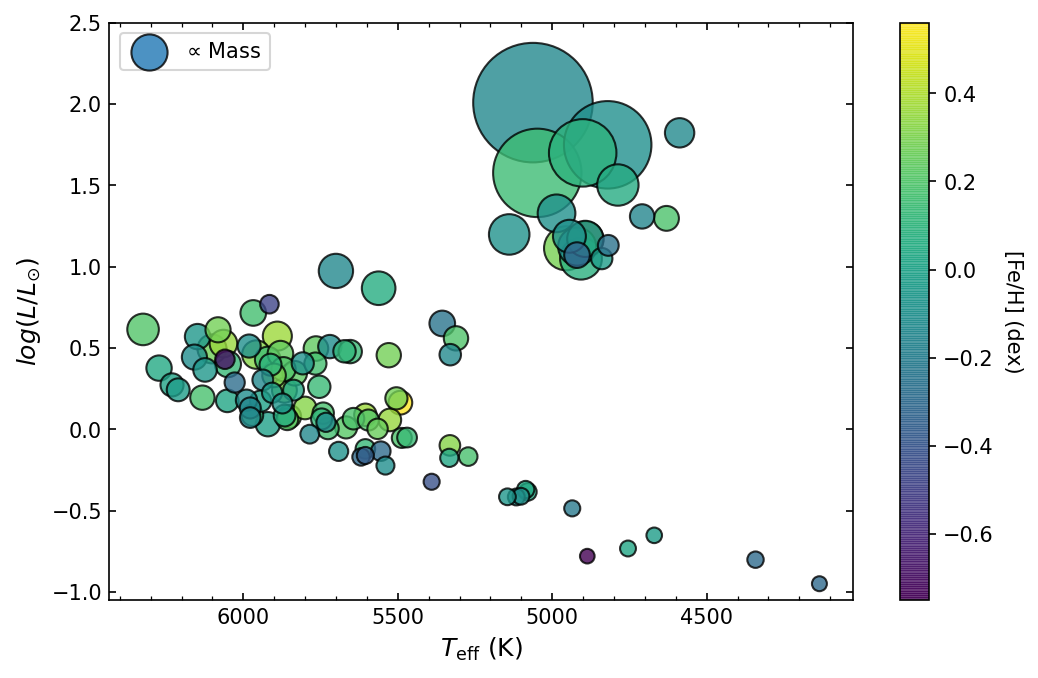}
    }
    \subcaptionbox{$M_{\rm p}\,{\rm sin}\,i\text{-}a$ diagram \label{subfig:msini_a}}[.495\linewidth]{
        \includegraphics[width=65mm]{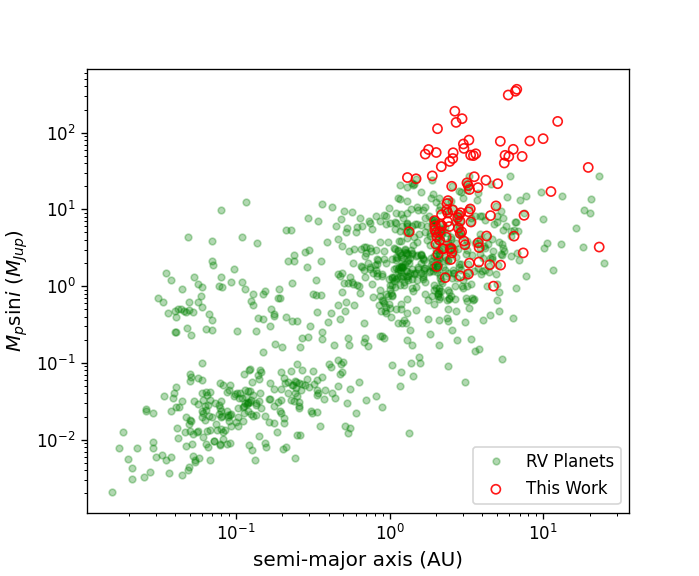}
    }
    \caption{\small (a) the Hertzsprung-Russell diagram of our 113 host stars color-coded by the stellar metallicity. The scatter size is proportional to the stellar mass; (b) the $M_{\rm p}\,{\rm sin}\,i\text{-}a$ distribution of our 115 RV-detected companions (red circles). The green circles represent all the RV-detected companions compiled from the \href{http://exoplanet.eu/catalog/}{exoplanet.eu} database.}
    \label{fig:samples}
\end{figure}

However, some companions do not get a reasonable solution with \texttt{orvara}. In fact, less than half of the companions can be accepted since we considered an additional criterion to further improve the reliability of orbital fit. For each case, the criterion is that the $1\sigma$ uncertainty of inclinations should be $<30\degr$ when the MCMC chains converge. Finally, our sample reduces to 115 companions (113 systems), including 65 planet candidates ($M_{\rm p}\,{\rm sin}\,i < 13.5\ M_{\rm Jup}$), 30 BD candidates ($13.5\leqslant M_{\rm p}\,{\rm sin}\,i < 80\ M_{\rm Jup}$) and 20 M dwarfs ($M_{\rm p}\,{\rm sin}\,i\geqslant 80\ M_{\rm Jup}$).
In Figure \ref{fig:samples}, we show the Hertzsprung-Russell ($H\text{-}R$) diagram and $M_{\rm p}\,{\rm sin}\,i\text{-}a$ diagram of 113 systems. Our sample spans spectral types F, G, K, and M with a median temperature of 5662\,K, some of which belong to evolved stars. Most stars have mass between 0.52 $M_{\tiny \sun}$ and 2.5 $M_{\tiny \sun}$ except HD\,175679 (2.7 $M_{\tiny \sun}$, \citealt{Wang2012}), GJ\,179 (0.357 $M_{\tiny \sun}$, \citealt{Howard2010}) and GJ\,832 (0.449 $M_{\tiny \sun}$, \citealt{Wittenmyer2014}), and have metallicity lower than 0.4 dex. Besides, Most stars are within the distance of about 100 pc and have apparent $V$ magnitudes ranging from 6 to 10. The detailed stellar parameters of 113 host stars can be found in the Appendix, Table \ref{Tab:stellar1}.

\section{Data}
\label{sect:data}

\subsection{Radial-Velocity Data}

As mentioned in section \ref{sect:Sample}, the RV time series we used for orbit fitting came from several long-term RV surveys based on high-resolution spectroscopy. In Table \ref{Tab:RV}, we list the details of instrument, observation count, time span, and mean RV uncertainty for each star. All the data can be found in corresponding literatures or the Vizier online catalogue \citep{Ochsenbein2000}. As we can see, 60\% of the targets have more than 30 data points, and 87\% of the stars have been monitored for at least 5 years. In addition, more than half of the stars are observed by multiple instruments. Considering the velocity offset between different instruments, the systematic zero-point was set as a free parameter in our joint orbital analysis. 
For some stars monitored by HIRES, the RV offset before and after the upgrade is negligible \citep{Tal-Or2019}. 
%In addition, HD\,211847 have few RV data points before and after CORALIE upgrade \citep{Sahlmann2011}, which may lead to incorrect orbital solutions due to the lack of sufficient phase coverage. We thus ignore its RV offset to ensure the accuracy of the result.

\begin{table}
\begin{center}
%\tabcolsep=0.25cm
%Derived Relative Astrometry Data from Gaia EDR3
\caption{Gaia EDR3 Absolute Astrometry of Companions
}\label{Tab:ASTEDR3}
%%Please Capitalize the First Letter of Each Notional Word in table's caption
\resizebox{\textwidth}{!}{
\begin{tabular}{lccccccc}
\hline \hline
 Name & Epoch& $\varpi$ &$\rho $ &PA &$\mu_{{\rm Gaia},\alpha \ast}$&$\mu_{{\rm Gaia},\delta}$&Corr \\
 & (yr) &(mas)&($\arcsec$) &(\degr)&(${\rm mas\ yr}^{-1}$)&(${\rm mas\ yr}^{-1}$)&\\
\hline

HD\,23596\,B&2016.0& $18.91\pm 0.22$ &$70.73 \pm 0.01$&$62.9 \pm 0.1$&$53.088 \pm 0.290$&$22.095 \pm 0.187$&0.054\\
HIP\,84056\,B&2016.0&$13.32\pm 0.02$ &$12.35 \pm 0.01$&$29.1 \pm 0.1$&$-10.944 \pm 0.029$&$-138.809 \pm 0.019$&$-0.084$\\
HD\,108341\,B&2016.0& $20.41\pm 0.01$&$7.82 \pm 0.01$&$7.5 \pm 0.1$&$-122.219 \pm 0.013$&$118.486 \pm 0.014$&0.035\\
HD\,142022\,B&2016.0& $29.20\pm 0.01$&$20.02 \pm 0.01$&$309.2 \pm 0.1$&$-339.651 \pm 0.041$&$-26.321 \pm 0.019$&$-0.168$\\
\hline
\end{tabular}
}
\end{center}
\tablecomments{\textwidth}{The errors of $\rho$ and PA are expanded to aid MCMC convergence \citep{Li2021}. The last column represents the correlation coefficient between R.A. and Dec. proper motion.}
\end{table}

\begin{table}
\centering
\caption{Compiled Relative Astrometry Data from Literatures}\label{Tab:ASTLI}
%%Please Capitalize the First Letter of Each Notional Word in table's caption
\begin{tabular}{lcccc}
\hline \hline
 Name & Epoch &$\rho $ &PA & Refs \\
 & (yr) &($\arcsec$) &(\degr)&\\
\hline
HD\,211847\,B&2015.4&$0.220 \pm 0.002$&$193.3 \pm 0.4$&\citet{Moutou2017}\\
HD\,43587\,B&2002.1&$0.5690 \pm 0.0025$&$15.86 \pm 0.28$&\citet{Catala2006}\\
&2004.7&$0.7200 \pm 0.0052$&$27.36 \pm 0.40$&\\
&2005.1&$0.7280 \pm 0.0043$&$28.62 \pm 0.51$&\\
HD\,5608\,B&2012.0&$0.627 \pm 0.009$&$58.9 \pm 0.4$&\citet{Ryu2016}\\
&2012.7&$0.627 \pm 0.022$&$59.9 \pm 1.0$&\\
&2014.8&$0.588 \pm 0.012$&$55.7 \pm 0.6$&\\
HD\,196050\,B&2000.5&$10.860 \pm 0.085$&$174.360 \pm 0.460$&\citet{Mugrauer2005}\\
&2003.5&$10.880 \pm 0.011$&$174.920 \pm 0.040$&\\
&2004.6&$10.875 \pm 0.011$&$174.872 \pm 0.040$&\\
HD\,126614\,B&2009.3&$0.4890 \pm 0.0019$&$56.1 \pm 0.3$&\citet{Howard2010}\\
&2011.0&$0.499 \pm 0.067$&$60.7 \pm 5.6$&\citet{Ginski2012}\\
&2015.0&$0.4861 \pm 0.0015$&$69.1 \pm 0.2$&\citet{Ngo2017}\\
&2015.5&$0.4853 \pm 0.0015$&$70.4 \pm 0.2$&\\
HD\,217786\,B&2011.6&$2.8105 \pm 0.0091$&$170.81 \pm 0.26$&\citet{Ginski2016}\\
&2013.5&$2.8327 \pm 0.0092$&$170.22 \pm 0.20$&\\
&2014.6&$2.8560 \pm 0.0069$&$170.34 \pm 0.16$&\\
\hline
\end{tabular}
\end{table}

\subsection{Absolute Astrometry Data}
Absolute astrometry of our host stars consists of parallax ($\varpi$), position ($\alpha,\ \delta$) and proper motion ($\mu_{\alpha},\ \mu_{\delta}$), including both Hipparcos and Gaia measurements. Since Hipparcos and Gaia astrometry have a temporal baseline of $\sim25$ years, the variation of proper motions might indicate the acceleration of a star which could be caused by an invisible companion. Furthermore, the difference in positions between two measurements can provide an additional proper motion. This third measurement of proper motion has been achieved by \citep{Brandt2021} and has been archived in the Hipparcos-Gaia Catalog of Accelerations (HGCA) \citep{Brandt2018, Brandt2021}. In HGCA, the reference frame of two satellites has been placed in a common inertial frame, and the systematic error has also been calibrated. The orbital fit tool \texttt{orvara} takes all three proper motions as observed values and compare them with model values to constrain the orbit of the host stars. Therefore, we directly use the absolute astrometry from HGCA (EDR3 version) to perform joint analysis with RV and relative astrometry.
\subsection{Relative Astrometry Data}
In our sample, seven systems (HD\,120066, HD\,142022\,A, HD\,108341, HIP\,8541, HD\,23596, HD\,213240, and HIP\,84056) have broad stellar or substellar companions measured in Gaia EDR3, but three of them (HD\,120066, HIP\,8541 and HD\,213240) have the projected separations of $\sim15372$ AU, $\sim2600$ AU and $\sim3898$ AU between the host star and the companion, respectively \citep{Blunt2019, Stassun2019, Mugrauer2005}. So we ignore the effect of the third star and regard them as 2-body systems when we perform orbital fits. As for the other four systems, we derived position angle east of north (PA) and projected separation ($\rho$) using the single epoch measurement of position in R.A and Dec. ($\alpha,\ \delta$) from Gaia EDR3. The derived relative astrometry data are listed in Table $\ref{Tab:ASTEDR3}$. 
For two companions (HD\,23596\,B and HD\,142022\,B) with $G$ magnitude below 13, we have inflated the error of their proper motions to account for the magnitude-dependent systematics characterized by \citet{Cantat-Gaudin2021}.

In addition, only two 2-body systems (HD\,43587 and HD\,211847) and four 3-body systems (HD\,5608, HD\,196050, HD\,126614\,A, and HD\,217786) with companion masses in the M-dwarf mass regime have additional direct imaging data in previous literatures.
HD\,5608 hosts a cool Jupiter-like planet HD\,5608\,b and a low-mass M dwarfs HD\,5608\,B \citep{Sato2012, Ryu2016}, and the latter one was imaged with the High Contrast Instrument for the Subaru Next Generation Adaptive Optics (HiCIAO; \citealt{Suzuki2010}) on the 8.2 m Subaru Telescope. HD\,217786 was found hosting a long-period brown dwarf and a wide stellar companion, HD\,217786\,B, which is located at a projected separation of $\sim150$ AU \citep{Ginski2016}. This system was imaged by the lucky imaging instrument at the Calar Alto 2.2 m telescope. HD\,126614\,A system contain a planet HD\,126614\,Ab with a minimum mass $M_{\rm p}\,{\rm sin}\,i=0.38\ M_{\rm Jup}$ \citep{Howard2010}, a faint M dwarf HD\,126614\,B separated from the primary star by $\sim33$ AU, and a second M dwarf NLTT\,37349 with a separation about 3070 AU which makes a negligible impact on our fit. HD\,196050 system was initially regarded as a binary system hosting a cold planet with $M_{\rm p}\,{\rm sin}\,i = 2.8\ M_{\rm Jup}$ \citep{Jones2002}, and the stellar companion HD\,196050\,B is 511 AU far away from the primary. However, \citet{Eggenberger2007} resolved HD\,196050\,B as a close pair of M dwarfs (HD\,196050\,Ba, Bb) with a separation of 20 AU by using the NACO facility \citep{Rousset2003} on the Very Large Telescope (VLT). For simplicity, we still regard HD\,196050 Ba and Bb as a single star. As a consequence, eight of our systems fulfill the requirements of 3-body fitting. The relative astrometry data we adopted can be found in Table $\ref{Tab:ASTLI}$. 

\section{ORBITAL FIT}
\label{sect:fit}
Since RV and astrometry can measure the orthogonal components of a star's motion, 
the combination of them makes it possible to determine the mass and inclination of RV-detected companions.
In this section, we will briefly describe the characteristics of the orbit fitting package \texttt{orvara} \citep{Brandt2021b} and some crucial equations used to perform full orbital analysis.
The \texttt{orvara} was designed to fit Keplerian orbits to any combination of radial velocity, relative and/or absolute astrometry data. It uses the built-in package \texttt{htof} \citep{GMBrandt2021} to parse the Intermediate Astrometry Data (IAD) of Hipparcos, and then constructs covariance matrices to yield best-fit positions and proper motions of a star relative to the barycenter. Since the Gaia epoch astrometry or the along-scan residuals have not been released in the EDR3 and DR3, \texttt{htof} tentatively uses the synthetic data from Gaia Observation Forecast Tool \footnote{\url{https://gaia.esac.esa.int/gost/index.jsp}} (GOST) that contains the predicted observation time and scan angles to fit 5-, 7-, and 9-parameter astrometric model.

A Keplerian orbit can be fully described by six parameters: the semi-major axis $a$, the eccentricity $e$, the orbital inclination $i$, the longitude of the ascending node $\Omega$, the argument of pericenter $\omega$, and the time of periastron passage $T_{p}$. In addition, given the total mass of the system, we can derive the orbital period $P$ through Kepler’s third law. In an inertial frame, when a celestial body is moving in an elliptical orbit, the true anomaly, $\nu(t)$, is related to the eccentric anomaly, $E(t)$, which is given by 
\begin{equation}
    {\rm tan}\frac{\nu(t)}{2}=\sqrt{\frac{1+e}{1-e}}\cdot {\rm tan}\frac{E(t)}{2},
\end{equation}
where $e$ is the eccentricity \citep{Perryman2011}. This relation can be derived geometrically.
The mean anomaly $M(t)$ at a specific time is then defined as
\begin{equation}
    M(t)=\frac{2\pi}{P}(t-T_{p}),
\end{equation}
where $P$ is the orbital period and has the same units as time $t$, $T_{p}$ is the epoch of periastron passage. According to Kepler’s equation, the relation between The mean anomaly $M(t)$ and the eccentric anomaly $E(t)$ is given by
\begin{equation}
    M(t)=E(t)-e\,{\rm sin}\,E(t).
\end{equation}
This transcendental equation can be solved iterative but inefficient. So \citet{Brandt2021b} developed a more efficient eccentric anomaly solver for \texttt{orvara} based on the approach of \citet{Raposo2017}.
The radial velocity is given by
\begin{equation}
    {\rm RV} = K\,[{\rm cos}(\omega+\nu)+e\,{\rm cos}(\omega)],
\end{equation}
where $K$ is the radial velocity semi-amplitude, which is given by
\begin{equation}
    K\equiv\frac{2\pi}{P}\frac{a_{\star}{\rm sin}\,i}{\sqrt{1-e^2}}.
\end{equation}
In Equation (5), $i$ is the orbital inclination, $a_{\star}$ is the semi-major axis of the primary star relative to the system's barycenter. Then, Kepler’s third law can be written as
\begin{equation}
    \frac{a_{\rm rel}^3}{P^2} = M_{\star} + M_{\rm p}
\end{equation}
\begin{equation}
    a_{\rm rel}=a_{\star}+a_{p}
\end{equation}
\begin{equation}
    \frac{a_{\star}}{a_{p}}=\frac{M_{\rm p}}{M_{\star}},
\end{equation}
where $a_{\rm rel}$ is the semi-major axis of the secondary relative to the primary star in units of AU, $a_{p}$ is the semi-major axis of the secondary star relative to the system's barycenter, $P$ is the period in units of year, $M_{\star}$ and $M_{\rm p}$ are masses (in units of solar mass) of the primary and secondary stars, respectively.

In rectangular coordinates, the Thiele-Innes coefficients $A$, $B$, $F$, $G$ \citep{Thiele1883, Binnendijk1960, Heintz1978} are defined as
\begin{equation}
    A=a\,({\rm cos}\,\omega\,{\rm cos}\,\Omega -{\rm sin}\,\omega\,{\rm sin}\,\Omega\,{\rm cos}\,i)
\end{equation}
\begin{equation}
    B=a\,({\rm cos}\,\omega\,{\rm sin}\,\Omega +{\rm sin}\,\omega\,{\rm cos}\,\Omega\,{\rm cos}\,i)
\end{equation}
\begin{equation}
    F=a\,(-{\rm sin}\,\omega\,{\rm cos}\,\Omega -{\rm cos}\,\omega\,{\rm sin}\,\Omega\,{\rm cos}\,i)
\end{equation}
\begin{equation}
    G=a\,(-{\rm sin}\,\omega\,{\rm sin}\,\Omega +{\rm cos}\,\omega\,{\rm cos}\,\Omega\,{\rm cos}\,i),
\end{equation}
where $a$ is the semi-major axis in angular units and can be written as $a=a_{\rm rel} \cdot \varpi$ 
in a relative orbit. $\varpi$ is the parallax in units of mas.
And the elliptical rectangular coordinates $X$ and $Y$ are functions of 
eccentric anomaly $E(t)$ and eccentricity $e$, which are given by
\begin{equation}
    X={\rm cos}\,E(t)-e
\end{equation}
\begin{equation}
    Y=\sqrt{1-e^2}\cdot {\rm sin}\,E(t).
\end{equation}
The projected offsets in the plane of the sky between the secondary and primary star are then given by
\begin{equation}
    \Delta\delta=AX+FY
\end{equation}
\begin{equation}
    \Delta\alpha\ast=BX+GY,
\end{equation}
where $\Delta\delta$  and $\Delta\alpha\ast=\Delta\alpha\,\rm{cos}\,\delta$ are the offset in declination and right ascension, respectively.  
Combined with Equation (7) and (8), the projected offsets of the primary star relative to the system's barycenter can be written as
\begin{equation}
    \Delta\alpha \ast_{\star}=(-\frac{M_{\rm p}}{M_{\star}+M_{\rm p}})\Delta\alpha \ast
\end{equation}
\begin{equation}
    \Delta\delta_{\star}=(-\frac{M_{\rm p}}{M_{\star}+M_{\rm p}})\Delta\delta.
\end{equation}
After getting the offset from Equation (17) and (18), which means a group of synthetic position time series was generated, \texttt{htof} will use singular value decomposition to solve for the best-fit astrometric parameters (e.g., parallax, positions, proper motions, acceleration, and jerk terms). Comparing with the values from HGCA, \texttt{orvara} then may find the best-fit orbital parameters with the combination of RV data. 

The basic astrometric model that \texttt{htof} adopted is \citep{GMBrandt2021}
\begin{equation}
    \alpha_{m,i}=\varpi f_{\varpi}[t_{i}]+\sum_{n=0}^{N} \frac{a_{n}}{n!}(t_{i}-t_{ref})^n
\end{equation}
\begin{equation}
    \delta_{m,i}=\varpi g_{\varpi}[t_{i}]+\sum_{n=0}^{N} \frac{b_{n}}{n!}(t_{i}-t_{ref})^n.
\end{equation}
The left-hand side is the model value in right ascension and declination, and $f_{\varpi}[t_{i}]$, $g_{\varpi}[t_{i}]$ are the parallax factors \citep{vanLeeuwen2007b}. $N$ is the fitting degree (e.g., $N=1$ represents a 5-parameter fitting, $N=2$ represents a 7-parameter fitting and $N=3$ represents a 9-parameter fitting), and $a_{n}$, $b_{n}$ are the astrometric parameters (e.g., $a_{1}$ represents proper motion, $a_{2}$ represents acceleration and $a_{3}$ represents jerk). 

In this study, we mainly use a 5-parameter model to fit the Hipparcos IAD and the Gaia GOST data for each star because of the lack of additional acceleration or jerk data.
However, the recent Gaia DR3 \footnote{\url{https://gea.esac.esa.int/archive/}} provides four non-single star (NSS) tables for the first time \citep{Holl2022, GaiaCollaboration2022}. One of the tables \texttt{gaiadr3.nss\_acceleration\_astro} uses NSS astrometric models (e.g., 7-parameter or 9-parameter model) for those stars whose proper motion is more compatible with an acceleration solution. As a result, the astrometric parameters in this table are slightly different from the main catalog or EDR3. In addition, \texttt{nss\_acceleration\_astro} also provides acceleration or jerk data. We found that six of our stars are contained in this table with \texttt{nss\_solution\_type=Acceleration7} and five stars have a solution type of \texttt{Acceleration9}. Therefore, we approximately use the newly released values to replace the corresponding values of HGCA and supplement the acceleration or jerk data. \texttt{orvara} will decide on which models to adopt to parse the Gaia GOST data. 
But only five systems (HD\,145428, HD\,154697, HD\,92987, HD\,156728 and HD\,87899) can converge well, and we therefore still use 5-parameter model for the rest.

For most of the 2-body systems in our sample, \texttt{orvara} adopts \texttt{ptemcee} to fit the nine parameters, including the primary star mass $M_{\star}$, the secondary star mass $M_{\rm sec}$, semi-major axis $a$, $\sqrt{e}\ {\rm sin}\ \omega$, $\sqrt{e}\ {\rm cos}\ \omega$, inclination $i$, ascending node $\Omega$, mean longitude $\lambda_{\rm ref}$ at a reference epoch (2010.0 yr or ${\rm JD}=2455197.50$) and RV jitter $\sigma_{\rm jit}$ (depends on the number of instruments). As for 3-body systems, additional six orbital elements and the mass of the third companion are required. For simplicity, \texttt{orvara} will ignore the interaction between companions and approximate star's motion by a superposition of each Keplerian orbit. Strictly speaking, this method is not entirely correct compared to N-body simulations, and has limited implication in multi-planetary systems. Besides, some nuisance parameters, such as RV zero point, parallax, and proper motion of system's barycenter, are marginalized by \texttt{orvara} in order to reduce computational costs \citep{Brandt2021b}. 

\begin{table}[h]
\centering
\caption{Basic Parameters and Adopted Priors}\label{Tab:prior}
%%Please Capitalize the First Letter of Each Notional Word in table's caption
\begin{tabular}{lc}
\hline \hline
 Parameter & Prior \\
\hline 
 ${\rm RV\ Jitter}\ \sigma_{\rm jit}$ &   $1/\sigma_{\rm jit}$  \text{(log-uniform)} \\
 \text{Primary Mass $M_{\star}$}  &   exp[$-\frac{1}{2}(M-M_{\rm prior})^2/\sigma_{M, {\rm prior}}^{2}$]\\
 \text{Secondary Mass $M_{\rm sec}$}  &   $1/M$ \text{(log-uniform)}\\
 \text{Semi-major axis $a$}  &   $1/a$ \text{(log-uniform)}\\
 $\sqrt{e}\ {\rm sin}\ \omega$  &   uniform\\
 $\sqrt{e}\ {\rm cos}\ \omega$  &   uniform\\
\text{Inclination $i$}  &   sin($i$), $0\degr<i<180\degr$ (geometric)\\ 
\text{Mean longitude at 2010.0 $\lambda_{\rm ref}$}  &   uniform\\
\text{Acending node $\Omega$}  &   uniform\\
\text{Parallax $\varpi$}  &   exp[$-\frac{1}{2}(\varpi-\varpi_{\rm Gaia})^2/\sigma_{\varpi, {\rm Gaia}}^{2}$]\\
\hline
\hline
\end{tabular}
\end{table}

Gaussian priors are used for the mass of the primary star, and default priors (i.e., log-uniform, uniform, geometric) are used for the rest of the fitting parameters (see Table \ref{Tab:prior}). For the purpose of quick convergence, we use the RV-only orbital parameters from literatures as initial guesses. We believe those parameters are sufficiently reliable and accurate. Then we run MCMC sampling twice for each system. Firstly, we use 6 temperatures, 100 walkers, and 50,000 steps per chain to generate posterior distributions of all fitted parameters. The best-fit parameters derived by the maximum a posterior (MAP) method from MCMC chains are saved as the initial values for the next sampling. Secondly, we set 20 temperatures and run MCMC sampling again. At last, 
we visually inspect the convergence of each MCMC chain. Those that fail to converge or satisfy our additional criteria (see Section \ref{sect:Sample}) are excluded from our sample. We discard the first 25,000 steps as burn-in in each convergent case and post-process the rest. 
By default, \texttt{orvara} chooses the median value from posterior distributions as the best-fit parameters and selects the 1$\sigma$ quantiles (the 16\% and 84\% quantiles)  as the uncertainties.

\section{Results}
\label{sect:result}

\begin{figure}
   \centering
   \includegraphics[width=\textwidth, angle=0]{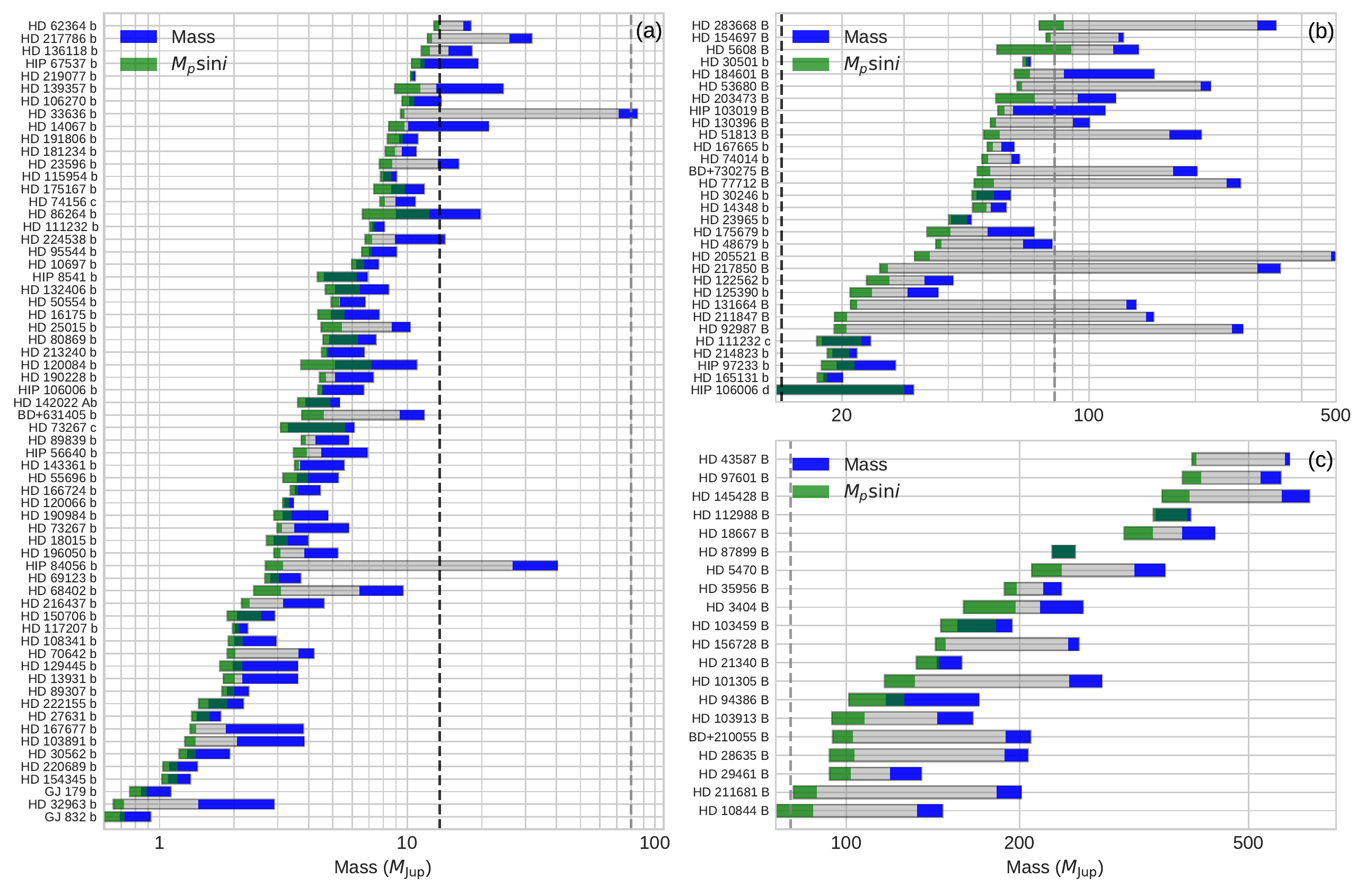}
   \caption{\small The comparison of masses and minimum masses. (a) 64 planet candidates with $M_{\rm p}\,{\rm sin}\,i < 13.5\ M_{\rm Jup}$. (b) 31 BD candidates with $13.5\leqslant M_{\rm p}\,{\rm sin}\,i < 80\ M_{\rm Jup}$. (c) 20 M dwarfs with $M_{\rm p}\,{\rm sin}\,i\geqslant 80 \ M_{\rm Jup}$. The green and blue rectangles represent the $1\sigma$ minimum mass and mass of each companion, respectively. The vertical black and grey dashed lines indicate the classical boundaries of BD ($\sim13.5$ and $\sim80$ $M_{\rm Jup}$, \citealt{Burrows1997, Spiegel2011}).} 
   \label{Fig:results}
\end{figure}

In this section, we present the general results of our orbital fits and demonstrate the details of some interesting cases. In Table \ref{Tab:orbparas}, we list nine best-fit orbital parameters from the posterior distributions of \texttt{orvara} and one parameter, $M_{\rm p}\,{\rm sin}\,i$, inferred from the values of $M_{\rm p}$ and $i$. Among our 115 companions, most of our fitted parameters and derived $M_{\rm p}\,{\rm sin}\,i$ agree well with the RV-only literature values within 1$\sigma$. The distributions of inclination for individual companions are bimodal, as the current HGCA astrometry can't distinguish whether the companion is in prograde ($0\leqslant i_{1}\leqslant90\degr$) or retrograde ($i_{2}=180\degr-i_{1}$) orbit \citep{Kervella2020, Li2021}. But for the wide stellar companions in our 3-body fits, the inclination have a single certain value due to the additional relative astrometry. Figure \ref{Fig:results} summarizes the mass and the minimum mass $M_{\rm p}\,{\rm sin}\,i$ measured with \texttt{orvara}. A total of 115 companions, including 55 planets, 24 BDs, and 36 M dwarfs, are finally presented in Table \ref{Tab:Statistics}. For two 3-body systems, HD\,5608 and HD\,126614\,A, we only provide the masses of the outer stellar binaries in Table \ref{Tab:orbparas}, since the masses of the inner planets seem to be less plausible. 
We can see that most companions with $M_{\rm p}\,{\rm sin}\,i<13.5\ M_{\rm Jup}$ remain in the planet mass domain (i.e. prefer edge-on inclination), and nearly half of the BD candidates with $13.5\leqslant M_{\rm p}\,{\rm sin}\,i\leqslant 80\ M_{\rm Jup}$ should be classified as low-mass M dwarfs. This trend may imply that the BD desert is more barren than previous researchers. In addition, 9 out of 115 companions are found to have an extremely face-on orbit ($i<10\degr\ {\rm or}\ i>170\degr$). For each stellar name, we add the lowercase letters (e.g., ``b'', ``c'' and ``d'') to refer to those companions with masses below the hydrogen burning limit, and add the capital letter (e.g., ``B'') to indicate low-mass M dwarfs.

\begin{table}
\centering
\caption{The Statistics of Three Type Companions}\label{Tab:Statistics}
%%Please Capitalize the First Letter of Each Notional Word in table's caption
\begin{tabular}{cccc}
\hline \hline
 Type & Giant planet & Brown Dwarf & M Dwarf \\
 &$M<13.5\ M_{\rm Jup}$&$13.5\leqslant M<80\ M_{\rm Jup}$&$80\ M_{\rm Jup}\leqslant M$\\
\hline 
$M_{\rm p}\,{\rm sin}\,i$ & 64 &31 &20 \\
$M_{\rm p}$&55&24&36\\
\hline
\hline
\end{tabular}
\end{table}

\subsection{Two Confirmed Planets}
\begin{figure}
    \centering
    \subcaptionbox{HD\,167677\label{subfig:HD167677b_gls}}[.495\linewidth]{
        \includegraphics[width=72mm]{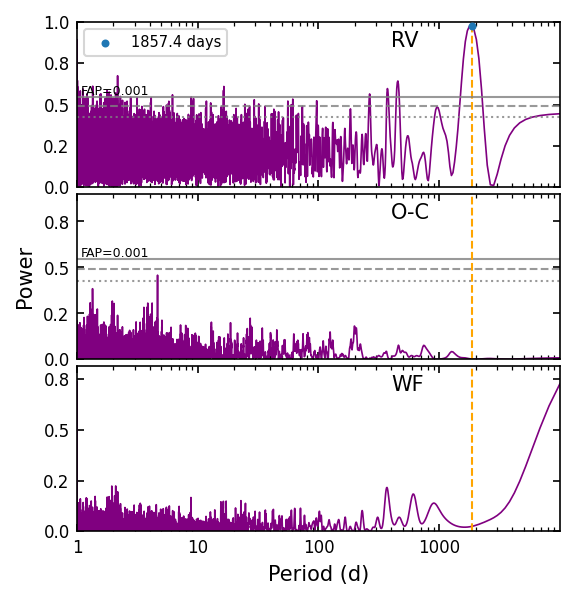}
    }
    \subcaptionbox{HD\,89839\label{subfig:HD89839b_gls}}[.495\linewidth]{
        \includegraphics[width=72mm]{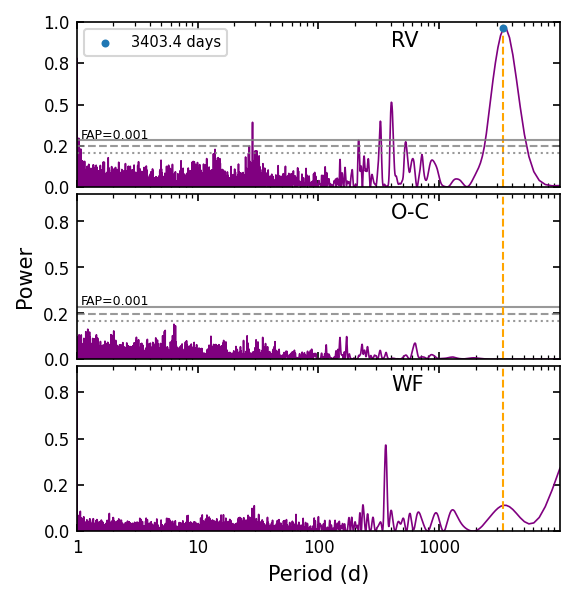}
    }
    \caption{\small Generalized Lomb-Scargle (GLS) periodograms for two stars. Top panel: the GLS periodograms of the observed RVs. Middle panel:  the residuals to single Keplerian orbital fit (after subtracting the planet solution). Bottom panel: window function of sampling. The horizontal grey lines indicate the 0.001, 0.01, 0.1 False Alarm
Probability (FAP) levels. The vertical orange dashed line represents the period of planet signal.}
    \label{fig:periodogram}
\end{figure}

\begin{figure}
   \centering
   \includegraphics[width=\textwidth, angle=0]{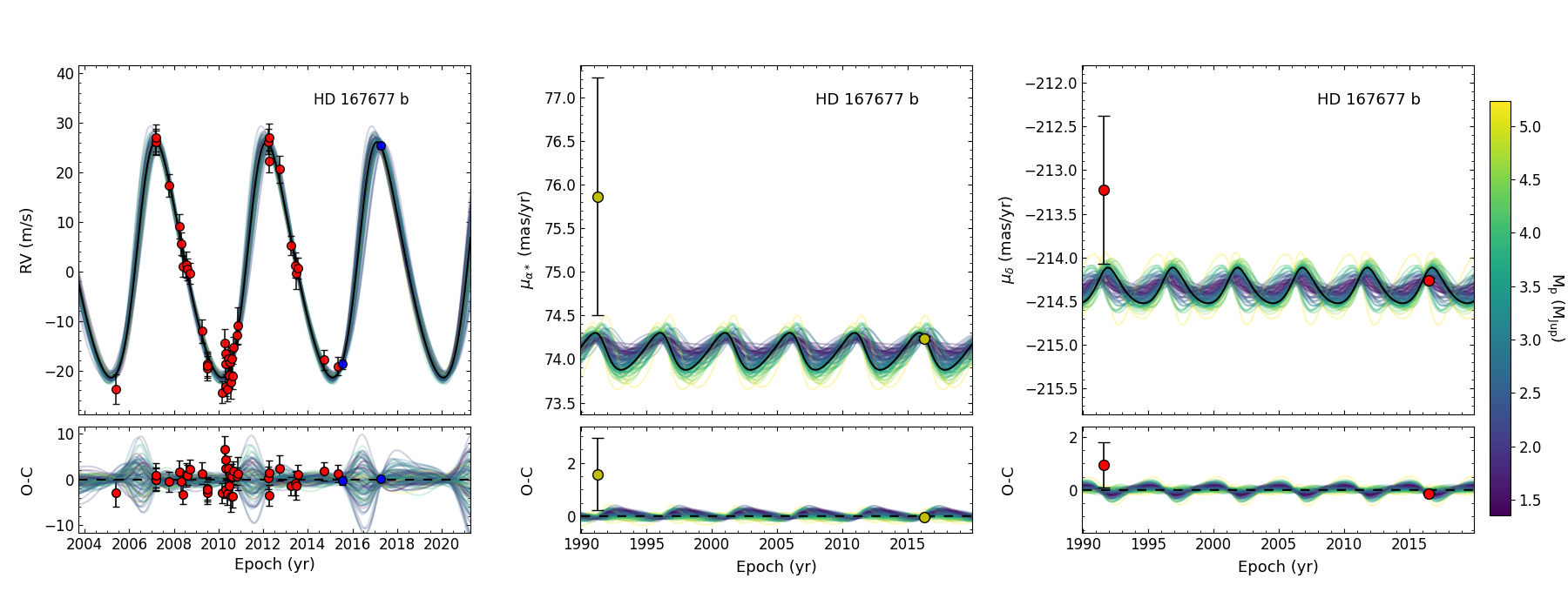}
   \caption{\small Left panel: radial velocity curve of HD\,167677\,b. The red and blue points represent the pre-upgrade and post-upgrade HARPS data reanalysed by \citet{Trifonov2020}. Middle and right panel: astrometric acceleration in right ascension and declination. The points near epoch 1991 are measured from Hipparcos, and the points near epoch 2016 are from Gaia EDR3. The black lines represent the best-fit orbit, and the colored lines, color-coded by the companion's mass, indicate the possible orbital solution randomly drawn from the MCMC chain. All figures are post-processed with \texttt{orvara}.} 
   \label{Fig:HD167677rv_pm}
\end{figure}
\begin{figure}
   \centering
   \includegraphics[width=\textwidth, angle=0]{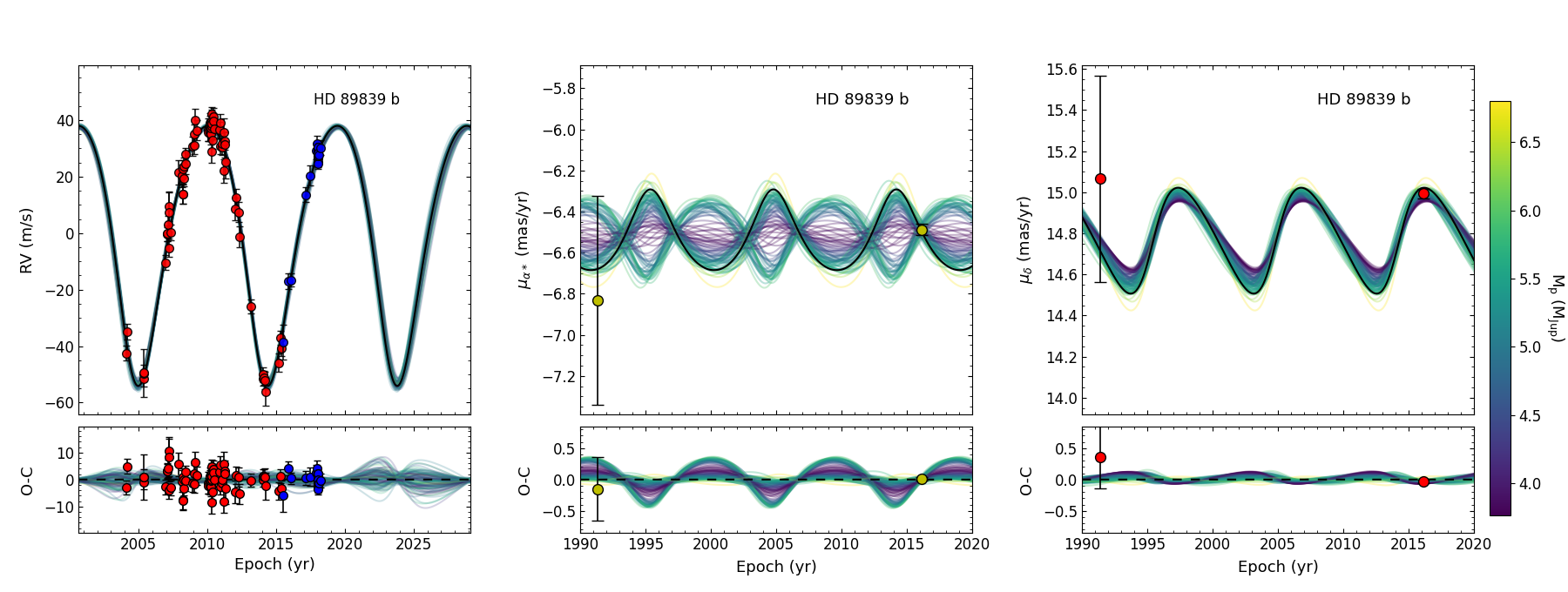}
   \caption{\small Radial velocity curve and astrometric acceleration of HD\,89839 b. Same as Figure \ref{Fig:HD167677rv_pm}} 
   \label{Fig:HD89839rv_pm}
\end{figure}

We find two planets, HD\,167677\,b and HD\,89839\,b,  are labelled as ``Unconfirmed'' in the \href{http://exoplanet.eu/catalog/}{exoplanet.eu} database \citep{Schneider2011}. Both planets are initially reported by \citet{Moutou2011} based on the measurements of the HARPS survey. The details of two systems are as follows:

$-$ HD\,167677\,b. The host HD\,167677, located at a distance of 58 pc away from our solar system, is a chromospherically quiet star (${\rm log}R_{\rm HK}^{'}=-4.99$) with apparent $V$ magnitude of 7.90. Using 26 HARPS measurements, \citet{Moutou2011} obtained a best-fit Keplerian solution of $P=1814\pm100\ {\rm days}$, $e=0.17\pm0.07$, $a=2.9\pm0.12\ {\rm AU}$ and $M_{\rm p}\,{\rm sin}\,i=1.36\pm0.12\ M_{\rm Jup}$. However, almost all of their RV data points centered on one half of the orbital phase, while only one point is on the other half, making it difficult to convince someone else. Fortunately, additional data have been collected by the follow-up HARPS survey and published by \citet{Trifonov2020}. The newly published RV data are more precise than previous ones. Therefore, we compile 42 RV data points and perform the joint fits with the absolute astrometry from HGCA. We obtain a relatively tight orbital solution for HD\,167677 b, $M_{\rm p}={2.85}_{-1.00}^{+0.95}\ M_{\rm Jup}$, $a={2.877}_{-0.025}^{+0.025}\ {\rm AU}$, $e= {0.182}_{-0.026}^{+0.031}$, $P={1815}_{-14}^{+15}\ {\rm days}$, and $i= {28.7}_{-7.5}^{+19}\degr$ (or ${151.3}_{-19}^{+7.5}\degr$). Our solution is in perfect agreement with \citet{Moutou2011} within 1$\sigma$ but more precise. In Figure \ref{subfig:HD167677b_gls}, we plot the generalized Lomb-Scargle periodogram (GLS) of HD\,167677. A significant signal near 1857.4 days can be found. In addition, to evaluate the effect of line profile asymmetry, we calculate the correlation coefficient between Bisector inverse span (BIS; \citealt{Dall2006}) and RVs. The value of $r=0.22$ suggests that BIS has a weak correlation with RVs. Similarly, the $H_{\alpha}$ index also shows no correlation with RVs ($r=-0.1$), which rules out the effect of stellar activities. We thus confirm and refine the orbit of HD\,167677\,b. The RV orbit and astrometric acceleration are plotted in Figure \ref{Fig:HD167677rv_pm}.

$-$ HD\,89839\,b. The primary is an F7V type star at a distance of 57 pc. The low ${\rm log}R_{\rm HK}^{'}\ (=-4.97)$ indicates itself as a quiet chromosphere. \citet{Dommanget2002} identified HD\,89839 as a double star in the Catalogue of the Components of Double and Multiple stars (CCDM). However, according to Gaia EDR3, we find the stellar companion who was thought to be at a separation of $10\arcsec$ from the primary should be a background star, with a distance of $\sim500$ pc away from our solar system. Based on 39 RVs, \citet{Moutou2011} revealed $P={6601}_{-3570}^{+4141}$ days, $e=0.32\pm0.2$, $a={6.8}_{-2.4}^{+3.3}$ AU and $M_{\rm p}\,{\rm sin}\,i$ = $3.9\pm0.4\ M_{\rm Jup}$ for HD\,89839\,b. The poor constraint is caused by incomplete orbital phase coverage. In this study, we select 91 measurements from the public HARPS RV database \citep{Trifonov2020}. In Figure \ref{subfig:HD89839b_gls}, the GLS periodogram of HD\,89839 is plotted. Our best-fit solution has $P={3440}_{-21}^{+22}\ {\rm days}$, $e={0.186}_{-0.013}^{+0.013}$, $a={4.76}_{-0.044}^{+0.044}\ {\rm AU}$ and $M_{\rm p}\,{\rm sin}\,i={3.808}_{-0.076}^{+0.077}\ M_{\rm Jup}$, and two extra parameters $M_{\rm p}={5.03}_{-0.75}^{+0.79}\ M_{\rm Jup}$ and $i= {49.2}_{-8.2}^{+14}\degr$ (or ${130.8}_{-14}^{+8.2}\degr$). The plots of RV orbit and astrometric acceleration can be found in Figure \ref{Fig:HD89839rv_pm}. In addition, the BIS and $H_{\alpha}$ index have no correlation with RVs ($r=0.14$ and $r=-0.15$, respectively). We therefore confirm the nature of HD\,89839 b and refine its orbit.

\subsection{Two Low-mass and Highly Eccentric BDs}
\begin{figure}
    \centering
    \subcaptionbox{HD\,165131\label{subfig:HD165131b_gls}}[.495\linewidth]{
        \includegraphics[width=72mm]{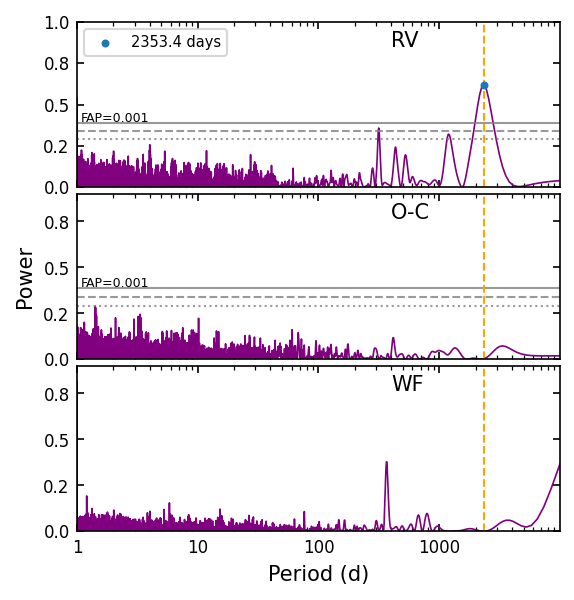}
    }
    \subcaptionbox{HD\,62364\label{subfig:HD62364b_gls}}[.495\linewidth]{
        \includegraphics[width=72mm]{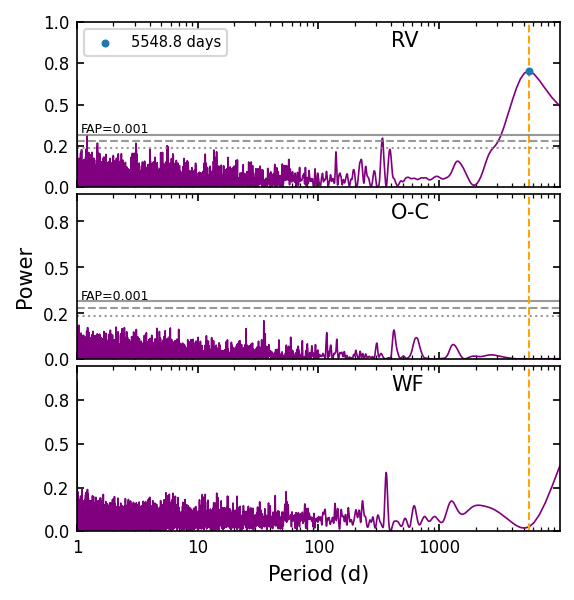}
    }
    \caption{\small Generalized Lomb-Scargle (GLS) periodograms for two stars. Same as Figure \ref{fig:periodogram}.}
    \label{fig:periodogram_2}
\end{figure}

\begin{figure}
   \centering
   \includegraphics[width=\textwidth, angle=0]{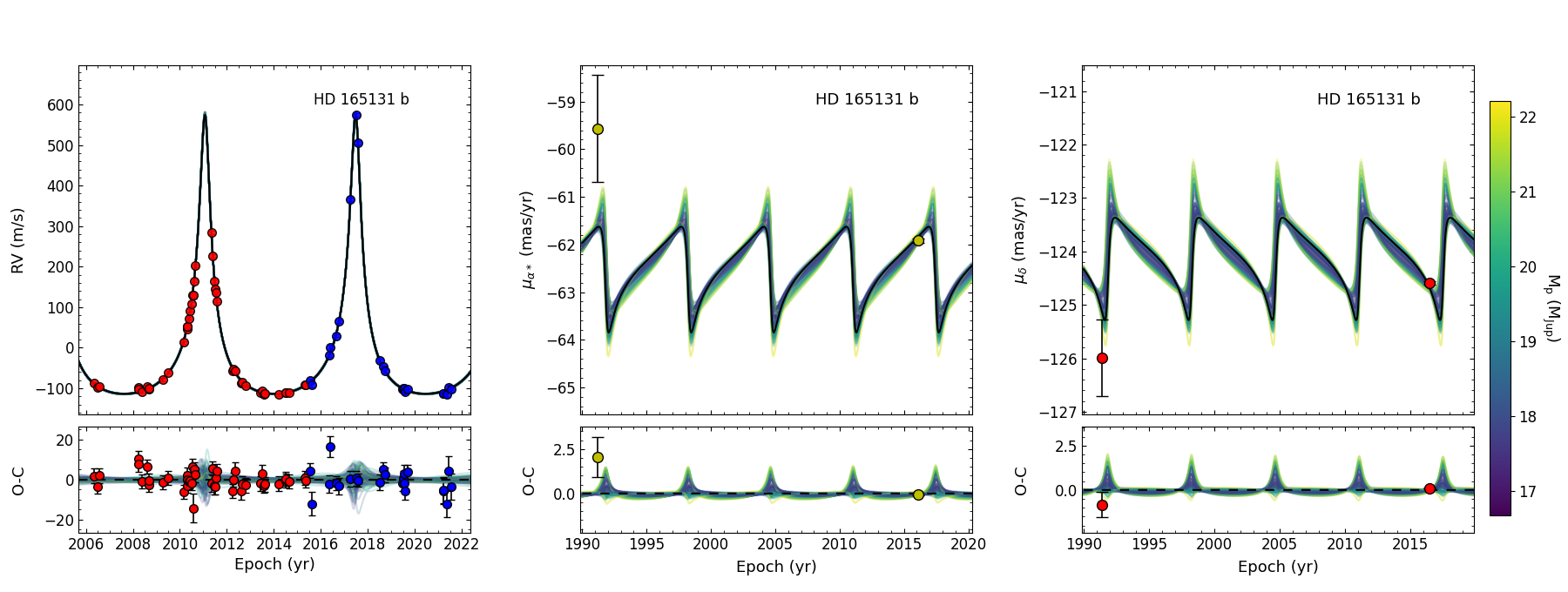}
   \caption{\small Left panel: radial velocity curve of HD\,165131\,b. The red points represent the pre-upgrade HARPS data reanalysed by \citet{Trifonov2020} and the blue points represent the post-upgrade RVs compiled from ESO archive data. Other symbols are the same as Figure \ref{Fig:HD167677rv_pm}.} 
   \label{Fig:HD165131rv_pm}
\end{figure}

We report the discovery of a new low-mass BD, HD\,165131\,b, and the parameter refining of a previously known BD, HD\,62364\,b, based on the published HARPS RV database \citep{Trifonov2020} and the ESO archive data \footnote{Based on data obtained from the ESO Science Archive Facility with DOI(s): \url{https://doi.org/10.18727/archive/33}.}. Both host stars show no significant emission in the core of Ca \uppercase\expandafter{\romannumeral2} HK line and thus belong to chromospheric quiet stars. The details of the two systems are as follows: 

$-$ HD\,165131\,b. The primary is a G3/5V main-sequence star with $T_{\rm eff}=5870\ {\rm K}$, ${\rm log }g=4.39\ {\rm cgs}$ and ${\rm [Fe/H]}=0.06\ {\rm dex}$ \citep{CostaSilva2020}. We then use \texttt{isochrones} package to derive $M_{\star}={1.06}_{-0.05}^{+0.05}\ M_{\tiny \sun}$, $R_{\star}={1.08}_{-0.01}^{+0.01}\ R_{\tiny \sun}$ and an age of ${4.04}_{-1.96}^{+1.88}$ Gyr for HD\,165131. For those 44 pre-upgrade HARPS RVs are taken from \citet{Trifonov2020} and 23 post-upgrade RVs are directly collected from ESO archive data. They show significant RV variations with an amplitude of ~300 ${\rm m\ s^{-1}}$ and a period near 2353 days (see Figure \ref{subfig:HD165131b_gls}). No correlations are found between RVs and BIS ($r=0.13$), and $H_{\alpha}$ index ($r=0.03$), which rule out the effect of stellar activity. Additionally, no outer stellar companions are found in the CCDM \citep{Dommanget2002} or in the Washington Double Star Catalog (WDS) \citep{Mason2001}. Our best-fit solution has $P={2342.6}_{-1.3}^{+1.3}\ {\rm days}$, $e={0.6708}_{-0.019}^{+0.019}$, $a={3.54}_{-0.054}^{+0.054}\ {\rm AU}$ and $M_{\rm p}\,{\rm sin}\,i={17.56}_{-0.54}^{+0.55}\ M_{\rm Jup}$, and two extra parameters $M_{\rm p}={18.7}_{-1.0}^{+1.4}\ M_{\rm Jup}$ and $i= {70}_{-8.5}^{+12}\degr$ (or ${110}_{-12}^{+8.5}\degr$). 
No additional long-term linear trend were found. The RV orbit and astrometric acceleration are plotted in Figure \ref{Fig:HD165131rv_pm}. We therefore confirm the discovery of this low-mass and highly eccentric BD.

\begin{figure}[h!]
   \centering
   \includegraphics[width=\textwidth, angle=0]{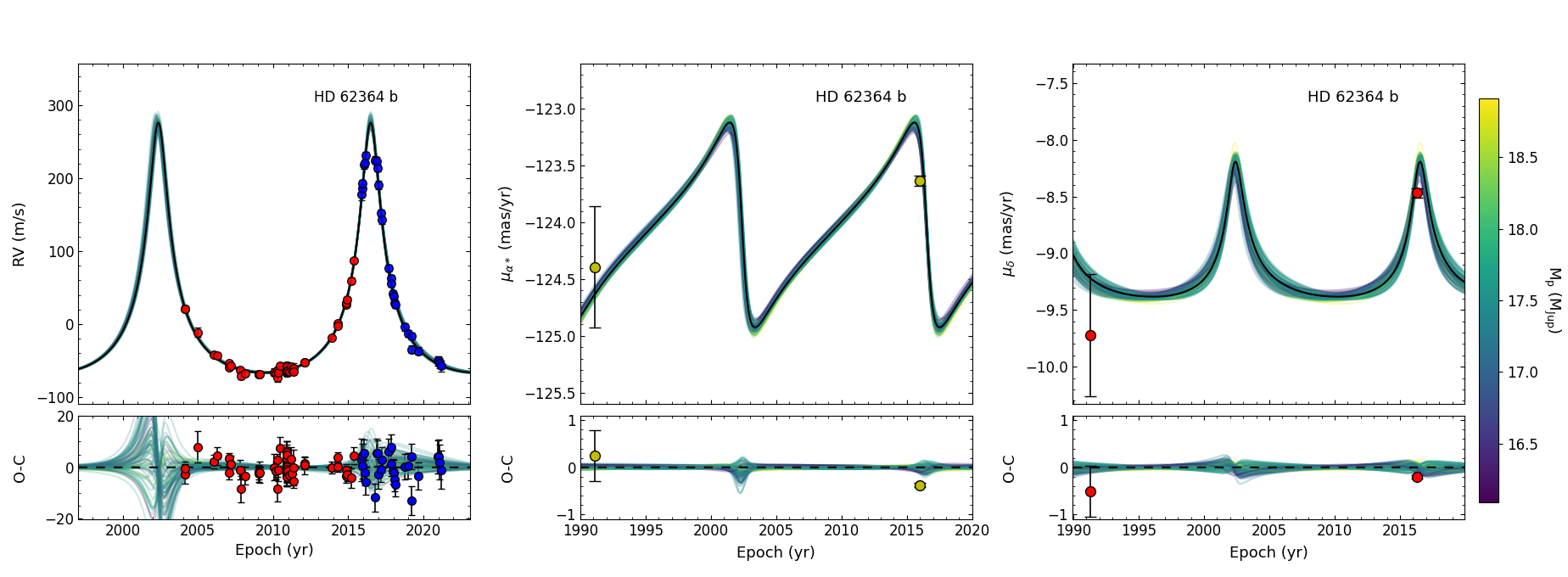}
   \caption{\small Radial velocity curve and astrometric acceleration of HD\,62364\,b. Same as Figure \ref{Fig:HD165131rv_pm}} 
   \label{Fig:HD62364rv_pm}
\end{figure}

$-$ HD\,62364\,b. This BD is orbiting a F7V type star with $T_{\rm eff}=6255\ {\rm K}$, ${\rm log }g=4.29\ {\rm cgs}$ and ${\rm [Fe/H]}=-0.11\ {\rm dex}$ (\citealt{CostaSilva2020}) on a long-period and eccentric orbit ($P={5170}_{-21}^{+22}\ {\rm days}$, $e={0.6092}_{-0.042}^{+0.042}$, $a={6.248}_{-0.072}^{+0.070}\ {\rm AU}$). Similar to HD\,165131\,b, no correlations are found between RVs and BIS ($r=-0.03$), and $H_{\alpha}$ index ($r=0.05$). In Figure \ref{subfig:HD62364b_gls}, we present the GLS periodogram of HD\,62364. 
Although only one orbital phase was sampled by HARPS (see Figure \ref{Fig:HD62364rv_pm}), the full orbital solution of HD\,62364\,b is well-constrained by the joint analysis of RV and absolute astrometry from HGCA.  
Our best-fit orbit also reveals a mass of $M_{\rm p}={17.46}_{-0.59}^{+0.62}\ M_{\rm Jup}$, 
an $M_{\rm p}\,{\rm sin}\,i={13.16}_{-0.33}^{+0.33}\ M_{\rm Jup}$, and $i= {48.9}_{-1.7}^{+1.8}\degr$ (or ${131.1}_{-1.8}^{+1.7}\degr$). 
We note that \citet{Feng2022} revealed this system as a multi-planetary system (HD\,62364\,b: $M_{\rm p}={17.43}_{-1.66}^{+1.63}\ M_{\rm Jup}$, $a={19.0}_{-1.2}^{+1.5}$ AU; HD\,62364\,c: $M_{\rm p}={24.8}_{-2.8}^{+2.8}\ M_{\rm Jup}$, $a={36.9}_{-2.7}^{+3.1}$ AU), based on the combined analysis of the Hipparcos-Gaia astrometry and the limited RV observations from \citet{Trifonov2020}. Thanks to additional data from ESO archive, we found that the semi-major axis of HD\,62364\,b should be 6.248 AU, rather than 19 AU. Additionally, we have not found any significant trend that can imply the existence of the outer companion according to Bayesian Information Criteria (BIC; \citealt{Schwarz1978}) and reduced Chi-square $\chi_{\rm red}^{2}$
(no trend: $\chi_{\rm red}^{2}=1.0, {\rm BIC}=521$; trend: $\chi_{\rm red}^{2}=0.95, {\rm BIC}=517$).
Along with HD\,165131\,b, the two low-mass and highly eccentric BDs may be interesting cases in studying planet-planet scattering mechanism (see section \ref{sect:discussion}).

\subsection{New BDs Previously Classified as Planets Candidates}
Among 64 companions with $M_{\rm p}\,{\rm sin}\,i<13.5\ M_{\rm Jup}$, the  masses of 9 companions are in the BD mass region and the rest are still in the planetary mass region. The details of 8 BDs (HD\,62364\,b has been discussed above) are described below:

$-$ HD\,14067\,b was found to be a long-period eccentric planet, orbiting an evolved intermediate-mass star. \citet{Wang2014} reported two possible solutions: short-period solution with a linear trend ($P=1455\ {\rm days}$) and long-period solution without a trend ($P=2850\ {\rm days}$). We find that the acceleration $\chi^2$ value from HGCA is only 13.89, which may rule out the presence of additional companions who can induce a significant trend. In addition, the extra RV monitoring with HIDES at Okayama Astrophysical Observatory (OAO) also confirms the latter solution (Teng et al. in prep). Thereby, we obtain a mass of ${14.9}_{-4.8}^{+6.4}\ M_{\rm Jup}$ and an inclination of ${38}_{-13}^{+27}\degr$ (or ${142}_{-27}^{+13}\degr$) without considering the linear trend. \citet{Feng2022} recently also reported a mass of ${15.74}_{-5.34}^{+7.03}\ M_{\rm Jup}$, consistent with this work within $1\sigma$.

$-$ HIP\,67537\,b. \citet{Jones2017} published the discovery of an $M_{\rm p}\,{\rm sin}\,i=11.1\ M_{\rm Jup}$ planet at the edge of BD desert around the giant star HIP\,67537. Our best-fit solution reveals a mass of ${13.7}_{-2.4}^{+5.6}\ M_{\rm Jup}$ which is comparable with the theoretical deuterium-burning limit, but the posterior distribution of mass presents a high-mass tail of up to $\sim30$ $M_{\rm Jup}$. Our estimated mass is in perfect agreement with the value of ${13.56}_{-2.38}^{+5.06}\ M_{\rm Jup}$ found by \citet{Feng2022}. Other parameters such as period and eccentricity agree well with \citet{Jones2017}.

$-$ HD\,33636\,b was first reported by \citet{Perrier2003}. It has a poor constraint for orbital parameters due to the incomplete coverage of the orbital phase. \citet{Butler2006} has revised its orbital solution with additional RV data from Lick, Keck, and HET survey. Their solution yields a semi-major axis of $3.27\pm0.19$ AU, an eccentricity of $0.4805\pm0.06$ and an $M_{\rm p}\,{\rm sin}\,i$ of $9.28\pm 0.77\ M_{\rm Jup}$, which are all in agreement with our results. \citet{Bean2007} indicated that HD\,33636\,b was a low-mass star with a mass of $142\pm11\ M_{\rm Jup}$ and an inclination of $4.1\pm 0.1\degr$ based on the joint analysis of HST astrometry and RV. They found $\varpi = 35.6\pm 0.2$ mas, $\mu_{\alpha}=169.0\pm0.3$ mas ${\rm yr}^{-1}$ and $\mu_{\delta}=-142.3\pm0.3$ mas ${\rm yr}^{-1}$ for HD\,33636, which are inconsistent with the Hipparcos values ($\varpi = 34.9\pm 1.3$ mas, $\mu_{\alpha}=180.8\pm1.1$ mas ${\rm yr}^{-1}$ and $\mu_{\delta}=-137.3\pm0.8$ mas ${\rm yr}^{-1}$). The significant difference in proper motion could be caused by 
the fewer observations of Hipparcos (only 16). Consequently, we simply use HST absolute astrometry to replace Hipparcos values in HGCA. Our solution roughly yields a mass of ${77.8}_{-6.6}^{+6.9}$ $M_{\rm Jup}$ and an inclination of ${7.07}_{-0.54}^{+0.62}\degr$ which indicates this system as an extremely face-on system. Future Gaia data releases will confirm the mass of HD\,33636\,b.

$-$ HD\,23596\,b. The 3-body fit presents a mass of ${14.6}_{-1.3}^{+1.5}$ $M_{\rm Jup}$ and an inclination of ${34}_{-2.9}^{+3.6}\degr$ (or ${146}_{-3.6}^{+2.9}\degr$), which puts itself at the transition between planets and brown dwarfs. \citet{Feng2022} found a relatively small value of ${11.91}_{-1.77}^{+0.99}$ $M_{\rm Jup}$, mainly because they adopted a smaller host star's mass (1.1 $M_{\tiny \sun}$) than this work (1.3 $M_{\tiny \sun}$).
Other fitted parameters agree with \citet{Perrier2003}, \citet{Wittenmyer2009} and \citet{Stassun2017}.

$-$ HD\,217786\,b was discovered by \citet{Moutou2011} with an $M_{\rm p}\,{\rm sin}\,i=13\ M_{\rm Jup}$. Then \citet{Ginski2016} reported a substellar companion to the host star HD\,217786 at a separation of 155 AU via direct imaging observations. Our 3-body fit suggests that HD\,217786\,b is a brown dwarf with a mass of ${28.3}_{-2.8}^{+3.1}$ $M_{\rm Jup}$, rather than a planet. The inclination is ${25.7}_{-2.5}^{+3.0}\degr$ or ${154.3}_{-3.0}^{+2.5}\degr$. For HD\,217786\,B, we obtain a semi-major axis of ${213}_{-64}^{+85}$ AU, which is consistent with \citet{Ginski2016}. In addition, the inclination has a single value of ${109}_{-18}^{+16}\degr$, which presents itself in a retrograde orbit. For HD\,217786\,b, \citet{Feng2022} estimated a significantly smaller mass of ${13.85}_{-1.31}^{+1.27}$ $M_{\rm Jup}$. This is mainly because they have ignored the reflex motion induced by the stellar companion HD\,217786\,B in their analysis.

$-$ HD\,136118\,b was identified as a brown dwarf by the HST/FGS astrometry and RV data from HET. \citet{Martioli2010} found an inclination of ${163.1}_{-3}^{+3}\degr$ and a mass of ${42}_{-18}^{+11}$ $M_{\rm Jup}$. From our orbit fits, we obtain a mass of ${16.5}_{-1.8}^{+1.7}$ $M_{\rm Jup}$ and an inclination of ${134}_{-7.5}^{+4.7}\degr$. Our solution is smaller than \citet{Martioli2010}’s results but has higher precision. Other orbital parameters agree well with theirs. In addition, \citet{Feng2022} also found a lower mass of 
${13.10}_{-1.27}^{+1.35}$ $M_{\rm Jup}$ for HD\,136118\,b.

$-$ HD\,139357\,b. \citet{Dollinger2009} announced the discovery of an $M_{\rm p}\,{\rm sin}\,i=9.76\pm 2.15\ M_{\rm Jup}$ planet on a $P=1125.7\pm9.0$ days and slightly eccentric ($e=0.1\pm0.02$) orbit around a K giant star. Our estimation of the mass of ${18.2}_{-5.1}^{+6.2}\ M_{\rm Jup}$ confirms that HD\,139357\,b should be a brown dwarf instead of a giant planet. \citet{Feng2022} also reported a comparable mass of ${19.87}_{-3.44}^{+4.42}\ M_{\rm Jup}$.

$-$ HIP\,84056\,b was classified as a giant planet with a minimum mass of $M_{\rm p}\,{\rm sin}\,i\sim2.6\ M_{\rm Jup}$ \citep{Jones2016,Wittenmyer2016}, whereas our 3-body fit reveals that it is a brown dwarf with $M_{\rm p}={31.9}_{-5.3}^{+8.5}\ M_{\rm Jup}$ on an extremely face-on orbit.

\subsection{New M Dwarfs Previously Classified as BD Candidates}
Among 31 BD candidates, we find that 16 BD candidates have a mass beyond the hydrogen burning limit that should be regarded as low-mass M dwarfs. 13 of the 16 BDs have published estimations of mass using other methods (e.g.,  \citealt{Kiefer2019,Sahlmann2011}). The three newfound M dwarfs are discussed as follows: 

$-$ HD\,203473\,B was regarded as a giant planet candidate on a 4.25-year orbit whose minimum mass is $7.8\pm1.1$ $M_{\rm Jup}$ found by \citet{Ment2018}. 
However, three pre-upgrade HIRES data used in their orbital fit are quite different from those reduced by \citet{Butler2017}, which leads to a relatively short-period orbit with a significant linear trend as well as an acceleration. We prefer the latter because it can simplify the orbital model. From our solution, we find
the period $P={2962.7}_{-3.3}^{+3.1}$ (about 8.11 years) and the minimum mass $M_{\rm p}\,{\rm sin}\,i={62.3}_{-7.8}^{+8.0}$ $M_{\rm Jup}$ are significantly underestimated by \citet{Ment2018}. The mass is ${106}_{-13}^{+13}$ $M_{\rm Jup}$, located on M-dwarf domain, and the inclination is either ${36.1}_{-1.3}^{+1.4}$ or ${143.9}_{-1.4}^{+1.3}$. \citet{Feng2022} reported a relatively lower mass of ${95.8}_{-8.8}^{+8.6}$ $M_{\rm Jup}$, agreeing well with our results.

$-$ HD\,283668\,B was found to be a brown dwarf candidate with $M_{\rm p}\,{\rm sin}\,i=53\pm4\ M_{\rm Jup}$ \citep{Wilson2016}. But our estimated mass of ${319}_{-19}^{+19}\ M_{\rm Jup}$ indicates it as a stellar companion. We find a higher eccentricity of ${0.698}_{-0.039}^{+0.047}$ than the value of $0.577\pm0.011$ provided by \citet{Wilson2016}.

$-$ HD\,184601\,B was reported by \citet{Dalal2021} based on 15 SOPHIE measurements. They obtained a minimum mass of $60.27\pm 2.15\ M_{\rm Jup}$ and a loose upper limit of the mass, $276\ M_{\rm Jup}$. Our best-fit solution has $M_{\rm p}={117}_{-32}^{+36}\ M_{\rm Jup}$ and $i={33.3}_{-7.6}^{+14}\degr$ or ${146.7}_{-14}^{+7.6}\degr$.

\section{Discussion}
\label{sect:discussion}

\subsection{Cross Validation with Gaia DR3}

Given the temporal baseline of the Gaia mission ($\sim$34 months), the HGCA astrometry may not be reliable for our 65 sample systems with orbital period less than  about 6 years. Therefore, the results derived from the HGCA data need to be validated with the add-on of Gaia DR3. We combine the Gaia DR3 astrometric excess noise ($\epsilon_{\rm {DR3}}$), Renormalised Unit Weight Error (RUWE), and semi-major axis of the primary star ($a_{0}$, in units of mas) or the astrometric-orbit solution from \texttt{gaiadr3.nss\_two\_body\_orbit} table to check the consistency of our results. 

The astrometric excess noise is the noise parameter used in the fit to the observed astrometry, while the RUWE is a renormalisation of this  which corrects for magnitude- and colour-related systematics \citep{Lindegren2018}.
In practice, astrometric excess noise can be used to constrain the astrometric amplitude of a signal, and RUWE is usually used to assess the quality of an astrometric solution. It is widely accepted that a value of $\rm{RUWE}<1.4$ indicates a good solution, while $\rm{RUWE}>1.4$ implies the multiplicity of a star~\citep{Lindegren2018, GaiaCollaboration2022}. 
A value of $\epsilon_{\rm {DR3}}>0.5\ \rm{mas}$ may absorb the potential binary astrometric motion for a 5-parameter model \citep{Kiefer2019}.

%RUWE is the square root of the normalized $\chi^{2}$ of the astrometric fit to the along-scan (AL) measurements, and it's usually used to assess the quality of an astrometric solution. It is widely accepted that a value of $\rm{RUWE}<1.4$ indicates a good solution, while $\rm{RUWE}>1.4$ implies the multiplicity of a star~\citep{Lindegren2018, GaiaCollaboration2022}. 

%Similar to the RV jitter, $\epsilon_{\rm {DR3}}$ measures the scatter of the 5-parameter astrometric solution \citep{Lindegren2012}, and it can be used to constrain the astrometric amplitude of a signal.

%\ww{\sout{Although the excess noise is quite different from the angular semi-major axis of a star's astrometric orbit, it still can be used to constrain the astrometric amplitude of a signal.}} 
 In addition, we find that 12 of the 65 sample stars are recorded in the \texttt{nss\_two\_body\_orbit} table, however, unfortunately without providing estimates of companion mass. Therefore, we use the \texttt{nsstools} code\footnote{\url{https://gitlab.obspm.fr/gaia/nsstools}} \citep{Halbwachs2022} to convert the preceding orbital elements from the \texttt{nss\_two\_body\_orbit} table into the Campbell orbital elements, and determine the companion mass by solving Kepler’s third law.

\begin{figure}
   \centering
   \includegraphics[width=10.0cm, angle=0]{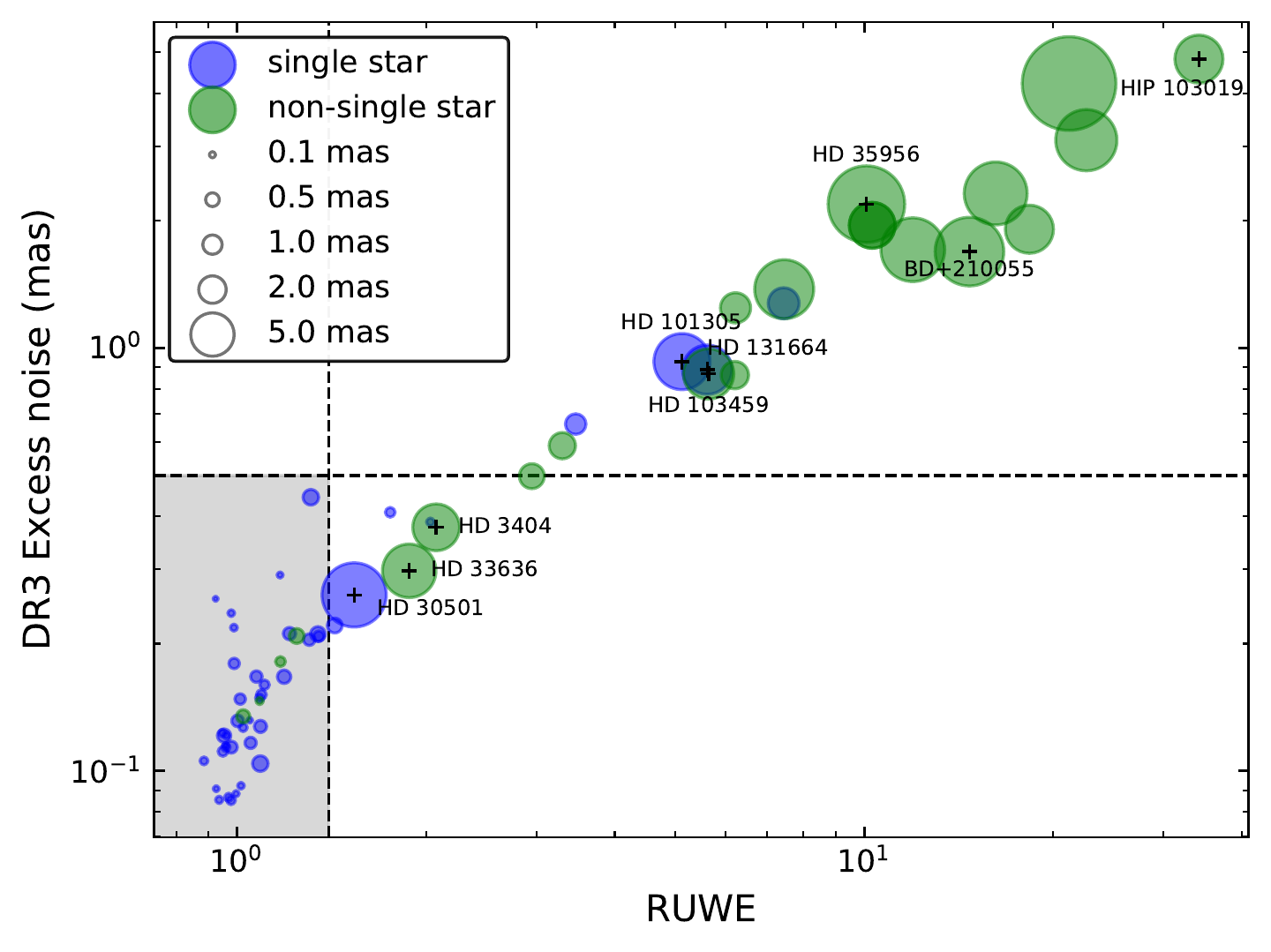}
   \caption{\small $\epsilon_{\rm DR3}$ against RUWE of our sample stars. The blue circles indicate the single stars resolved in Gaia DR3, while the green circles represent the non-single stars (e.g., astrometric binary, spectroscopic binary and eclipsing binary). The semi-major axis (in units of mas) of primary star is represented by the sizes of the circles. The horizontal and vertical dashed lines represent the noise value of 0.5\,mas and $\rm{RUWE}=1.4$, respectively.} 
   \label{Fig:ruwe}
\end{figure}

\begin{figure}
   \centering
   \includegraphics[width=10.0cm, angle=0]{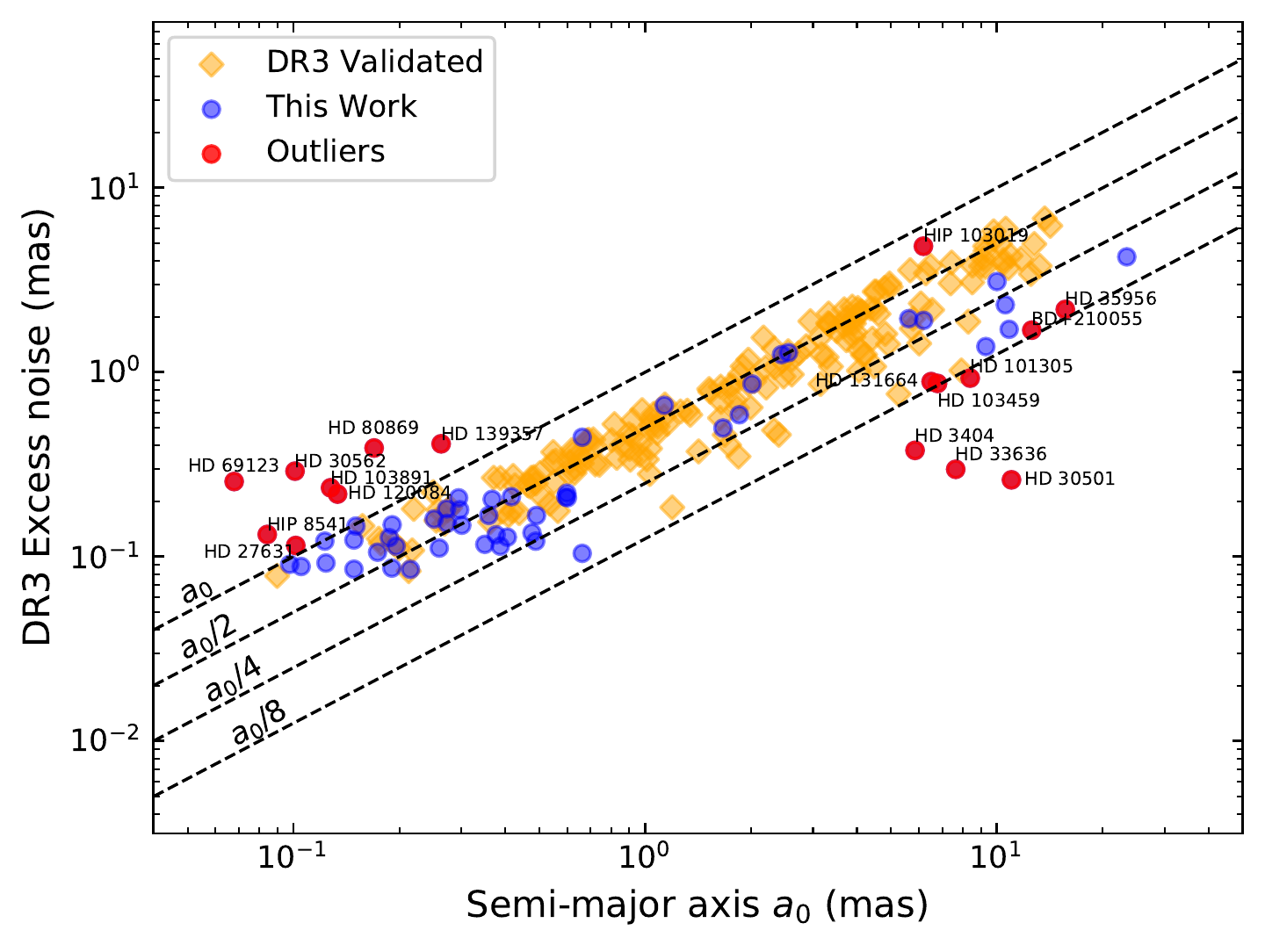}
   \caption{\small $\epsilon_{\rm DR3}$ against $a_{0}$ of the primary star. The filled orange diamonds represent the DR3 validated sample of 188 stars, while the blue and red (possible outliers) circles stands for the 65 stars from our sample. The four dashed lines represent $\epsilon_{\rm DR3}=a_{0},\ a_{0}/2,\ a_{0}/4$ and $a_{0}/8$, respectively.} 
   \label{Fig:dr3_vali}
\end{figure}

Figure~\ref{Fig:ruwe} shows the comparisons of $\epsilon_{\rm {DR3}}$ and RUWE of the 65 sample stars with the filled blue and green circles, where the blue represents for single stars while the red for multiple systems. The circle size is scaled by the semi-major axis $a_{0}$ of the primary star. It is quite obvious that $\epsilon_{\rm {DR3}}$ has a close-to-linear relationship with RUWE with a non-unity slope, suggesting that the two parameters are well consistent with each other. 
In addition, $a_{0}$ seems to be positively correlated with $\epsilon_{\rm {DR3}}$ and RUWE as well, despite of several obvious outliers, for example HD\,30501, 33636 and 3404, that will be spotted out in Figure~\ref{Fig:dr3_vali}.

Figure~\ref{Fig:dr3_vali} shows the comparison of $\epsilon_{\rm DR3}$ and $a_{0}$ for our sample (blue and red circles) and the 188 stars (yellow diamonds) compiled from \texttt{nss\_two\_body\_orbit}. The  astrometric-orbit solutions of 188 stars were derived from the Gaia DR3 exoplanet pipeline by \citet{Holl2022}, and validated using significance test, the available Gaia radial velocity, as well as literature radial velocity and astrometric data~\citep[cf. more details in][]{Holl2022}. They reported an interesting relation between $\epsilon_{\rm DR3}$ and $a_{0}$, in the sense that $\epsilon_{\rm DR3}$ is typically about half of $a_{0}$. This is confirmed in Figure~\ref{Fig:dr3_vali}, where the yellow diamonds are mainly concentrated around the $\epsilon_{\rm DR3}=a_{0}/2$ line. Our sample stars are distributed mainly between the boundary of $\epsilon_{\rm DR3}=a_{0}$ and $\epsilon_{\rm DR3}=a_{0}/8$ with a median ratio of $\epsilon_{\rm DR3}/a_{0}\sim0.71\pm0.09$, with a dozen of outliers (marked in red). We will discuss the above outliers in details in the following subsections.

\begin{table}
\centering
\caption{Comparisons between our results using \texttt{orvara} and those from the Gaia Two-body solutions and from literature.}\label{Tab:tb}
%%Please Capitalize the First Letter of Each Notional Word in table's caption
\resizebox{\textwidth}{!}{
\begin{tabular}{lcccccccc}
\hline \hline
\noalign{\smallskip}
\noalign{\smallskip}
 Name & $M_{\rm p}\,{\rm sin}\,i$ &$M_{\rm p, Gaia}$ & $i_{\rm Gaia}$ & $M_{\rm p}$ & $i$& Within $1\sigma$
 &$M_{\rm lit}$&Refs\\
 &($M_{\rm Jup}$)&($M_{\rm Jup}$)& (\degr)& ($M_{\rm Jup}$) &(\degr) &&($M_{\rm Jup}$)&\\
 \noalign{\smallskip}
\hline
\noalign{\smallskip}
\multicolumn{9}{c}{${\rm RUWE}\,<\,1.4$}\\
\noalign{\smallskip}
HD\,132406\,b&${5.25}_{-0.57}^{+1.2}$&$6.8\pm4.2$&$122.3\pm14.7$&${6.2}_{-1.1}^{+2.2}$&${116.0}_{-18.0}^{+19.0}$&$\surd$&&\\
 \noalign{\smallskip}
HD\,175167\,b&${8.1}_{-0.77}^{+1.7}$&$10.0\pm4.2$&$28.2\pm19.4$&${9.8}_{-1.2}^{+1.9}$&${60.0}_{-13.0}^{+17.0}$&$\surd$&&\\
 \noalign{\smallskip}
HD\,111232\,b&${7.24}_{-0.17}^{+0.17}$&$7.7\pm0.6$&$96.6\pm3.6$&${7.47}_{-0.26}^{+0.6}$&${102.9}_{-9.1}^{+12.0}$&$\surd$&${8.26}_{-0.78}^{+0.83}$&\citet{Feng2022}\\
 \noalign{\smallskip}
HD\,111232\,c& & & &${20.7}_{-3.2}^{+3.4}$ & ${102.9}_{-8.9}^{+12.0}$ & $-$ & ${19.15}_{-2.70}^{+3.13}$ & \citet{Feng2022}  \\
\hline
\noalign{\smallskip}
\multicolumn{9}{c}{${\rm RUWE}\,>\,1.4$}\\
\noalign{\smallskip}
HD\,184601\,B&${64.7}_{-3.2}^{+3.5}$&$119.4\pm7.2$&$151.4\pm3.1$&${117.0}_{-32.0}^{+36.0}$&${146.7}_{-14.0}^{+7.6}$&$\surd$&&\\
 \noalign{\smallskip}
HD\,130396\,B&${95.1}_{-5.1}^{+5.3}$&$107.5\pm26.1$&$27.3\pm6.3$&${95.1}_{-5.1}^{+5.3}$&${34.3}_{-1.8}^{+2.0}$&$\surd$&&\\
 \noalign{\smallskip}
HD\,30246\,b&${49.7}_{-3.0}^{+4.2}$&$42.1\pm9.5$&$78.0\pm2.2$&${51.8}_{-3.8}^{+8.1}$&${79.8}_{-21.0}^{+7.3}$&$\surd$&&\\
 \noalign{\smallskip}
BD+730275\,B&${50.4}_{-2.1}^{+2.1}$&$195.3\pm8.9$&$15.5\pm1.4$&${187.0}_{-14.0}^{+15.0}$&${15.63}_{-0.88}^{+1.0}$&$\surd$&${210}_{-31}^{+31}$&\citet{Wilson2016}\\
 \noalign{\smallskip}
HD\,51813\,B&${53.2}_{-2.8}^{+2.7}$&$219\pm20$&$152.3\pm5.4$&${188.0}_{-19.0}^{+20.0}$&${163.6}_{-1.7}^{+1.5}$&$\surd$&${282}_{-73}^{+73}$&\citet{Wilson2016}\\
 \noalign{\smallskip}
HD\,35956\,B&${193.1}_{-4.8}^{+4.7}$&$173.2\pm38.6$&$84.8\pm1.1$&${228.1}_{-8.3}^{+8.3}$&${57.8}_{-2.0}^{+2.2}$&&&\\
 \noalign{\smallskip}
BD+210055\,B&${98.8}_{-4.1}^{+3.9}$&$129.6\pm14.3$&$162.8\pm11.8$&${199.0}_{-10.0}^{+10.0}$&${150.22}_{-0.84}^{+0.8}$&&&\\
 \noalign{\smallskip}
HIP\,103019\,B&${56.3}_{-1.1}^{+1.4}$&$129.8\pm1.3$&$158.3\pm0.6$&${83.0}_{-22.0}^{+28.0}$&${137.0}_{-23.0}^{+11.0}$&&${188.1}_{-26.4}^{+26.5}$&\citet{Sahlmann2011}\\
 \noalign{\smallskip}
HD\,48679\,B&${37.59}_{-0.73}^{+0.7}$&$110.3\pm6.5$&$26.4\pm5.2$&${71.6}_{-6.6}^{+7.0}$&${31.6}_{-2.9}^{+3.5}$&&${51.0}_{-5.6}^{+6.0}$&\citet{Feng2022}\\
 \noalign{\smallskip}
&&&&&&&$41\sim55$&\citet{Kiefer2019}\\
 \noalign{\smallskip}
\hline
\end{tabular}
}
\end{table}

\subsubsection{RUWE \textless\ 1.4}
In Figure~\ref{Fig:ruwe}, we find that the 40 stars with $\rm{RUWE}<1.4$ have $\epsilon_{\rm DR3}<0.5$ mas (the grey-shaded area) and $a_{0}<0.7\ \rm{mas}$. The fact that $a_{0}$ is in general do not overlarge much than $\epsilon_{\rm DR3}$ suggests that the derived masses for them should not be significantly overestimated. However, there are 8 of the 40 stars exhibiting possible underestimations according to the left part of Figure~\ref{Fig:dr3_vali} (the red circles), implying a probability of 20\% for yielding underestimated solutions in this regime. Interestingly, for such small $a_{0}$, there are still three systems, namely HD\,132406, HD\,175167 and HD\,111232, resolved as non-single stars and archived in the \texttt{nss\_two\_body\_orbit} table. Their masses and inclination angles derived from the Gaia DR3 solution and \texttt{orvara} are listed in Table \ref{Tab:tb} for comparisons. All of them are consistent with each within $1\sigma$. For HD\,111232, we additionally determine a mass of ${20.7}_{-3.2}^{+3.4}\ M_{\rm{Jup}}$ for the outer companions, HD\,111232\,c, in agreement with that from \citet{Feng2022}.

\subsubsection{RUWE \textgreater\ 1.4}
Among the 25 stars with $\rm{RUWE}>1.4$, nine stars can be found in the \texttt{nss\_two\_body\_orbit} table, whose masses are thus able to be derived from this table by solving Kepler’s third law. The results are listed in the third column of Table~\ref{Tab:tb}. We find that four of them have significantly different mass values derived from the two different approaches, as discussed below:

$-$ HD\,35956. The companion mass derived from Gaia is $173.2\pm38.6\ M_{\rm{Jup}}$ and the inclination is $85\pm1\degr$, which disagree with the mass $M_{\rm p}={228.1}_{-8.3}^{+8.3}\ M_{\rm Jup}$ and the inclination $i={57.8}_{-2.0}^{+2.2}\degr$ yielded by \texttt{orvara}. The mass determined from \texttt{orvara} is larger than that from the Gaia DR3 data.

$-$ BD+210055. The companion mass derived from Gaia is $129.6\pm14.3\ M_{\rm{Jup}}$ and the inclination is $163\pm12\degr$, which disagree with $M_{\rm p}={199}_{-10}^{+10}\ M_{\rm Jup}$ and $i={150.22}_{-0.84}^{+0.80}\degr$ from \texttt{orvara}. However, as shown in Table~\ref{Tab:tb}, the orbital solution using the Gaia astrometry alone seems not conforming with RV solution, with the latter yielding $M_{\rm p}\,{\rm sin}\,i$ significantly smaller than that from astrometry alone. Further Gaia data with longer temporal baseline is required to check it's nature.

$-$ HIP\,103019. The Gaia two-body solution yields a companion mass of $129.8\pm1.3\ M_{\rm{Jup}}$ and an inclination of $158.28\pm0.55\degr$, disagreeing with $M_{\rm p}={83}_{-22}^{+28}\ M_{\rm Jup}$ and $i={137}_{-23}^{+11}\degr$ from \texttt{orvara}. \citet{Sahlmann2011} reported a higher mass of $M_{\rm p}={188.1}_{-26.4}^{+26.5}\ M_{\rm Jup}$ based on Hipparcos IAD.

$-$ HD\,48679. The companion mass is $110.3\pm6.5\ M_{\rm{Jup}}$ and the inclination is $26\pm5\degr$ according to Gaia two-body solution, disagreeing with the mass $M_{\rm p}={71.6}_{-6.6}^{+7.0}\ M_{\rm Jup}$ and the prograde inclination $i={31.6}_{-2.9}^{+3.5}\degr$ found by \texttt{orvara}. \citet{Feng2022} reported a much lower mass of $M_{\rm p}={51.0}_{-5.6}^{+6.0}\ M_{\rm Jup}$, and \citet{Kiefer2019} estimated a comparable mass of $41\sim55\ M_{\rm Jup}$ with the \texttt{GASTON} tool.

For the remaining 16 stars that the \texttt{nss\_two\_body\_orbit} table does not provide   astrometric-orbit solution, we can only compare between our results and the DR3 results for verification. As shown in Figures~\ref{Fig:ruwe} and \ref{Fig:dr3_vali}, masses of six stars may be over-estimated. They are:

$-$ HD\,30501. \citet{Sahlmann2011} obtained a companion mass of ${89.6}_{-12.5}^{+12.3}\ M_{\rm Jup}$ and an inclination of ${49.1}_{-7.8}^{+10.1}\degr$ via the combination of CORALIE RV measurements with Hipparacos IAD. While our solution yields $M_{\rm p}={67.3}_{-1.1}^{+1.1}\ M_{\rm Jup}$ and $i={79.6}_{-2.2}^{+2.9}\degr$ (or ${100.4}_{-2.9}^{+2.2}\degr$), which implies the brown dwarf nature of HD\,30501\,b. In Figures~\ref{Fig:ruwe} and \ref{Fig:dr3_vali}, our $a_{0}$ seems to be larger than the reported astrometric signals as represented by $\epsilon_{\rm DR3}$. Note that the minimum mass deduced from RV alone is $M_{\rm p}\,{\rm sin}\,i={62.3}_{-2.1}^{+2.1}\ M_{\rm Jup}$, comparable to our estimated mass, and the orbital period ($\sim$2074 days) is nearly twice the Gaia time baseline. Therefore, it is likely our assessment is more reliable, as $\epsilon_{\rm DR3}$ might be less reliable for those long-period and edge-on systems. Further individual epoch astrometric measurement is needed to further characterize this system.

$-$ HD\,33636. We have discussed the result in the previous section (see Section \ref{sect:result}). In Figure \ref{Fig:ruwe} and \ref{Fig:dr3_vali}, the low $\epsilon_{\rm {DR3}}$ implies a significant overestimation of our estimated mass, which also contradict the mass derived by HST astrometry \citep{Bean2007}. Further Gaia release with longer temporal baseline is required to check it's nature.

$-$ HD\,101305. We obtain a companion mass of $M_{\rm p}={261}_{-17}^{+17}\ M_{\rm Jup}$ and an inclination of $i={28.36}_{-0.64}^{+0.65}\degr$ (or ${151.64}_{-0.65}^{+0.64}\degr$), agreeing well with the mass ($M_{\rm p}=220\sim270\ M_{\rm Jup}$) reported by \citet{Kiefer2019}. Being an outlier in Figures~\ref{Fig:ruwe} and \ref{Fig:dr3_vali} may suggest an overestimation of our mass. 

$-$ HD\,3404. Our estimated mass for the companion is $M_{\rm p}={239}_{-22}^{+19}\ M_{\rm Jup}$ and the inclination is $i={48.8}_{-2.1}^{+2.1}\degr$ (or ${131.21}_{-2.2}^{+2.1}\degr$). The relatively low  $\epsilon_{\rm {DR3}}$ may suggest an overestimation of our mass.

$-$ HD\,103459. Our estimated mass for the companion is $M_{\rm p}={176}_{-20}^{+18}\ M_{\rm Jup}$, which may have been overestimated by \texttt{orvara}.

$-$ HD\,131664. Our estimated mass for HD\,131664\,B is $M_{\rm p}={131.8}_{-4.1}^{+4.1}\ M_{\rm Jup}$ , which may exhibit a significant overestimation. This agrees with the value of $127.8\pm17.9\ M_{\rm Jup}$ found by \citet{Feng2021}, and disagrees with \citet{Sozzetti2010} who determined a mass of ${23.0}_{-5.0}^{+26.0}\ M_{\rm Jup}$ based on Hipparcos astrometry alone.

\subsection{Comparison with Previous Studies }

\begin{figure}
    \centering
    \subcaptionbox{Comparison with \citet{Feng2022}. \label{subfig:comp_feng}}[.495\linewidth]{
        \includegraphics[width=72mm]{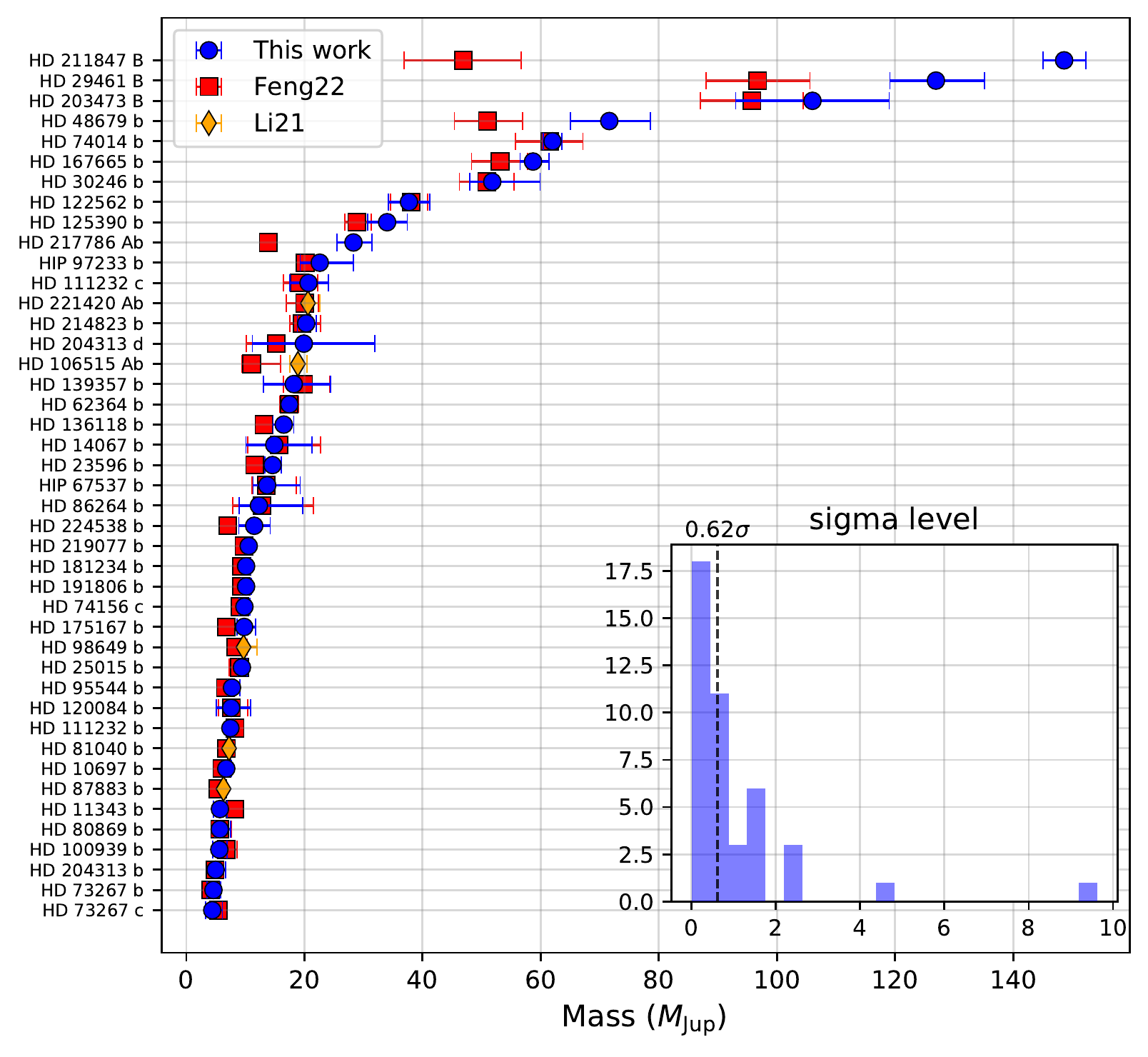}
    }
    \subcaptionbox{Comparison with other literatures. \label{subfig:comp_others}}[.495\linewidth]{
        \includegraphics[width=72mm]{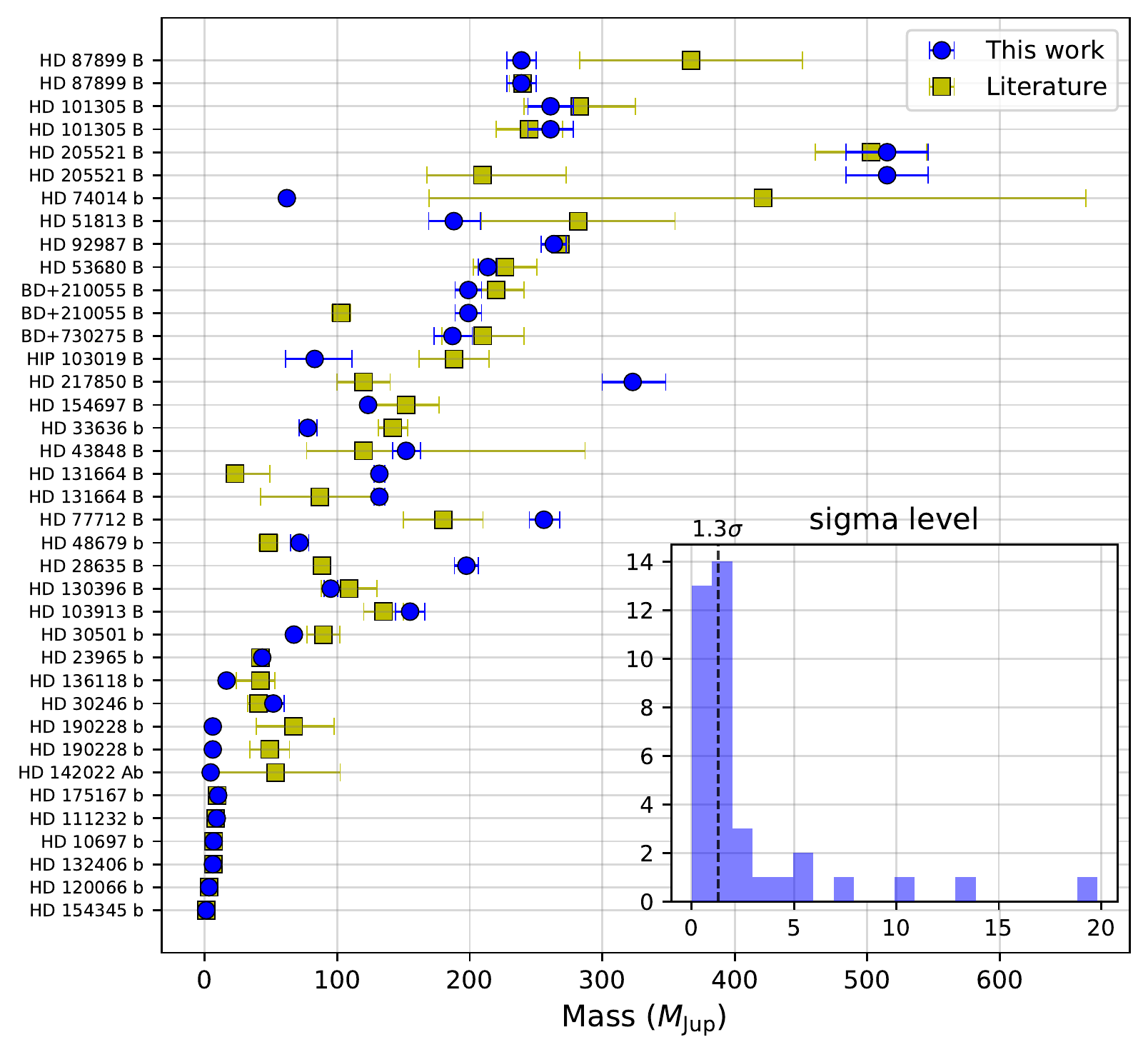}
    }
    \caption{\small (a) The masses of 38 companions (blue circles) in this work and 5 companions (orange diamonds) from \citet{Li2021} (Li21) are derived with \texttt{orvara}. The red squares represent the same companions whose masses are estimated by \citet{Feng2022} (Feng22). 
    (b) The comparison of 33 companions in this work and other works (yellow squares). The lower right histograms represent the differences (sigma levels) between our results and references, and the vertical dashed line indicates the median of the differences.}
    \label{fig:comp_feng_other}
\end{figure}

We find that 38 of our sample stars were included in the recent work by \cite{Feng2022}, and 33 companions with mass determined in other literature studies~\citep{Zucker2001, Wilson2016,Kiefer2019,Sahlmann2011,Dalal2021,Venner2021,Venner2021b,Martioli2010,Sozzetti2010,Reffert2011,Feng2021,Simpson2010,GaiaCollaboration2022,Bean2007}. Most of them were derived through joint analysis using both ground-based RV data and astrometry data obtained by Hipparcos, HST, and Gaia.
%\ww{\sout{We decided to separately compare these two subsamples, although there are some overlaps between them}}. 
We first compare our results with those from \citet{Feng2022}. As presented in Figure~\ref{subfig:comp_feng}, both work yield consistent results at $1\sigma$ level, strongly suggesting that the method used in this work, i.e., \texttt{orvara}, is reliable. We then add the five systems from \citet{Li2021}, who used exactly the same method as ours, to the comparison. It is evident from Fig.~\ref{subfig:comp_feng}) that their results (the yellow diamonds) are also in good agreement with those from \citet{Feng2022} (the red squares). This is a supporting proof for the validation of our employed method. The histogram of sigma levels of mass difference is shown in the lower right corner, which further confirms that the two studies yield very similar results with only $0.62\sigma$ median offset. Note that there are four exceptions, namely HD\,29461\,B, HD\,211847\,B, HD\,48679\,b, and HD\,217786\,Ab, showing obvious discrepancies in the determined masses. The discrepancies may be caused by the differences of these two studies in the data quality and RV baseline of these systems. 
For example, using VLT/SPHERE, \citet{Moutou2017} revealed HD\,211847\,B as a substellar companion with a mass of $155 \pm 9\,M_{\rm Jup}$, consistent well with the value $M_{\rm p}={148.6}_{-3.6}^{+3.7}\,M_{\rm Jup}$ found by \texttt{orvara} (Feng: ${46.9}_{-10.1}^{+9.7}\,M_{\rm Jup}$). Our joint fit has taken into account the additional imaging data, but \citet{Feng2022} hadn't.

%we compare our result along with five companions from \citet{Li2021} to the estimated masses reported by \citet{Feng2022}. We find that 37 companions are in perfect agreement with \citet{Feng2022} within $1\sigma$, which suggests great consistency between these two methods. 
%For three companions (HD\,29451 B, HD\,48679 b, and HD\,217786 Ab), our results are quite different from \citet{Feng2022}. This may be caused by data quality and RV baseline. In general, the result (see the lower right histogram in Figure \ref{subfig:comp_feng}) from \citet{Feng2022} is slightly smaller than those derived with \texttt{orvara}, which may imply the presence of system error between these two methods.

In Figure~\ref{subfig:comp_others}, we compare the \texttt{orvara} results with the ones from the above-mentioned literature excluding \citet{Feng2022}. It is shown that the \texttt{orvara} masses of 27 companions are in agreement with the values from literature within $3\sigma$.
Particularly, the four companions, HD\,175167\,b, HD\,132406\,b, HD\,111232\,b, and HD\,30246 \,b, they show strikingly good agreement with the results provided by Gaia DR3 \citep{GaiaCollaboration2022}, even though they have orbital periods comparable to the 34-month baseline. 
%\citet{Simpson2010} determined the masses of HD\,10697\,b and HD\,154345\,b by decoupling plant orbit inclination angle $i$ from $M_{\rm P}\,{\rm sin}\,i$, where $i$ is represented approximately by the measured rotation axis angle of the host star~\citep{Simpson2010}, assuming the planetary orbital planes are coplanar with their host star's equator planes. The \citet{Simpson2010} masses are well in line with our results. 
\citet{Simpson2010} estimated the masses of HD\,10697\,b and HD\,154345\,b assuming that their orbital inclinations are equal to the rotational inclination of the stars. Their measurements are well in line with our results.

Significant discrepancies are observed for 7 systems, namely HD\,33636\,b, HD\,190228\,b, BD+210055\,B, HD\,131664\,B, HD\,217850\,B, HD\,28635\,B, HIP\,103019\,B, whose masses are derived based on Hipparcos and HST astrometry, subject to relatively large uncertainties. For example, \citet{Zucker2001} identified HD\,190228\,b as a BD with a mass of $67\pm 29$ $M_{\rm Jup}$ using the original Hipparcos astrometry (hereafter HIP-1; \citealt{Perryman1997}), and \citet{Sahlmann2011} reported a smaller value of $49\pm 18$ $M_{\rm Jup}$ based on the re-reduced Hipparcos astrometry (hereafter HIP-2; \citealt{vanLeeuwen2007}), whereas our solution yields a planet-regime mass of $M_{\rm p}={6.1}_{-1.0}^{+1.2}\  M_{\rm Jup}$. Given a relatively large uncertainty in Hipparcos astrometry, and the low value of $\rm{RUWE}=0.88$ and $\epsilon_{\rm {DR3}}= 0.11\ \rm{mas}$, the high-mass solution is less preferred. 

Therefore, the above comparison analysis suggest that the \texttt{orvara} package is able to yield reasonably good results for most cases, although caution should be still taken for short-period systems.

\subsection{Mass-Period Diagram}
\begin{figure}
   \centering
   \includegraphics[width=8.0cm, angle=0]{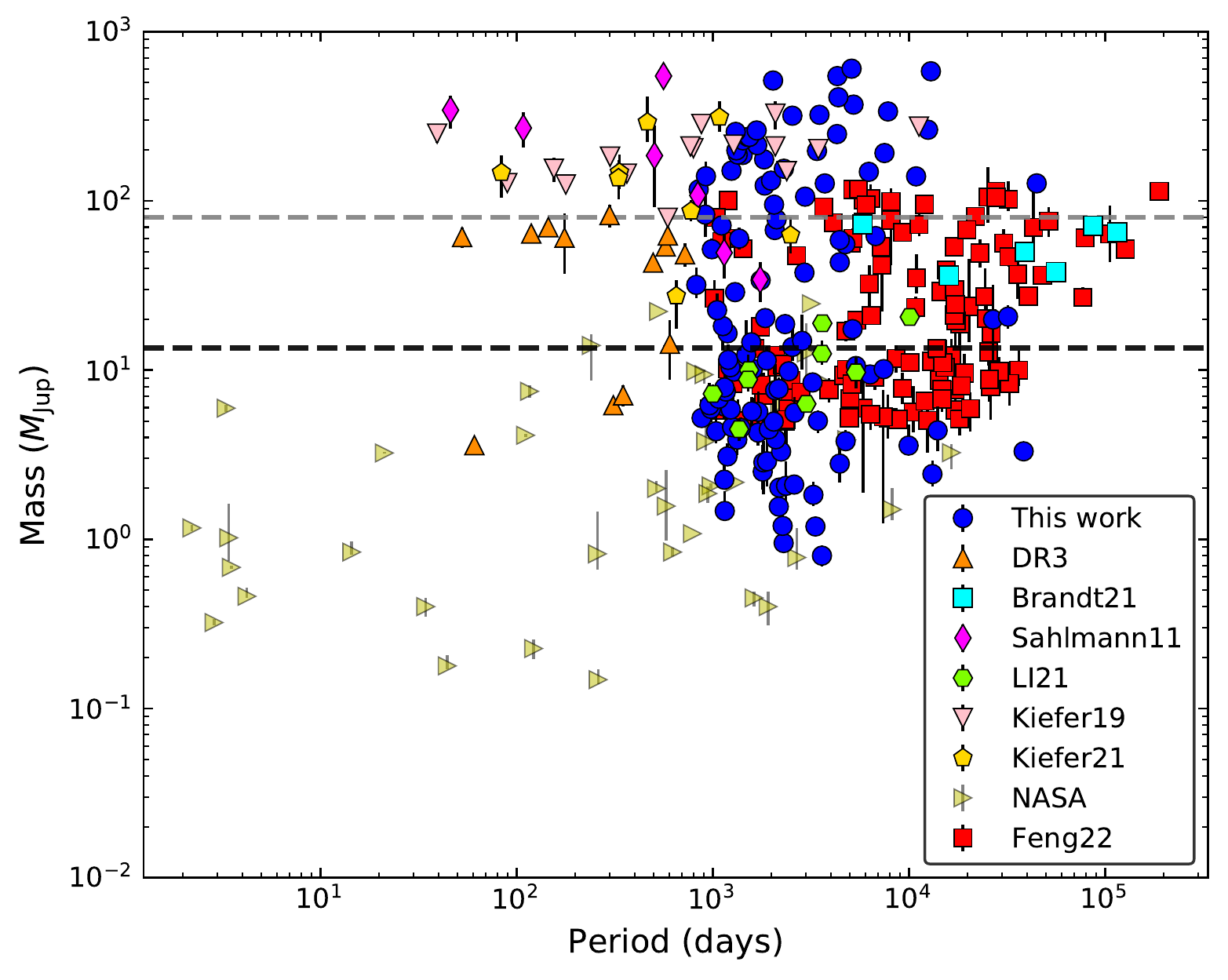}
   \caption{\small Mass-period diagram of the sample from planets up to M dwarfs. The blue circles and red squares represent the data from this work and \citet{Feng2022}, respectively. The data compiled from the NASA Exoplanet Archive \citep{Akeson2013} are plotted in yellow rightward-pointing triangles. The massive companions derived with the \texttt{GASTON} method by \citet{Kiefer2019} and \citet{Kiefer2021} are plotted in pink downward-pointing triangles and golden pentagons, respectively. The masses measured by \citet{Sahlmann2011} based on HIP-2 data are shown as fuchsia diamonds, and the Gaia DR3 companions \citep{GaiaCollaboration2022} are represented as orange upward-pointing triangles. Finally, the chartreuse hexagons represent the planets from \citet{Li2021}, and the cyan squares represent the BDs measured by \citet{GMBrandt2021b}. The horizontal black and grey dashed line indicate the classical boundaries of BDs ($\sim13.5$ and $\sim80$ $M_{\rm Jup}$, \citealt{Burrows1997, Spiegel2011}).} 
   \label{Fig:mass_period}
   \end{figure}

In Figure \ref{Fig:mass_period}, we plot the mass-period ($M\text{-}P$) diagram of planets, BDs, and low-mass M dwarfs. Besides the masses derived in the present work, we also complement 38 companions from NASA Exoplanet Archive \citep{Akeson2013}, 121 companions from \citet{Feng2022}, and 60 companions from other literatures \citep{Kiefer2019, Kiefer2021, Sahlmann2011, GaiaCollaboration2022, Li2021, Brandt2021b}. For some repetitive companions, we give priority to utilise those data derived by \texttt{orvara}. 
These systems are almost discovered by radial velocity method and have well-constrained mass measurements ($>0.1\,M_{\rm Jup}$).  Unlike \citet{Kiefer2021} who only selected the mass below 150 $M_{\rm Jup}$, we include a wider range of up to 450 $M_{\rm Jup}$ to construct a relatively complete sample of M dwarfs (or stellar binaries). 
%Some companions with ambiguous mass (e.g., HD\,190228 b, we adopted the mass of $49\pm 18$ $M_{\rm Jup}$ found by \citealt{Sahlmann2011}) in our work were excluded to minimize bias. 
The wide stellar binaries (e.g., HD\,126614\,B, HD\,217786\,B, HD\,108341\,B, HIP\,84056\,B, HD\,142022\,B, HD\,196050\,B and HD\,23596\,B) in 3-body system are also excluded in the following analysis. 

Recently, some studies have reported the estimation of the BD desert boundaries \citep{Ma2014,Kiefer2019,Kiefer2021}. \citet{Kiefer2021} found an empty region bounded by masses 20 $\sim$ 85 $M_{\rm Jup}$ and periods $0\sim100$ days (see their figure.16). In Figure \ref{Fig:mass_period}, it seems that the $M\text{-}P$ distribution of the combined sample presents an unapparent cut in the BD region at $\sim100$ days (orange triangles). We can't draw a similar conclusion independently from our samples (blue circles) due to the strong selection effects that \texttt{orvara} is not sensitive to the short-period companions ($P<1000\ {\rm days}$). However, we note, quite interestingly, there is also an empty region bounded by masses 13 $\sim$ 50 $M_{\rm Jup}$ and periods 0 $\sim$ 400 days. Considering the fact that most RV-detected planets with $M_{\rm p}\,{\rm sin}\,i>1\ M_{\rm Jup}$ are concentrated around P $\sim$ 500 days (e.g., \citealt{Liu2008}), the vacancies may be filled by potential BDs whose mass is currently not well constrained. We expect the further Gaia data releases will provide comprehensive measurements for these companions.

\subsection{Formation Scenario of Companions}
According to the current core-accretion paradigm \citep{Santos2004}, giant planets form preferentially around metal-rich stars, whose disks harbor more solids or planetary building materials, and less frequently around metal-poor host stars \citep{Gonzalez1997, Santos2001b, Zhao2002, Fischer2005, Mordasini2012}. As for stellar binaries, the formation of gravitational instability is less sensitive to the stellar metallicity, implying that the metallicity distribution for these companions is homogeneous. However, there is still a puzzle about the boundary between the two formation processes. For instance, whether low-mass BDs have the same formation channels as massive planets and whether BDs have multiple populations remains unknown.

By combining the radial velocity and Hipparcos astrometric measurements, \citet{Sahlmann2011} broke the ${\rm sin}i$ degeneracy and determined the mass of companions. They found a clear separation between low-mass BDs ($M_{\rm p}=13\sim25\ M_{\rm Jup}$) and high-mass BDs ($M_{\rm p}>45\ M_{\rm Jup}$), and claimed that the companion-mass distribution
function of the low-mass BDs might represent the high-mass tail of the planetary distribution function. Based on the distribution of eccentricity, \citet{Ma2014} suggested that the low-mass BDs with $M_{\rm p}\,{\rm sin}\,i<42.5\ M_{\rm Jup}$ formed by disc gravitational instability, while the high-mass BDs with $M_{\rm p}\,{\rm sin}\,i> 42.5\ M_{\rm Jup}$ formed stellar-like binaries mainly through molecular cloud fragmentation. They also showed that there was no correlation between the occurrence of BDs and host star  metallicity, which is different from giant planets. While \citet{Schlaufman2018} suggested that selection effects and contamination from low-mass stars may have affected their results, \cite{Maldonado2017} then confirmed that BD hosts do not show the giant planet metallicity correlations. They concluded that the core-accretion mechanism might efficiently form low-mass BDs on metal-rich discs, while low-mass BDs orbiting metal-poor hosts could form by gravitational instability. \citet{Santos2017} explored the properties of the minimum mass (or mass) and metallicity distribution of giant planets discovered through RV and transit methods. They only selected planets with $M_{\rm p}\,{\rm sin}\,i< 15\ M_{\rm Jup}$, 10 days $<P<5$ yrs, and with homogeneous stellar parameters listed in SWEET-Cat database\footnote{\url{http://www.astro.up.pt/resources/sweet-cat}} \citep{Sousa2021}. Their results suggested that giant planets with $M_{\rm p}\,{\rm sin}\,i< 4\ M_{\rm Jup}$, are formed by a core-accretion mechanism, while giant planets with $M_{\rm p}\,{\rm sin}\,i> 4\ M_{\rm Jup}$ are formed by gravitational instability.  \citet{Schlaufman2018} found a limit of $M_{\rm p}\sim10\ M_{\rm Jup}$ using a homogeneous sample with masses derived by transit and Doppler technique,
then he divided his samples into two parts, thereby inferring different formation scenarios. He suggested that planets formed by core accretion have a maximum mass of no more than 10 $M_{\rm Jup}$, while companions with masses above 10 $M_{\rm Jup}$ may have formed through gravitational instability. More recently, \citet{Kiefer2021} applied the \texttt{GASTON} method to constrain the inclination and mass of published RV exoplanet candidates. When studying the distribution of eccentricity with mass, they did not find a well-defined transition at 42.5 $M_{\rm Jup}$, but they reported that some BDs with $M_{\rm p}> 45\ M_{\rm Jup}$ can even stand above $e\ =\ 0.7$, while all BDs with $M_{\rm p}< 45\ M_{\rm Jup}$ have $e<0.7$, which seems to agree with \citet{Ma2014}.

\begin{figure}
   \centering
   \includegraphics[width=8.0cm, angle=0]{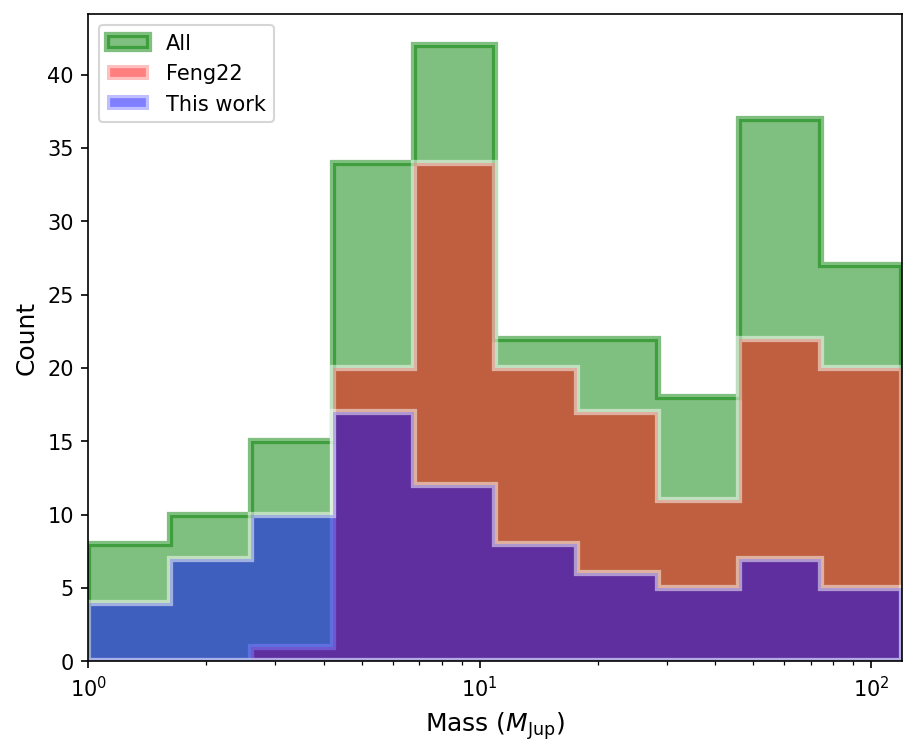}
   \caption{\small Mass distribution diagram (${\rm bins}=10$). The blue histogram represents the data from the current work. The red histogram represents the data from \citet{Feng2022}, and the green histogram represents all the combined data.} 
   \label{Fig:mass_distribution}
\end{figure}

In our study, we also explore companions' formation scenarios through the distributions of metallicity and eccentricity with respect to masses. We only select FGK star systems with masses above $0.52\ M_{\tiny \sun}$ to exclude M-type host stars whose stellar atmospheric parameters might be unreliable. For the sample from \citet{Feng2022}, we select those with $5<M_{\rm p}<120\ M_{\rm Jup}$, period $P>1000$ days and $\sigma_{i}<30\degr$.
As a result, a total of 309 companions are included (see Table \ref{Tab:comb_samp}). Figure \ref{Fig:mass_distribution} shows the distribution of masses. A clear valley can be found near 40 $M_{\rm Jup}$, which seems to match that of \citet{Feng2022}.
\begin{table}
\centering
\caption{The Statistics of Our Combined Sample}\label{Tab:comb_samp}
%%Please Capitalize the First Letter of Each Notional Word in table's caption
\begin{tabular}{lcccc}
\hline \hline
 Data & This Work & \citet{Feng2022} & NASA Exoplanet Archive & Other work \\
 \hline
 Counts&113&100&38&59\\
\hline
\end{tabular}
\end{table}

\begin{figure}
    \centering
    \subcaptionbox{Mass-metallicity diagram\label{subfig:mass_feh}}[.495\linewidth]{
        \includegraphics[width=72mm]{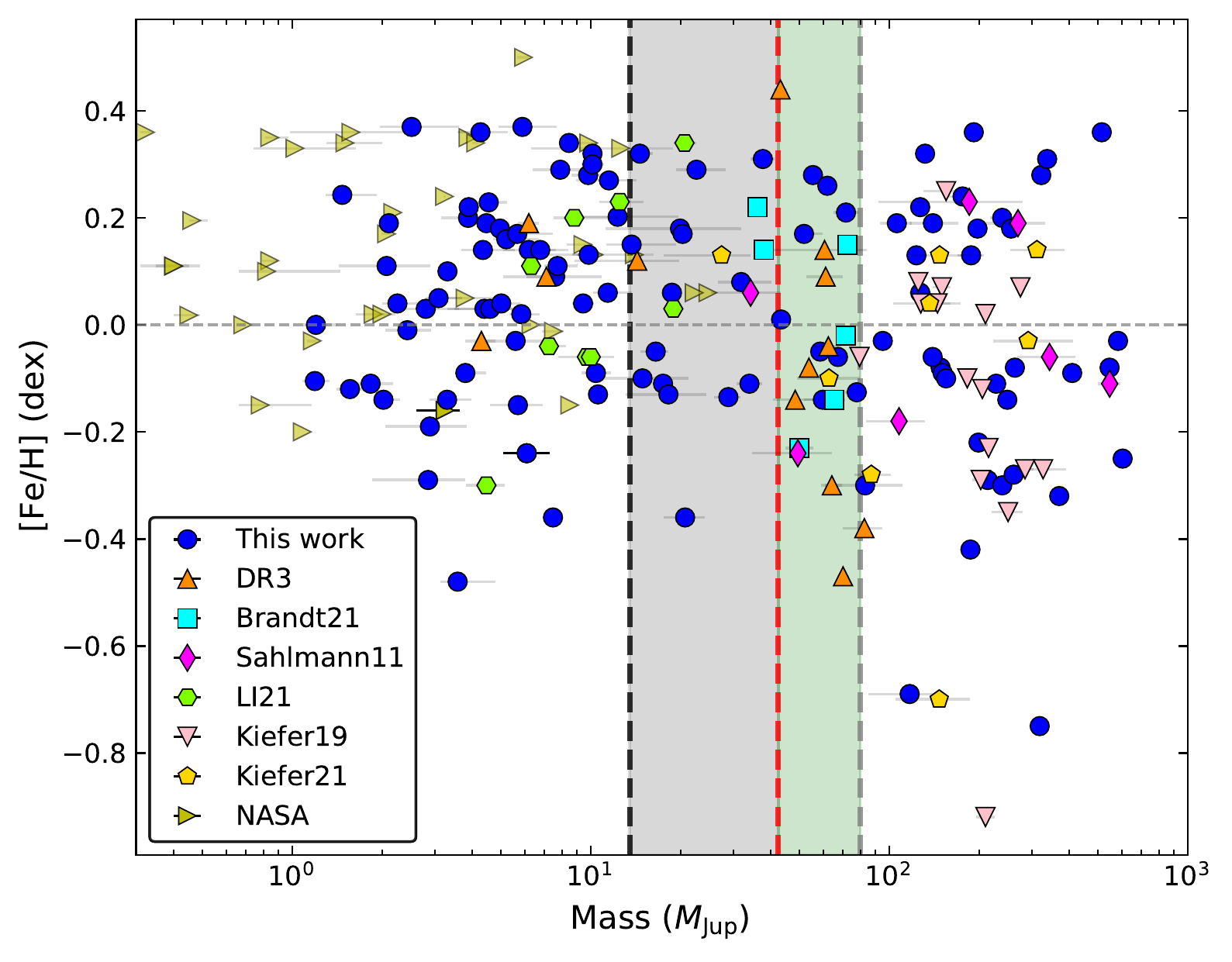}
    }
    \subcaptionbox{Mass-eccentricity diagram\label{subfig:mass_ecc}}[.495\linewidth]{
        \includegraphics[width=72mm]{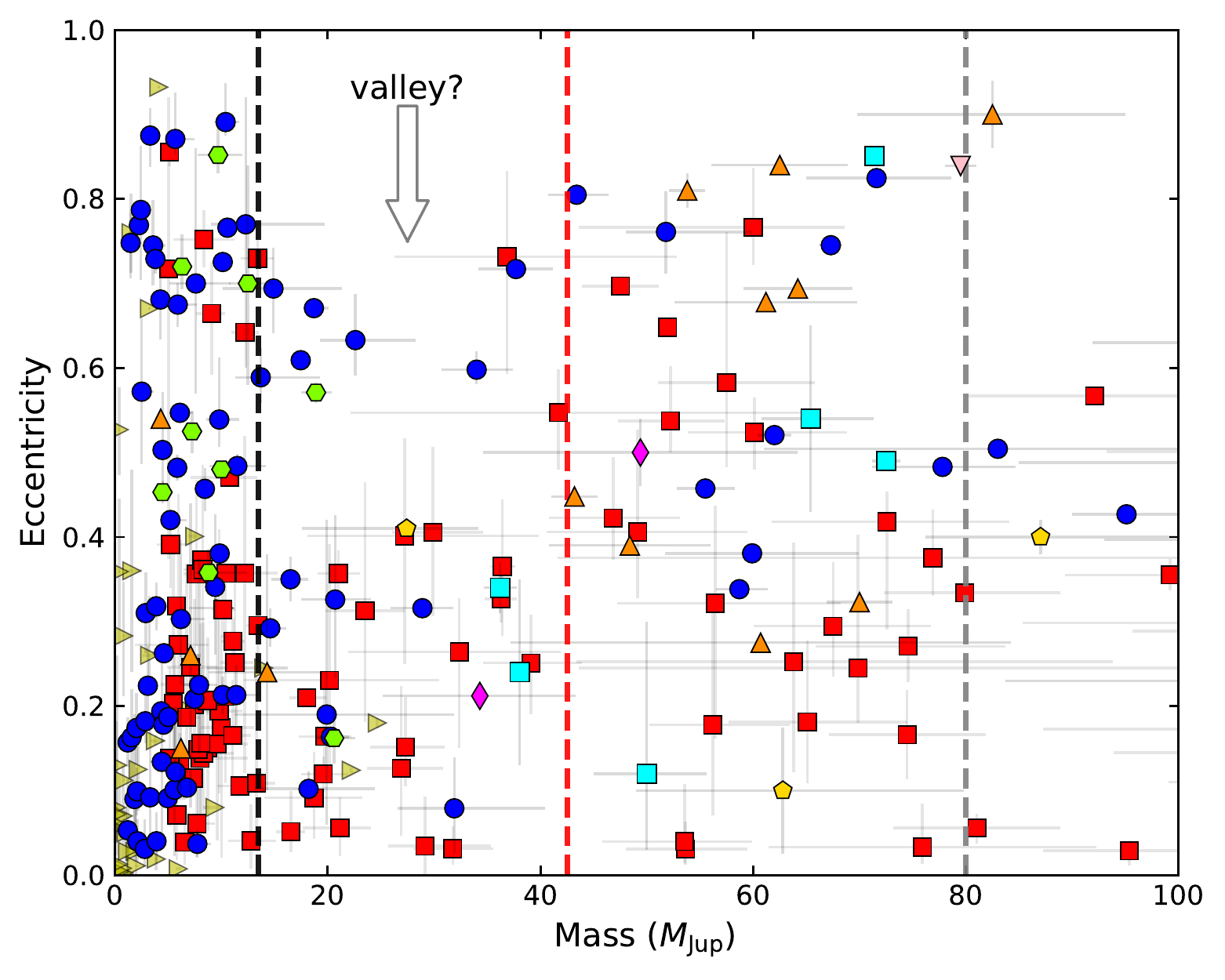}
    }
    \caption{\small (a) Mass-metallicity diagram.  The vertical black and grey dashed line indicates the classical boundaries of BDs ($\sim13.5$ and $\sim80$ $M_{\rm Jup}$), and the red dashed line represents the 42.5 $M_{\rm Jup}$ mass limit derived by \citet{Ma2014}. The grey and green shaded areas represent low-mass and high-mass BD regimes, respectively. (b) Mass-eccentricity diagram. A clear valley can be seen in the low-mass BD regime. The symbols are the same as in Figure \ref{Fig:mass_period}. 
    %(A complete sample will be available online as supplementary after the publication.)
    }
    \label{fig:mass_feh_ecc}
\end{figure}

Figure \ref{subfig:mass_feh} plots the mass-metallicity distribution for our samples along with literatures values, spanning planets, BDs, and the domain of low-mass M dwarfs. The data from \citet{Feng2022} are not included due to the lack of available metallicity. Figure \ref{subfig:mass_ecc} presents the eccentricity distributions with respect to companion's masses. The black and grey dashed line indicate the classical boundaries of BDs ($\sim13.5$ and $\sim80$ $M_{\rm Jup}$), and the red dashed line represents the 42.5 $M_{\rm Jup}$ mass limit derived by \citet{Ma2014}.

In Figure \ref{subfig:mass_feh}, We can see that the so-called brown dwarf desert is located in the transition region between giant planets and low-mass stellar binaries. Planets are inclined to orbit hosts with super-solar metallicity ($0.09\pm 0.19$ dex), while stellar binaries have a subsolar metallicity of $-0.07\pm 0.27$ dex spanning a larger range than planets. In the BD domain, the mass limit of 42.5 $M_{\rm Jup}$ seems to divide BDs into two groups, which may imply two different formation scenarios. One group preferentially orbits metal-rich stars like giant planets, and the other group spans a large metallicity range like stellar binaries. Thus we simply split the overall sample into four groups: $M_{\rm p}<13.5\ M_{\rm Jup}$, $13.5\leqslant M_{\rm p}<42.5\ M_{\rm Jup}$, $42.5\leqslant M_{\rm p}<80\ M_{\rm Jup}$ and $M_{\rm p}\geqslant80\ M_{\rm Jup}$. The two-sample Kolmogorov-Smirnov (K-S) test \footnote{Using the python \texttt{scipy.stats.ks\_2samp} library} is then applied to explore whether each two metallicity distributions are derived from the same parent distribution. The results are presented in Table \ref{Tab:ks_feh}. We find that BDs with $13.5\leqslant M_{\rm p}<42.5\ M_{\rm Jup}$ show strong evidence that they are part of the planetary population that primarily formed in the protoplanetary disc (core accretion or disc instability), whereas BDs with $M_{\rm p}\geqslant 42.5\ M_{\rm Jup}$ appear to belong to the stellar binary population that primarily formed through gravitational instability of molecular cloud like stars. Statistically, 
our result may also imply that core accretion can occur in low-mass and metal-rich BD regime ($13.5\leqslant M_{\rm p}<42.5\ M_{\rm Jup}$), which seems to agree with that of \citet{Maldonado2017}. However, it is worth noting that selection effects may have affected our results, as our samples are drawn from various RV surveys and some metal-poor planetary systems might be unexpectedly omitted by us for some reason (Teng et al. in prep).
\begin{table}
\begin{center}
\caption[]{ K-S Test of the Metallicity Distributions of the Star with Companions in Four Mass Regimes}\label{Tab:ks_feh}
%%Please Capitalize the First Letter of Each Notional Word in table's caption
 \begin{tabular}{|c|c|c|c|c|c|c|c|c|} \hline
\multirow{2}{*}{No} &  \multirow{2}{*}{Sample} & \multirow{2}{*}{N} & \multirow{2}{*}{\textless [Fe/H]\textgreater} & \multirow{2}{*}{STD} & \multicolumn{4}{c|}{K-S $p$-value}  \\
\cline{6-9} &&&&&A &B &C &D  \\ 
\hline
A  & $M_{\rm p}<13.5\ M_{\rm Jup}$ & 98     & 0.09  & 0.19 &- &0.72 &$9 \times 10^{-3} $ &$9 \times 10^{-4} $ \\ % new variable 
\hline
B  & $13.5 \leqslant M_{\rm p}<42.5\  M_{\rm Jup}$ &   25 &0.07 &0.17 &- &- &0.04 &0.04                 \\
\hline
C  & $42.5 \leqslant M_{\rm p}<80\  M_{\rm Jup}$     &   24     & $-0.02$ & 0.20  &- &- &- &0.51                \\
\hline
D  & $M_{\rm p}\geqslant80\ M_{\rm Jup}$     &   63     & $-0.07$ & 0.27  &\multicolumn{4}{c|}{-}                 \\
\hline
\end{tabular}
\end{center}
\end{table}

\begin{table}
\begin{center}
\caption[]{ K-S Test of the Eccentricity Distributions of the  Companions in Four Mass Regimes}\label{Tab:2d_ks_pecc}
%%Please Capitalize the First Letter of Each Notional Word in table's caption
 \begin{tabular}{|c|c|c|c|c|c|c|c|c|} \hline
\multirow{2}{*}{No} &  \multirow{2}{*}{Sample} & \multirow{2}{*}{N} & \multirow{2}{*}{\textless $e$\textgreater} & \multirow{2}{*}{STD} & \multicolumn{4}{c|}{K-S $p$-value}  \\
\cline{6-9} &&&&&A &B &C &D  \\ 
\hline
A  & $M_{\rm p}<13.5\ M_{\rm Jup}$ & 136& 0.32  & 0.25 &- &0.62 &$4 \times 10^{-4} $ &$4 \times 10^{-7} $ \\ % new variable 
\hline
B  & $13.5 \leqslant M_{\rm p}<42.5\  M_{\rm Jup}$ &46 &0.31 &0.20 &- &- &$7 \times 10^{-3} $ &$7 \times 10^{-4} $       \\
\hline
C  & $42.5 \leqslant M_{\rm p}<80\  M_{\rm Jup}$ & 46 & 0.46 & 0.24  &- &- &- &0.84        \\
\hline
D  & $M_{\rm p}\geqslant80\ M_{\rm Jup}$& 76& 0.46 & 0.22  &\multicolumn{4}{c|}{-}                 \\
\hline
\end{tabular}
\end{center}
\tablecomments{\textwidth}{\small The mean eccentricity of group B and C are significantly different.}
\end{table}

In Figure \ref{subfig:mass_ecc}, by the side of the apparent transition of $\sim42.5$ $M_{\rm Jup}$ reported by \citet{Ma2014} (see their figure 5), we find a relatively empty valley bounded by the upper profile of the BDs eccentricity distribution. The local minimum value seems to locate at $M_p =35\ M_{\rm Jup}$ and $e=0.6$. The decreased trend for BDs with masses below $35\ M_{\rm Jup}$ seems to agree with \citet{Ma2014} and \citet{Kiefer2021}, and may be explained by the planet-planet scattering model \citep{Ford2008} that the low-mass BDs formed in the protoplanetary disc might be pumped to higher eccentricity by other companions, while the heavier BDs are harder to scatter. According to the current core-accretion model, it is difficult to form more than one massive BDs to invoke the scattering mechanism. Still, under the disc gravitational instability assumption, it is possible to form such systems \citep{Ma2014,Forgan2013}. The existence of highly eccentric BDs with masses below $42.5\ M_{\rm Jup}$ may imply that disc gravitational instability also contributes to the formation of low-mass BDs \citep{Maldonado2017, Goda2019}. In addition, a two-sample Kolmogorov-Smirnov (K-S) test is performed on the eccentricity distribution of BDs with masses lower and greater than 42.5 $M_{\rm Jup}$. The low $p$-value of $7\times10^{-3}$ means that these two samples are unlikely to show the same distribution. Furthermore, we compare the distribution of BDs with giant planets and stellar binaries. In Table \ref{Tab:2d_ks_pecc}, our result also suggests that BDs with masses below 42.5 $M_{\rm Jup}$ show a similar eccentricity distribution to giant planets ($p=0.62$). 

Overall, considering the metallicity and eccentricity distributions of BDs, we propose that low-mass BDs with masses below 42.5 $M_{\rm Jup}$ may form in the protoplanetary disc through core accretion or disc gravitational instability, while high-mass BDs with masses above 42.5 $M_{\rm Jup}$ seem to unambiguously show similar formation mechanism with stellar binaries who primarily formed through gravitational instability of molecular cloud. 
However, given that almost all of our systems (blue circles) have been found to harbour only one companion, if the planet-planet scattering mechanism does work at highly eccentric and low-mass BDs, where did the other objects go? Have they fallen into the host star or been scattered into more distant orbits that exceed the current detection capability? Therefore, a larger BD sample is required to draw a more convincing conclusion in the future.

\section{Conclusion}
\label{sect:conclusion}
In this paper, we use the orbit fitting package \texttt{orvara} to measure the masses and inclinations of 115 RV-detected companions, including 55 planets, 24 brown dwarfs and 36 low-mass M dwarfs. Among them, nine planets are verified as BDs, and 16 BD candidates should be classified as M dwarfs. Our results show that  majority of planets with $M_{\rm p}\,{\rm sin}\,i<13.5\ M_{\rm Jup}$ are still planets, while half of the BD candidates have masses above the hydrogen-burning limit implying that the BD desert is more deficient than previous observations. 

We have updated the mass-period diagram using our mass measurements instead of the published minimum mass. In our $M$-$P$ distribution, we are still unable to verify the boundary of BDs at $\sim100$ days. Finally, the distributions of metallicity and eccentricity suggest that companions with mass smaller than 42.5 $M_{\rm Jup}$ may primarily form in the protoplanetary disc through core accretion or disc gravitational instability, and highly eccentric BD systems are more consistent with the prediction of the disc-instability model, while companions with mass above 42.5 $M_{\rm Jup}$ dominantly formed through gravitational instability like stars.

Again, it is important to note that our current sample can't completely rule out the existence of selection effects. The tools we used to measure the mass of companions may affect our results to some extent. Further observational evidence and detailed analysis are required to confirm our results. 
In addition to Gaia, there are other space-based astrometry missions proposed to operate in the next decade or so, such as the Closeby Habitable Exoplanet Survey (CHES: \citealt{Ji2022}) mission and the Theia mission \citep{Malbet2021}. They will make great efforts to detect Earth-like habitable exoplanets of nearby stars via microarcsecond astrometry, and directly measure their true masses. It is expected that these missions will lead to a thorough investigation of nearby habitable worlds and provide insights into the formation, evolution, and habitability of those planets.

\normalem
\begin{acknowledgements}
We thank the anonymous referee for providing great suggestions to improve the paper.
We thank Fabo Feng for providing their latest orbital parameters of 167 substellar companions in private.
This research is supported by the National Key R\&D
Program of China No. 2019YFA0405102 and 2019YFA0405502. This research is
also supported by the National Natural Science Foundation of China (NSFC) under Grant No. 12073044 and 11988101.

This research has made use of the SIMBAD database,
operated at CDS, Strasbourg, France \citep{Wenger2000}.
This research has made use of data obtained from or tools provided by the portal \href{http://exoplanet.eu/catalog/}{exoplanet.eu} of The Extrasolar
Planets Encyclopaedia. This research has made use of the NASA Exoplanet Archive, which is operated by the California Institute of Technology, under contract with the National Aeronautics and Space Administration under the Exoplanet Exploration Program. This work presents results from the European Space Agency (ESA) space mission Gaia. Gaia data are being processed by the Gaia Data Processing and Analysis Consortium (DPAC). Funding for the DPAC is provided by national institutions, in particular the institutions participating in the Gaia MultiLateral Agreement (MLA). The Gaia mission website is \href{https://www.cosmos.esa.int/gaia}{https://www.cosmos.esa.int/gaia}. The Gaia archive website is \href{https://archives.esac.esa.int/gaia}{https://archives.esac.esa.int/gaia}.

This research has made use of the following Python packages for scientific
calculation: \texttt{Numpy} \citep{Harris2020}, \texttt{Scipy} \citep{Virtanen2020}, \texttt{astropy} \citep{Astropy2013}, \texttt{pandas} \citep{Reback2022}, \texttt{matplotlib} \citep{Hunter2007}, \texttt{isochrones} \citep{Morton2015}, and \texttt{PyMultiNest} \citep{Feroz2019}.

\end{acknowledgements}

\appendix                  %%appendicial material is supported

\section{The additional tables}

\begin{table}
\centering
\caption{Stellar Parameters}\label{Tab:stellar1}
%%Please Capitalize the First Letter of Each Notional Word in table's caption
\resizebox{\textwidth}{!}{
\begin{tabular}{lccccccccccc}
\hline \hline
 Name & HIP\,ID &Spectral Type &$V$ &$B-V$&$\varpi  $&$T_{\rm eff}$ &${\rm log}g$ &[Fe/H] &$M_{\star}$ &$\chi _{\rm HGCA}^{2}$ &Refs\\
 &&& (mag) &(mag) &(mas)&(K)&(cgs)&(dex)&($M_{\tiny \sun}$)&&\\
\hline
GJ 179&22627&M2V&11.96&1.59&$80.56\pm 0.02$&$3370\pm 100$&$4.83$&$0.30\pm 0.10$&$0.357\pm 0.030$&35.1&1\\
HD\,126614 A&70623&G8IV&8.81&0.81&$13.66\pm 0.02$&$5585\pm 44$&$4.39\pm 0.08$&$0.56\pm 0.04$&$1.145\pm 0.030$&1976.69&1\\
HD\,13931&10626&G0 &7.61&0.642&$21.19\pm 0.03$&$5829\pm 44$&$4.3\pm 0.08$&$0.03\pm 0.04$&$1.022\pm 0.020$&12.87&1\\
HD\,120084&66903& G7 III&5.91&1.0&$9.63\pm 0.03$&$4892\pm 22$&$2.71\pm 0.08$&$0.09\pm 0.05$&$2.15\pm 0.21^\ast$&9.98&2\\
HD\,30562&22336&G2IV &5.77&0.63&$38.25\pm 0.04$&$5861\pm 44$&$4.09\pm 0.10$&$0.243\pm 0.04$&$1.219\pm 0.04$&2.42&3\\
HD\,86264&48780&F7V&7.42&0.46&$14.86\pm 0.02$&$6210\pm 44$&$4.02\pm 0.10$&$0.202\pm 0.04$&$1.42\pm 0.05$&29.92&3\\
HD\,89307&50473&G0V&7.06&0.64&$31.41\pm 0.02$&$5950\pm 44$&$4.414\pm 0.10$&$-0.14\pm 0.04$&$1.028\pm 0.04$&9.47&3\\
HD\,129445&72203&G6 V&8.8&0.756&$14.91\pm 0.01$&$5646\pm 42$&$4.28\pm 0.10$&$0.37\pm 0.03$&$1.09\pm 0.09$&7.47&4, 56\\
HD\,175167&93281&G5 IV/V&8.0&0.78&$14.04\pm 0.02$&$5635\pm 28$&$4.09\pm 0.09$&$0.28\pm 0.02$&$1.17\pm 0.09$&36.73&4, 56\\
HD\,136118&74948&F7V&6.93&0.55&$19.81\pm 0.03$&$6097\pm 44$&$4.053\pm 0.06$&$-0.05\pm 0.03$&$1.21\pm 0.06^\ast$&71.32&5, 57\\
HD\,196050&101806&G3V&7.5&0.67&$19.79\pm 0.02$&$5892\pm 44$&$4.267\pm 0.06$&$0.229\pm 0.03$&$1.18\pm 0.04^\ast$&47.28&5, 57\\
HD\,33636&24205&G0V&7.0&0.59&$33.80\pm 0.05$&$5904\pm 44$&$4.429\pm 0.06$&$-0.126\pm 0.03$&$1.01\pm 0.04^\ast$&55.6&5, 57\\
HD\,120066&67246&G0V&6.402&0.694&$31.78\pm 0.03$&$5794\pm 100$&$4.02\pm 0.10$&$0.10\pm 0.06$&$1.07\pm 0.04^\ast$&22.71&6\\
HD\,35956 A&25662&G0V&6.76&0.59&$33.79\pm 0.29$&$5932\pm 52$&$4.38\pm 0.03$&$-0.11\pm 0.04$&$1.04\pm 0.04$&565.24&7, 47\\
HD\,43587&29860&G0V&5.769&0.66&$51.62\pm 0.12$&$5914\pm 63$&$4.29\pm 0.03$&$-0.03\pm 0.03$&$0.99\pm 0.04^\ast$&22570.82&7, 47\\
HD\,142022 A&79242&G9IV-V&7.7&0.79&$29.20\pm 0.02$&$5499\pm 27$&$4.36\pm 0.04$&$0.19\pm 0.04$&$0.98\pm 0.04$&62.37&8\\
HD\,5608&4552&K0III&5.99&1.0&$17.07\pm 0.03$&$4854\pm 25$&$3.03\pm 0.08$&$0.06\pm 0.04$&$1.55\pm 0.11$&148.69&9\\
HD\,131664&73408&G3V&8.13&0.667&$19.14\pm 0.08$&$5886\pm 21$&$4.44\pm 0.10$&$0.32\pm 0.02$&$1.10\pm 0.03$&1822.68&10\\
HD\,73267&42202&K0V&8.9&0.806&$19.94\pm 0.01$&$5317\pm 34$&$4.28\pm 0.10$&$0.03\pm 0.02$&$0.89\pm 0.03$&0.31&10\\
HD\,167677&89583&G5V&7.9&0.705&$18.28\pm 0.02$&$5474\pm 65$&$4.43\pm 0.10$&$-0.290\pm 0.043$&$0.96\pm 0.02$&22.46&11\\
HD\,217786&113834&F8V&7.8&0.578&$18.19\pm 0.11$&$5966\pm 65$&$4.35\pm 0.11$&$-0.135\pm 0.043$&$1.02\pm 0.03$&97.7&11\\
HD\,89839&50653&F7V&7.64&0.523&$17.50\pm 0.02$&$6314\pm 65$&$4.49\pm 0.12$&$0.04\pm 0.043$&$1.21\pm 0.03$&56.44&11\\
HD\,108341&60788&K2 V&9.36&0.93&$20.43\pm 0.01$&$5122\pm 79$&$4.45\pm 0.14$&$0.04\pm 0.06$&$0.843\pm 0.024$&7.96&12\\
HD\,190228&98714&G5IV&7.296&0.8&$15.90\pm 0.02$&$5360\pm 40$&$4.02\pm 0.10$&$-0.24\pm 0.06$&$1.22\pm 0.05^\ast$&36.33&13\\
HD\,23596&17747&F8&7.24&0.61&$19.32\pm 0.03$&$6125\pm 50$&$4.29\pm 0.15$&$0.32\pm 0.05$&$1.32\pm 0.02^\ast$&126.66&13\\
HD\,50554&33212&F8V&6.84&0.57&$32.19\pm 0.02$&$6050\pm 50$&$4.59\pm 0.15$&$0.02\pm 0.06$&$1.1\pm 0.04^\ast$&19.99&13\\
HD\,117207&65808&G7IV-V&7.24&0.727&$30.94\pm 0.03$&$5732\pm 53$&$4.371\pm 0.039$&$0.19\pm 0.03$&$1.053\pm 0.028$&17.81&14\\
HD\,143361&78521&G6V&9.2&0.792&$14.55\pm 0.02$&$5507\pm 10$&$4.472\pm 0.043$&$0.14\pm 0.06$&$0.968\pm 0.027$&20.75&14\\
HD\,216437&113137&G1V&6.057&0.677&$37.46\pm 0.03$&$5909\pm 31$&$4.188\pm 0.026$&$0.20\pm 0.10$&$1.165\pm 0.046$&46.46&14\\
HD\,70642&40952&G6V&7.169&0.8&$34.15\pm 0.02$&$5732\pm 23$&$4.458\pm 0.017$&$0.22\pm 0.02$&$1.078\pm 0.015$&129.62&14\\
HD\,204313&106006&G5V&7.99&0.697&$20.77\pm 0.03$&$5767\pm 17$&$4.37\pm 0.05$&$0.18\pm 0.02$&$1.045\pm 0.033$&112.6&15\\
HD\,181234&95015&G8IV&8.59&0.841&$20.73\pm 0.03$&$5386\pm 60$&$4.25\pm 0.11$&$0.32\pm 0.05$&$1.01\pm 0.06$&339.82&16\\
HD\,25015&18527&K2V&8.87&0.899&$26.75\pm 0.02$&$5160\pm 63$&$4.40\pm 0.14$&$0.04\pm 0.04$&$0.86\pm 0.05$&191.61&16\\
HD\,92987&52472& G2/3V&7.03&0.641&$22.98\pm 0.05$&$5770\pm 36$&$4.00\pm 0.15$&$-0.08\pm 0.08$&$1.08\pm 0.06$&79814.91&16\\
HD\,10844&8285&F8V&8.13&0.63&$18.43\pm 0.03$&$5845\pm 37$&$4.43\pm 0.05$&$-0.06\pm 0.03$&$0.98\pm 0.07$&5230.83&17\\
HD\,14348&10868&F5V&7.19&0.6&$16.59\pm 0.02$&$6237\pm 47$&$4.51\pm 0.07$&$0.28\pm 0.03$&$1.20\pm 0.08$&872.32&17\\
HD\,29461&21654&G5&7.945&0.648&$18.39\pm 0.30$&$5868\pm 25$&$4.47\pm 0.028$&$0.22\pm 0.01$&$1.06\pm 0.07$&431.37&17, 49\\
BD+210055&2397&K2&9.24&0.94&$27.29\pm 0.22$&$4833\pm 73$&$4.38\pm 0.24$&$-0.22\pm 0.09$&$0.77\pm 0.05$&2304.75&18\\
HD\,101305&56859&F6V&8.33&0.54&$14.13\pm 0.12$&$6040\pm 27$&$4.12\pm 0.20$&$-0.28\pm 0.02$&$0.99\pm 0.03$&2167.04&18\\
HD\,103913&58364&F8&8.28&0.52&$11.54\pm 0.10$&$5964\pm 27$&$3.93\pm 0.20$&$-0.10\pm 0.02$&$1.12\pm 0.04$&1469.41&18\\
HD\,130396&72336&F8V&7.45&0.5&$22.06\pm 0.24$&$6349\pm 26$&$4.18\pm 0.21$&$-0.03\pm 0.02$&$1.11\pm 0.03$&214.34&18\\
HD\,156728&84520&G5&8.03&0.64&$24.10\pm 0.20$&$5777\pm 21$&$4.35\pm 0.20$&$-0.14\pm 0.02$&$0.92\pm 0.03$&2833.68&18\\
HD\,211681&109169&G5&8.09&0.74&$13.83\pm 0.02$&$5793\pm 30$&$4.00\pm 0.20$&$0.36\pm 0.02$&$1.23\pm 0.09$&2907.44&18\\
HD\,217850&113789&G8V&8.5&0.8&$16.16\pm 0.52$&$5605\pm 30$&$4.13\pm 0.20$&$0.28\pm 0.02$&$1.08\pm 0.04$&342.36&18\\
HD\,23965&17928&F7&7.27&0.54&$23.21\pm 0.02$&$6423\pm 52$&$4.34\pm 0.21$&$0.01\pm 0.04$&$1.12\pm 0.03$&40.37&18\\
HD\,28635&21112&F9V&7.75&0.55&$20.28\pm 0.17$&$6238\pm 22$&$4.14\pm 0.20$&$0.18\pm 0.02$&$1.17\pm 0.08$&3364.93&18\\
HD\,48679&33548&G0&8.85&0.75&$14.93\pm 0.08$&$5621\pm 25$&$4.21\pm 0.20$&$0.21\pm 0.02$&$1.03\pm 0.03$&122.6&18\\
HD\,5470&4423&G0&8.33&0.64&$14.91\pm 0.04$&$6047\pm 29$&$4.12\pm 0.20$&$0.31\pm 0.02$&$1.19\pm 0.08$&13146.32&18\\
HD\,77712&44520&K1/2(V)&8.93&0.85&$20.50\pm 0.28$&$5309\pm 44$&$4.37\pm 0.20$&$0.18\pm 0.03$&$0.91\pm 0.04$&244.17&18\\
HD\,87899&49738&G5&8.88&0.65&$19.01\pm 0.19$&$5581\pm 23$&$4.38\pm 0.19$&$-0.30\pm 0.02$&$0.85\pm 0.06$&777.8&18\\
HD\,69123&40344& K1III&5.77&1.02&$13.30\pm 0.03$&$4842\pm 41$&$2.86\pm 0.11$&$0.05\pm 0.03$&$1.68\pm 0.09$&0.92&19\\
HD\,150706&80902&G0V&7.016&0.61&$35.48\pm 0.01$&$5961\pm 27$&$4.5\pm 0.10$&$-0.01\pm 0.04$&$1.17\pm 0.12$&4.7&20\\
HD\,222155&116616&G2V&7.188&0.714&$19.80\pm 0.02$&$5765\pm 22$&$4.1\pm 0.13$&$-0.11\pm 0.05$&$1.13\pm 0.11$&15.51&20\\
HD\,190984&99496&F8V&9.27&0.579&$6.71\pm 0.02$&$5988\pm 25$&$4.02\pm 0.22$&$-0.48\pm 0.06$&$0.91\pm 0.10$&1.9&21\\
HD\,224538&118228&F8/G0IV/V&8.06&0.581&$12.64\pm 0.02$&$6097\pm 100$&$4.19\pm 0.04$&$0.27\pm 0.10$&$1.34\pm 0.05$&34.84&22\\
HD\,68402&39589&G5IV/V&9.11&0.66&$12.72\pm 0.01$&$5950\pm 100$&$4.37\pm 0.05$&$0.29\pm 0.10$&$1.12\pm 0.05$&35.66&22\\
HD\,154697&83770&G6V&7.97&0.73&$29.40\pm 0.20$&$5648\pm 50$&$4.42\pm 0.10$&$0.13\pm 0.06$&$0.96\pm 0.02$&1209.55&23\\
HD\,167665&89620&F9V&6.48&0.54&$32.40\pm 0.09$&$6224\pm 50$&$4.44\pm 0.10$&$-0.05\pm 0.06$&$1.14\pm 0.03$&2387.88&23\\
HD\,211847&110340&G5V&8.78&0.66&$20.52\pm 0.03$&$5715\pm 50$&$4.49\pm 0.10$&$-0.08\pm 0.06$&$0.94\pm 0.04$&4972.69&23\\
HD\,30501&22122&K2V&7.73&0.88&$49.18\pm 0.03$&$5223\pm 50$&$4.56\pm 0.10$&$-0.06\pm 0.06$&$0.81\pm 0.02$&8431.95&23\\
HD\,53680&34052&K6V&8.61&0.9&$57.79\pm 0.34$&$5167\pm 94$&$5.37\pm 0.29$&$-0.29\pm 0.08$&$0.79\pm 0.02$&-&23\\
\hline
\end{tabular}
}
\end{table}

\begin{table}
\begin{center}

\caption{Stellar Parameters}\label{Tab:stellar2}
%%Please Capitalize the First Letter of Each Notional Word in table's caption
\resizebox{\textwidth}{!}{
\begin{tabular}{lccccccccccc}
\hline \hline
 Name & HIP\,ID &Spectral Type &$V$ &$B-V$&$\varpi  $&$T_{\rm eff}$ &${\rm log}g$ &[Fe/H] &$M_{\star}$ &$\chi _{\rm HGCA}^{2}$ &Refs\\
 &&& (mag) &(mag) &(mas)&(K)&(cgs)&(dex)&($M_{\tiny \sun}$)&&\\
\hline
HD\,74014&42634&K0III&7.73&0.76&$28.73\pm 0.03$&$5662\pm 55$&$4.39\pm 0.10$&$0.26\pm 0.08$&$1.00\pm 0.03$&593.19&23\\
HIP\,103019&103019&K6.5V&10.39&1.33&$35.70\pm 0.63$&$4913\pm 115$&$4.45\pm 0.28$&$-0.30\pm 0.06$&$0.70\pm 0.01$&26.1&23\\
HD\,103891&58331&F8V&6.55&0.567&$18.22\pm 0.04$&$6072\pm 20$&$3.79\pm 0.03$&$-0.19\pm 0.01$&$1.28\pm 0.01$&15.41&24\\
HD\,106270&59625&G5IV&7.58&0.74&$10.53\pm 0.03$&$5567\pm 11$&$3.76\pm 0.03$&$0.06\pm 0.01$&$1.48\pm 0.06^\ast$&5.65&25, 48\\
HD\,10697&8159&G3V&6.279&0.7&$30.15\pm 0.04$&$5641\pm 28$&$4.05\pm 0.05$&$0.14\pm 0.04$&$1.13\pm 0.04^\ast$&14.55&25, 56\\
HD\,112988&63458&G0&7.76&0.92&$8.62\pm 0.03$&$4906\pm 11$&$3.16\pm 0.03$&$-0.32\pm 0.01$&$1.22\pm 0.17^\ast$&23031.55&25, 48\\
HD\,125390&69888&G7III&8.59&0.69&$6.42\pm 0.02$&$4882\pm 29$&$3.04\pm 0.04$&$-0.11\pm 0.02$&$1.60\pm 0.07^\ast$&182.81&25, 48\\
HD\,145428&79364&K0III&7.75&1.02&$8.09\pm 0.05$&$4836\pm 32$&$3.05\pm 0.07$&$-0.25\pm 0.02$&$1.02\pm 0.10^\ast$&10202.66&25, 48\\
HD\,18015&13467&G6IV&7.9&0.67&$8.01\pm 0.02$&$5643\pm 15$&$3.63\pm 0.07$&$-0.14\pm 0.01$&$1.50\pm 0.03^\ast$&0.47&25, 48\\
HD\,18667&13989&G6/8IV&8.3&0.97&$5.60\pm 0.03$&$4928\pm 25$&$3.11\pm 0.06$&$-0.09\pm 0.01$&$1.46\pm 0.16^\ast$&932.12&25, 48\\
HD\,21340&15969&K0III&7.39&0.96&$7.23\pm 0.05$&$4948\pm 25$&$3.07\pm 0.03$&$-0.09\pm 0.01$&$1.59\pm 0.14^\ast$&616.75&25, 48\\
HD\,97601&54908&G5&7.45&0.89&$8.33\pm 0.08$&$5112\pm 25$&$3.20\pm 0.09$&$-0.08\pm 0.02$&$1.66\pm 0.16^\ast$&3796.65&25, 48\\
HIP\,97233&97233&K0/1III&7.34&1.0&$9.84\pm 0.03$&$5020\pm 100$&$3.26\pm 0.20$&$0.29\pm 0.13$&$1.74\pm 0.20^\ast$&22.03&26\\
HIP\,84056&84056& K1III&6.81&1.03&$13.37\pm 0.03$&$4960\pm 100$&$3.17\pm 0.20$&$0.08\pm 0.07$&$1.69\pm 0.14$&106.13&27\\
HIP\,8541&8541& K2III/ IV&7.88&1.08&$6.50\pm 0.02$&$4670\pm 100$&$2.70\pm 0.20$&$-0.15\pm 0.08$&$1.17\pm 0.28$&2.75&27\\
HIP\,67537&67537& K1III&6.44&0.99&$8.42\pm 0.03$&$4985\pm 100$&$2.85\pm 0.20$&$0.15\pm 0.08$&$2.41\pm 0.16$&12.09&28\\
HIP\,56640&56640& K1III&7.93&1.09&$8.22\pm 0.02$&$4769\pm 55$&$2.91\pm 0.12$&$-0.03\pm 0.05$&$1.04\pm 0.07$&15.56&29\\
HD\,111232&62534&G8V&7.59&0.701&$34.61\pm 0.02$&$5494\pm 100$&$4.50\pm 0.10$&$-0.36\pm 0.10$&$0.83\pm 0.03^\ast$&170.46&30\\
HD\,139357&76311&K4III&5.964&1.2&$8.81\pm 0.05$&$4700\pm 70$&$2.90\pm 0.15$&$-0.13\pm 0.05$&$1.35\pm 0.24$&14.53&31\\
HD\,203473&105521&G6V&8.284&0.75&$13.74\pm 0.04$&$5780\pm 25$&$4.11\pm 0.028$&$0.19\pm 0.01$&$1.12\pm 0.21$&1435.64&10, 49\\
HD\,103459&58093&G5V&7.6&0.68&$16.89\pm 0.12$&$5721\pm 25$&$4.03\pm 0.028$&$0.24\pm 0.01$&$1.18\pm 0.20$&1700.51&32, 49\\
HD\,214823&111928&G0&8.06&0.631&$9.88\pm 0.03$&$6215\pm 30$&$4.05\pm 0.10$&$0.17\pm 0.02$&$1.22\pm 0.13$&91.87&32, 52\\
HD\,3404&2902&G2V&7.92&0.82&$12.62\pm 0.05$&$5339\pm 25$&$3.81\pm 0.028$&$0.20\pm 0.01$&$1.17\pm 0.22$&1018.64&32, 49\\
HD\,55696&34801&G0V&7.93&0.61&$12.81\pm 0.01$&$6012\pm 25$&$4.15\pm 0.028$&$0.36\pm 0.01$&$1.29\pm 0.20$&8.4&32, 49\\
HD\,213240&111143&G0/1V&6.81&0.603&$24.42\pm 0.02$&$5975\pm 100$&$4.32\pm 0.10$&$0.16\pm 0.10$&$1.22\pm 0.05^\ast$&15.3&33\\
HD\,115954&65042&G5V&8.34&0.64&$11.45\pm 0.03$&$5957\pm 26$&$4.15\pm 0.04$&$0.34\pm 0.02$&$1.18\pm 0.06$&5.87&34\\
HD\,80869&46022&G5&8.45&0.68&$11.81\pm 0.02$&$5837\pm 15$&$4.18\pm 0.03$&$0.17\pm 0.01$&$1.08\pm 0.05$&4.3&34\\
HD\,95544&54203&G0&8.39&0.72&$11.39\pm 0.02$&$5722\pm 15$&$4.07\pm 0.03$&$0.11\pm 0.01$&$1.09\pm 0.07$&42.91&34\\
BD+730275&24329&G5&8.98&0.76&$21.82\pm 0.31$&$5260\pm 25$&$4.43\pm 0.04$&$-0.42\pm 0.02$&$0.77\pm 0.05$&229.49&35\\
HD\,122562&68578&G5&7.69&1.01&$18.82\pm 0.03$&$4958\pm 67$&$3.74\pm 0.14$&$0.31\pm 0.04$&$1.13\pm 0.13$&316.12&35\\
HD\,283668&20834&K2&9.478&1.04&$33.99\pm 0.30$&$4845\pm 66$&$4.35\pm 0.12$&$-0.75\pm 0.12$&$0.68\pm 0.06$&-&35\\
HD\,51813&33608&G&8.67&0.6&$16.57\pm 0.27$&$6012\pm 32$&$4.53\pm 0.05$&$0.13\pm 0.02$&$1.06\pm 0.07$&148.65&35\\
HD\,94386&53259&K2III&6.34&1.21&$12.43\pm 0.23$&$4558\pm 100$&$2.80\pm 0.10$&$0.19\pm 0.10$&$1.19\pm 0.19$&31.96&36\\
HD\,132406&73146&G0 V&8.45&0.65&$14.18\pm 0.02$&$5766\pm 23$&$4.19\pm 0.03$&$0.14\pm 0.02$&$1.09\pm 0.05$&4.53&37, 58\\
HD\,16175&12191&G2&7.291&0.66&$16.67\pm 0.03$&$6022\pm 34$&$4.21\pm 0.06$&$0.37\pm 0.03$&$1.34\pm 0.14$&9.07&38\\
HD\,191806&99306&K0&8.093&0.64&$15.20\pm 0.02$&$6010\pm 30$&$4.45\pm 0.03$&$0.30\pm 0.02$&$1.14\pm 0.12$&19.17&38\\
HD\,30246&22203&G5&8.28&0.67&$20.50\pm 0.08$&$5833\pm 44$&$4.39\pm 0.04$&$0.17\pm 0.10$&$1.05\pm 0.04$&57.34&38\\
BD+631405&88617& K0&9.065&0.98&$26.24\pm 0.01$&$5000\pm 53$&$4.20\pm 0.17$&$-0.09\pm 0.03$&$0.816\pm 0.083$&173.41&39\\
HD\,184601&96049&G0&8.28&0.47&$12.96\pm 0.11$&$6035\pm 50$&$4.17\pm 0.04$&$-0.69\pm 0.03$&$0.954\pm 0.070$&21.31&39\\
HD\,205521&105906& G5&8.129&0.91&$20.65\pm 0.38$&$5570\pm 36$&$4.20\pm 0.07$&$0.36\pm 0.03$&$1.10\pm 0.082$&991.04&39\\
HD\,154345&83389&G9&6.76&0.73&$54.74\pm 0.02$&$5468\pm 44$&$4.537\pm 0.060$&$-0.105\pm 0.030$&$0.88\pm 0.09$&12.81&40\\
HD\,14067&10657&G9III&6.51&1.04&$7.11\pm 0.02$&$4815\pm 100$&$2.61\pm 0.10$&$-0.10\pm 0.08$&$2.4\pm 0.2$&13.89&41\\
HD\,175679&92968& G8III&6.14&0.96&$5.95\pm 0.04$&$4844\pm 100$&$2.59\pm 0.10$&$-0.14\pm 0.10$&$2.7\pm 0.3$&39.91&42\\
HD\,166724&89354&K0 IV/V&9.33&0.861&$22.03\pm 0.02$&$5127\pm 52$&$4.43\pm 0.08$&$-0.09\pm 0.03$&$0.81\pm 0.02$&27.07&43\\
HD\,219077&114699&G8V+&6.12&0.787&$34.25\pm 0.02$&$5362\pm 18$&$4.00\pm 0.03$&$-0.13\pm 0.01$&$1.05\pm 0.02$&6.9&43\\
HD\,220689&115662&G3V&7.74&0.603&$21.31\pm 0.02$&$5921\pm 26$&$4.32\pm 0.03$&$0.00\pm 0.03$&$1.04\pm 0.03$&0.56&43\\
HD\,27631&20199&G3 IV/V&8.26&0.682&$19.93\pm 0.02$&$5737\pm 36$&$4.48\pm 0.09$&$-0.12\pm 0.05$&$0.94\pm 0.04$&0.72&43\\
HD\,32963&23884&G5IV&7.59&0.6&$26.13\pm 0.02$&$5727\pm 32$&$4.41\pm 0.03$&$0.11\pm 0.05$&$1.03\pm 0.05$&23.47&44\\
GJ 832&106440&M1V&8.672&1.5&$201.33\pm 0.02$&$3472$&$4.7$&$-0.3$&$0.45\pm 0.05$&278.28&45\\
HD\,74156&42723&G0&7.6&0.58&$17.42\pm 0.02$&$6068\pm 44$&$4.259\pm 0.060$&$0.131\pm 0.030$&$1.238\pm 0.042$&112.85&46\\
HD\,165131&88595&G3/5V&8.41&0.65&$17.27\pm 0.03$&$5870\pm 100$&$4.39\pm 0.10$&$0.06\pm 0.10$&$1.06\pm 0.05^\ast$&265.0&55\\
HD\,62364&36941&F7V&7.31&0.53&$18.88\pm 0.02$&$6255\pm 100$&$4.29\pm 0.10$&$-0.11\pm 0.10$&$1.20\pm 0.04^\ast$&363.0&55\\
\hline
\end{tabular}
}
\end{center}
\tablecomments{\textwidth}{The stellar masses labelled $^\ast$ are determined by \texttt{isochrones} in this work. (A machine-readable table will be available online as supplementary
after the publication.)}
\tablerefs{\textwidth}{(1) \citet{Howard2010}; (2) \citet{Sato2013}; (3) \citet{Fischer2009}; (4) \citet{Arriagada2010}; (5) \citet{Butler2006}; (6) \citet{Blunt2019}; (7) \citet{Vogt2002}; (8) \citet{Eggenberger2006}; (9) \citet{Sato2012}; (10) \citet{Moutou2009}; (11) \citet{Moutou2011}; (12) \citet{Moutou2015}; (13) \citet{Perrier2003}; (14) \citet{Barbato2018}; (15) \citet{Segransan2010}; (16) \citet{Rickman2019}; (17) \citet{Bouchy2016}; (18) \citet{Kiefer2019}; (19) \citet{Ottoni2022}; (20) \citet{Boisse2012}; (21) \citet{Santos2010}; (22) \citet{Jenkins2017}; (23) \citet{Sahlmann2011}; (24) \citet{Sreenivas2022}; (25) \citet{Luhn2019}; (26) \cite{Jones2015}; (27) \citet{Jones2016}; (28) \citet{Jones2017}; (29) \citet{Jones2021}; (30) \citet{Mayor2004}; (31) \citet{Dollinger2009}; (32) \citet{Ment2018}; (33) \citet{Santos2001}; (34) \citet{Demangeon2021}; (35) \citet{Wilson2016}; (36) \citet{Wittenmyer2016}; (37) \citet{daSilva2007}; (38) \citet{Diaz2012}; (39) \citet{Dalal2021}; (40) \citet{Wright2007};(41) \citet{Wang2014}; (42) \citet{Wang2012}; (43) \citet{Marmier2013}; (44) \citet{Rowan2016}; (45) \citet{Wittenmyer2014}; (46) \citet{Feng2015}; (47) \citet{Aguilera2018}; (48) \citet{Ghezzi2018}; (49) \citet{Brewer2016}.}
\end{table}

\begin{table}
\centering
\caption{Published RV Data for Our Sample}\label{Tab:RV}
%%Please Capitalize the First Letter of Each Notional Word in table's caption
\resizebox{\textwidth}{!}{
\begin{tabular}{llccccllcccc}
\hline \hline
 Name & Instrument &$N_{\rm obs} $ &$\langle \sigma_{\rm RV}\rangle$ &Time span &Refs & Name & Instrument &$N_{\rm obs} $ &$\langle \sigma_{\rm RV}\rangle$ &Time span &Refs\\
 &&& (${\rm m\ s}^{-1}$) &(days) & &&&& (${\rm m\ s}^{-1}$) &(days) &\\
\hline
GJ 179&${\rm HIRES}^a$&30&3.3&4774&1, 51&&CORALIE07&8&4.9&&\\
&HET&14&8.4&&&HD\,53680&CORALIE98&36&8.5&3337&23\\
&HARPS&22&2.7&&&&CORALIE07&15&6.5&&\\
HD\,126614&${\rm HIRES}^a$&89&1.4&5508&1&HD\,154697&CORALIE98&48&8.0&3959&23, 50\\
HD\,13931&HIRES&17&1.4&5837&1, 50&&CORALIE07&3&3.6&&\\
&HIRES+&36&1.5&&&&${\rm HIRES}^a$&4&1.7&&\\
HD\,120084&HIDES&33&4.4&3530&2&HD\,167665&CORALIE98&28&5.8&5104&23, 50\\
HD\,30562&Lick&45&5.4&3690&3&&CORALIE07&12&6.6&&\\
HD\,86264&Lick&37&18.6&2951&3&&${\rm HIRES}^a$&24&3.9&&\\
HD\,89307&Lick&59&6.5&4818&3, 5, 20&HD\,74014&CORALIE98&109&5.6&4930&23, 51\\
&SOPHIE&11&4.4&&&&CORALIE07&10&4.0&&\\
&ELODIE&46&12.7&&&&HARPS&26&0.3&&\\
HD\,129445&MIKE&17&4.8&2153&4&HIP\,103019&HARPS&30&1.7&2179&23, 51\\
HD\,175167&MIKE&13&4.2&1828&4&HD\,103891&HARPS&66&1.9&5153&24\\
HD\,136118&Lick&37&16.1&1617&5&&HARPS+&21&1.6&&\\
HD\,196050&CORALIE98&31&5.5&3139&5, 51&HD\,10697&${\rm HIRES}^a$&81&1.4&7439&25, 50\\
&AAT&44&4.7&&&&HET&40&9.1&&\\
&HARPS&37&0.4&&&&HJS&32&7.6&&\\
HD\,33636&Lick&12&10.8&3289&5, 7, 50&HD\,145428&HIRES&10&1.4&1838&25\\
&${\rm HIRES}^a$&26&3.4&&&HD\,106270&HIRES&29&1.6&3195&25\\
&HET&67&3.3&&&HD\,112988&HIRES&20&1.4&2258&25\\
HD\,120066&TULL&175&5.0&8212&6&HD\,125390&HIRES&15&1.6&2268&25\\
&APF&104&3.1&&&HD\,18015&HIRES&25&1.9&3044&25\\
&${\rm HIRES}^a$&21&1.4&&&HD\,18667&HIRES&15&1.5&2194&25\\
&HIRES+&57&1.6&&&HD\,21340&HIRES&12&1.2&2188&25\\
HD\,35956&HIRES&14&4.2&1822&7&HD\,97601&HIRES&16&1.4&3501&25\\
HD\,43587&HIRES&14&4.2&1637&7&HIP\,97233&CHIRON&19&5.3&1519&26\\
HD\,142022&CORALIE98&70&8.4&2161&8&&FEROS&22&6.4&&\\
&HARPS&6&1.0&&&HIP\,8541&CHIRON&23&4.6&2194&27\\
HD\,5608&HIRES&9&1.1&4366&9, 50&&FEROS&9&4.1&&\\
&HIDES&43&4.0&&&&AAT&6&2.0&&\\
HD\,131664&HARPS&60&1.9&4707&10, 51&HIP\,84056&CHIRON&22&4.2&1888&27, 36\\
HD\,73267&HARPS&65&1.4&6340&10, 51&&FEROS&19&4.7&&\\
&HARPS+&13&1.0&&&&AAT&21&2.1&&\\
HD\,167677&HARPS&40&1.9&5204&11, 51&HIP\,67537&CHIRON&18&4.8&4568&28\\
&HARPS+&3&1.1&&&&FEROS&20&4.4&&\\
HD\,217786&HARPS&27&1.6&2982&11, 51&HIP\,56640&AAT&5&2.3&3027&29\\
HD\,89839&HARPS&70&2.8&6191&11, 51&&FEROS&22&3.8&&\\
&HARPS+&27&1.4&&&HD\,111232&CORALIE98&38&6.0&7798&30, 54\\
HD\,108341&HARPS&52&1.8&5167&12, 51&&MIKE&15&3.7&&\\
HD\,190228&ELODIE&51&8.7&6557&13, 25, 50&&HARPS&50&0.4&&\\
&${\rm HIRES}^a$&33&1.3&&&&HARPS+&41&0.7&&\\
HD\,23596&ELODIE&39&9.1&3603&13, 5, 50&HD\,139357&TLS&49&8.1&1286&31\\
&HRS&63&10.2&&&HD\,203473&${\rm HIRES}^a$&36&1.3&4548&32, 51\\
HD\,50554&Lick&29&11.5&5939&13&&HARPS&12&0.5&&\\
&${\rm HIRES}^a$&38&1.7&&&HD\,103459&${\rm HIRES}^a$&32&1.6&3624&32, 50\\
&ELODIE&41&10.0&&&HD\,55696&${\rm HIRES}^a$&28&3.1&4701&32, 50\\
HD\,117207&HARPS&31&0.3&7151&14, 5, 50, 51&HD\,3404&${\rm HIRES}^a$&14&1.4&3509&32\\
&${\rm HIRES}^a$&51&1.5&&&HD\,214823&ELODIE&5&21.2&4014&32, 25, 50, 52\\
&HARPS2&56&0.4&&&&SOPHIE&13&6.0&&\\
HD\,70642&HARPS&25&0.8&6274&14, 5, 51&&SOPHIE+&11&5.8&&\\
&${\rm HIRES}^a$&28&3.2&&&&HIRES&28&1.8&&\\
HD\,143361&HARPS&55&1.3&4416&14, 51&HD\,213240&AAT&30&4.4&2541&33, 5, 51\\
&CORALIE07&45&14.4&&&&CORALIE98&72&6.7&&\\
&MIKE&17&3.5&&&&HARPS&3&0.5&&\\
HD\,216437&HARPS&33&0.3&6123&14, 51&HD\,115954&ELODIE&4&15.5&5065&34\\
&AAT&21&5.3&&&&SOPHIE&6&6.5&&\\
&CORALIE98&39&3.9&&&&SOPHIE+&38&4.7&&\\
HIP\,106006&HARPS&93&0.6&5162&15, 52&HD\,80869&ELODIE&22&17.6&5443&34\\
&CORALIE98&48&5.2&&&&SOPHIE&4&6.4&&\\
&CORALIE07&52&3.4&&&&SOPHIE+&33&2.5&&\\
&TULL&36&5.2&&&HD\,95544&SOPHIE+&23&3.5&2278&34\\
\hline
\end{tabular}
}
\end{table}

\begin{table}
\begin{center}
\caption{Published RV Data for Our Sample}\label{Tabc}
%%Please Capitalize the First Letter of Each Notional Word in table's caption
\resizebox{\textwidth}{!}{
\begin{tabular}{llccccllcccc}
\hline \hline
 Name & Instrument &$N_{\rm obs} $ &$\langle \sigma_{\rm RV}\rangle$ &Time span &Refs & Name & Instrument &$N_{\rm obs} $ &$\langle \sigma_{\rm RV}\rangle$ &Time span &Refs\\
 &&& (${\rm m\ s}^{-1}$) &(days) & &&&& (${\rm m\ s}^{-1}$) &(days) &\\
\hline
HD\,181234&CORALIE07&20&4.5&7223&16&BD+730275&SOPHIE&25&3.5&3282&35\\
&CORALIE14&59&4.4&&&HD\,122562&SOPHIE&17&2.4&2961&35\\
&CORALIE98&15&6.2&&&&SOPHIE+&12&2.9&&\\
&${\rm HIRES}^a$&19&1.2&&&HD\,283668&SOPHIE&12&6.3&2751&35\\
HD\,25015&CORALIE07&32&5.2&6355&16&&SOPHIE+&4&6.3&&\\
&CORALIE14&56&5.8&&&HD\,51813&SOPHIE&3&3.6&2300&35\\
&CORALIE98&22&9.1&&&&SOPHIE+&11&6.8&&\\
HD\,92987&CORALIE07&18&3.4&7373&16, 53&HD\,94386&AAT&14&1.8&1511&36\\
&CORALIE14&29&3.5&&&HD\,132406&ELODIE&17&12.0&1078&37\\
&CORALIE98&53&4.6&&&&SOPHIE&4&3.8&&\\
HD\,10844&ELODIE&25&17.5&4668&17&HD\,30246&SOPHIE&23&6.1&1436&38\\
&SOPHIE&27&3.2&&&HD\,191806&ELODIE&6&12.9&3878&38\\
HD\,14348&ELODIE&61&10.8&6153&17&&SOPHIE&27&4.0&&\\
&SOPHIE&38&2.9&&&&SOPHIE+&17&4.6&&\\
HD\,29461&${\rm HIRES}^a$&20&1.5&4796&17&HD\,16175&ELODIE&3&9.0&3986&38, 50\\
BD+210055&SOPHIE+&20&6.4&2303&18&&SOPHIE+&25&4.5&&\\
HD\,101305&SOPHIE+&21&7.4&1551&18&&HIRES&6&1.7&&\\
HD\,103913&SOPHIE&13&7.7&4112&18&&Lick&44&6.1&&\\
&SOPHIE+&9&6.5&&&BD+631405&SOPHIE+&18&2.4&1064&39\\
HD\,130396&SOPHIE+&30&5.5&2155&18&HD\,184601&SOPHIE+&16&4.8&2277&39\\
HD\,156728&SOPHIE+&12&4.4&2264&18&HD\,205521&SOPHIE+&19&2.3&2537&39\\
HD\,23965&SOPHIE&19&10.4&3779&18&HD\,154345&ELODIE&49&8.6&7245&40, 20, 50\\
&SOPHIE+&65&10.7&&&&SOPHIE&10&4.2&&\\
HD\,48679&SOPHIE+&26&4.9&1288&18&&${\rm HIRES}^a$&212&1.5&&\\
HD\,77712&SOPHIE+&19&3.5&1508&18&HD\,14067&HIDES&27&3.9&2301&41\\
HD\,87899&SOPHIE&20&8.4&1168&18&&HRS&22&8.3&&\\
HD\,211681&ELODIE&12&13.4&5942&18, 50&&subaru&3&4.6&&\\
&SOPHIE&23&5.8&&&HD\,175679&Xinglong&23&35.1&1949&42\\
&SOPHIE+&7&3.4&&&&OAO&22&5.6&&\\
&${\rm HIRES}^a$&14&1.6&&&&Xinglong+&8&15.9&&\\
HD\,217850&SOPHIE&9&5.8&4480&18, 50, 32&HD\,166724&HARPS&55&0.7&4018&43\\
&SOPHIE+&32&3.4&&&&CORALIE98&35&12.8&&\\
&HIRES&29&1.3&&&&CORALIE07&33&5.3&&\\
HD\,5470&${\rm HIRES}^a$&24&2.5&5812&18, 50, 51&HD\,219077&HARPS&33&0.3&4852&43\\
&SOPHIE+&3&5.1&&&&CORALIE98&34&5.2&&\\
&HARPS&11&1.1&&&&CORALIE07&27&3.5&&\\
HD\,28635&ELODIE&3&21.7&4679&18, 51&HD\,220689&HARPS&31&0.6&5168&43\\
&SOPHIE+&13&5.5&&&&CORALIE98&22&5.4&&\\
&HARPS&4&1.3&&&&CORALIE07&34&3.5&&\\
HD\,69123&CORALIE98&3&1.8&4506&19&HD\,27631&HARPS&23&0.5&5598&43\\
&CORALIE14&19&2.7&&&&CORALIE98&34&5.5&&\\
&CORALIE07&14&3.4&&&&CORALIE07&38&3.8&&\\
HD\,222155&ELODIE&44&8.8&4847&20&HD\,32963&${\rm HIRES}^a$&202&1.4&5838&44\\
&SOPHIE&67&4.2&&&GJ 832&AAT&39&2.6&5569&45\\
HD\,150706&ELODIE&48&10.4&5835&20, 50&&PFS&16&0.9&&\\
&SOPHIE&53&4.3&&&&HARPS&54&0.4&&\\
&${\rm HIRES}^a$&58&1.8&&&HD\,74156&CORALIE98&44&8.5&5852&46, 50\\
HD\,190984&HARPS&58&1.7&3383&21&&ELODIE&76&16.3&&\\
HD\,224538&CORALIE14&24&12.7&4127&22, 51&&HIRES+&78&2.1&&\\
&HARPS&21&1.0&&&&HRS&82&8.3&&\\
&MIKE&6&3.2&&&&HIRES&9&2.1&&\\
HD\,68402&CORALIE&17&12.1&2050&22, 51&HD\,165131&HARPS&44&2.0&5539&51\\
&HARPS&5&1.6&&&&HARPS+&23&2.9&&\\
HD\,211847&CORALIE98&18&10.3&2634&23&HD\,62364&HARPS&58&3.2&6233&51\\
&CORALIE07&14&3.6&&&&HARPS+&28&2.3&&\\
HD\,30501&CORALIE98&40&8.4&4134&23&&&&&&\\
\hline
\end{tabular}
}
\end{center}
\tablecomments{\textwidth}{The symbol '+' represents the instrument has been upgraded. $^{a}$The systematic velocity offset between pre-upgrade and post-upgrade has been ignored in our analysis.}
\tablerefs{\textwidth}{The references 1-46 are the same as Table \ref{Tab:stellar1}. (50) \citet{Butler2017}; (51) \citet{Trifonov2020}; (52) \citet{Diaz2016}; (53) \citet{Kane2019}; (54) \citet{Minniti2009}; (55) \citet{CostaSilva2020}; (56) \citet{Santos2013}; (57) \citet{Valenti2005}; (58) \citet{Sousa2015}.
}
\end{table}

\begin{table}
\centering
\caption{ Posteriors of RV Companions, Ordered by the Value of ${ M_{\rm p}}$}\label{Tab:orbparas}
%%Please Capitalize the First Letter of Each Notional Word in table's caption
\resizebox{\textwidth}{!}{
\begin{tabular}{lccccccccccc}
\hline \hline
 Name & $M_{\rm p}$ &$i<90^{\circ}$ &$i>90^{\circ}$ &$a$ &$e$ &$P$ &$\Omega$ &$\omega$ &$a_{rel}$ &$T_{P}-2450000$ &$M_{\rm p}\,{\rm sin}\,i $ \\
 & (${\rm M_{Jup}}$) &(\degr) &(\degr)&(AU)&&(yr)&(\degr)&(\degr)&(mas)&(day)&(${\rm M_{Jup}}$)\\
 \hline
GJ\,832\,b&${0.8}_{-0.11}^{+0.12}$&${54.9}_{-4.9}^{+6.6}$&${125.1}_{-6.6}^{+4.9}$&${3.53}_{-0.16}^{+0.15}$&${0.069}_{-0.027}^{+0.026}$&${9.88}_{-0.33}^{+0.34}$&${41.0}_{-23.0}^{+77.0}$&${213.0}_{-33.0}^{+33.0}$&${710.0}_{-32.0}^{+31.0}$&${7470.0}_{-294.0}^{+327.0}$&${0.657}_{-0.063}^{+0.066}$\\
GJ\,179\,b&${0.95}_{-0.11}^{+0.16}$&${61.0}_{-13.0}^{+16.0}$&${119.0}_{-16.0}^{+13.0}$&${2.424}_{-0.075}^{+0.071}$&${0.179}_{-0.044}^{+0.048}$&${6.306}_{-0.086}^{+0.094}$&${62.0}_{-44.0}^{+99.0}$&${129.0}_{-19.0}^{+21.0}$&${195.3}_{-6.0}^{+5.7}$&${7301.0}_{-150.0}^{+125.0}$&${0.821}_{-0.064}^{+0.067}$\\
HD\,154345\,b&${1.19}_{-0.11}^{+0.14}$&${69.0}_{-12.0}^{+13.0}$&${111.0}_{-13.0}^{+12.0}$&${4.2}_{-0.15}^{+0.14}$&${0.157}_{-0.029}^{+0.03}$&${9.15}_{-0.11}^{+0.11}$&${77.0}_{-53.0}^{+38.0}$&${319.6}_{-8.4}^{+8.7}$&${229.7}_{-8.3}^{+7.7}$&${8428.0}_{-116.0}^{+72.0}$&${1.103}_{-0.081}^{+0.08}$\\
HD\,220689\,b&${1.2}_{-0.11}^{+0.22}$&${71.0}_{-18.0}^{+13.0}$&${109.0}_{-13.0}^{+18.0}$&${3.433}_{-0.064}^{+0.065}$&${0.053}_{-0.037}^{+0.06}$&${6.23}_{-0.14}^{+0.15}$&${93.0}_{-62.0}^{+58.0}$&${105.0}_{-68.0}^{+183.0}$&${73.1}_{-1.4}^{+1.4}$&${6138.0}_{-764.0}^{+1117.0}$&${1.106}_{-0.074}^{+0.076}$\\
HD\,30562\,b&${1.47}_{-0.18}^{+0.45}$&${65.0}_{-22.0}^{+17.0}$&${115.0}_{-17.0}^{+22.0}$&${2.299}_{-0.033}^{+0.032}$&${0.748}_{-0.042}^{+0.036}$&${3.158}_{-0.042}^{+0.039}$&${92.0}_{-67.0}^{+64.0}$&${78.2}_{-6.4}^{+6.7}$&${87.9}_{-1.3}^{+1.2}$&${5914.0}_{-18.0}^{+17.0}$&${1.3}_{-0.1}^{+0.1}$\\
HD\,27631\,b&${1.56}_{-0.15}^{+0.2}$&${74.0}_{-15.0}^{+11.0}$&${106.0}_{-11.0}^{+15.0}$&${3.22}_{-0.064}^{+0.065}$&${0.163}_{-0.057}^{+0.057}$&${5.95}_{-0.12}^{+0.13}$&${91.0}_{-66.0}^{+63.0}$&${128.0}_{-27.0}^{+28.0}$&${64.2}_{-1.3}^{+1.3}$&${6110.0}_{-147.0}^{+158.0}$&${1.47}_{-0.12}^{+0.12}$\\
HD\,222155\,b&${1.83}_{-0.26}^{+0.35}$&${68.0}_{-16.0}^{+15.0}$&${112.0}_{-15.0}^{+16.0}$&${4.48}_{-0.18}^{+0.19}$&${0.09}_{-0.062}^{+0.1}$&${8.94}_{-0.33}^{+0.34}$&${50.0}_{-27.0}^{+36.0}$&${163.0}_{-70.0}^{+86.0}$&${88.8}_{-3.6}^{+3.8}$&${6788.0}_{-727.0}^{+786.0}$&${1.64}_{-0.2}^{+0.24}$\\
HD\,89307\,b&${2.02}_{-0.15}^{+0.27}$&${72.0}_{-15.0}^{+13.0}$&${108.0}_{-13.0}^{+15.0}$&${3.331}_{-0.053}^{+0.052}$&${0.174}_{-0.043}^{+0.041}$&${5.991}_{-0.078}^{+0.081}$&${151.0}_{-133.0}^{+20.0}$&${22.0}_{-14.0}^{+320.0}$&${104.6}_{-1.7}^{+1.6}$&${6758.0}_{-96.0}^{+96.0}$&${1.89}_{-0.1}^{+0.11}$\\
HD\,32963\,b&${2.07}_{-0.64}^{+0.83}$&${19.3}_{-5.7}^{+9.2}$&${160.7}_{-9.2}^{+5.7}$&${3.409}_{-0.064}^{+0.063}$&${0.099}_{-0.028}^{+0.028}$&${6.483}_{-0.061}^{+0.064}$&${70.0}_{-56.0}^{+30.0}$&${105.0}_{-19.0}^{+18.0}$&${89.1}_{-1.7}^{+1.7}$&${5462.0}_{-114.0}^{+119.0}$&${0.684}_{-0.033}^{+0.035}$\\
HD\,117207\,b&${2.106}_{-0.089}^{+0.16}$&${76.6}_{-12.0}^{+9.3}$&${103.4}_{-9.3}^{+12.0}$&${3.773}_{-0.035}^{+0.036}$&${0.04}_{-0.024}^{+0.026}$&${7.136}_{-0.035}^{+0.034}$&${42.0}_{-18.0}^{+17.0}$&${186.0}_{-47.0}^{+48.0}$&${116.7}_{-1.1}^{+1.1}$&${6669.0}_{-339.0}^{+341.0}$&${2.032}_{-0.06}^{+0.06}$\\
HD\,108341\,b&${2.25}_{-0.25}^{+0.71}$&${65.0}_{-19.0}^{+17.0}$&${115.0}_{-17.0}^{+19.0}$&${2.029}_{-0.022}^{+0.018}$&${0.769}_{-0.01}^{+0.021}$&${3.142}_{-0.011}^{+0.01}$&${64.0}_{-48.0}^{+95.0}$&${192.7}_{-1.7}^{+1.4}$&${41.45}_{-0.44}^{+0.38}$&${6175.2}_{-7.2}^{+10.0}$&${1.97}_{-0.074}^{+0.2}$\\
HD\,150706\,b&${2.43}_{-0.38}^{+0.48}$&${70.0}_{-17.0}^{+14.0}$&${110.0}_{-14.0}^{+17.0}$&${11.5}_{-2.4}^{+5.0}$&${0.787}_{-0.083}^{+0.076}$&${36.0}_{-11.0}^{+26.0}$&${143.0}_{-124.0}^{+24.0}$&${84.0}_{-16.0}^{+17.0}$&${408.0}_{-85.0}^{+178.0}$&${15210.0}_{-3931.0}^{+9518.0}$&${2.21}_{-0.33}^{+0.36}$\\
HD\,129445\,b&${2.51}_{-0.54}^{+1.1}$&${52.0}_{-19.0}^{+24.0}$&${128.0}_{-24.0}^{+19.0}$&${2.984}_{-0.054}^{+0.039}$&${0.572}_{-0.086}^{+0.087}$&${4.933}_{-0.13}^{+0.093}$&${105.0}_{-74.0}^{+47.0}$&${163.8}_{-8.9}^{+8.8}$&${44.51}_{-0.81}^{+0.58}$&${6705.0}_{-77.0}^{+57.0}$&${1.93}_{-0.18}^{+0.23}$\\
HD\,13931\,b&${2.8}_{-0.64}^{+0.81}$&${43.0}_{-11.0}^{+19.0}$&${137.0}_{-19.0}^{+11.0}$&${5.338}_{-0.075}^{+0.082}$&${0.031}_{-0.022}^{+0.035}$&${12.18}_{-0.22}^{+0.26}$&${116.0}_{-86.0}^{+46.0}$&${155.0}_{-109.0}^{+152.0}$&${113.1}_{-1.6}^{+1.7}$&${7042.0}_{-1279.0}^{+1722.0}$&${1.91}_{-0.1}^{+0.1}$\\
HD\,167677\,b&${2.85}_{-1.0}^{+0.95}$&${28.7}_{-7.5}^{+19.0}$&${151.3}_{-19.0}^{+7.5}$&${2.877}_{-0.025}^{+0.025}$&${0.182}_{-0.026}^{+0.031}$&${4.97}_{-0.038}^{+0.04}$&${54.0}_{-18.0}^{+16.0}$&${302.0}_{-10.0}^{+13.0}$&${52.6}_{-0.45}^{+0.45}$&${5770.0}_{-61.0}^{+80.0}$&${1.369}_{-0.037}^{+0.039}$\\
HD\,103891\,b&${2.89}_{-0.84}^{+0.94}$&${27.4}_{-7.1}^{+13.0}$&${152.6}_{-13.0}^{+7.1}$&${3.255}_{-0.026}^{+0.024}$&${0.31}_{-0.047}^{+0.048}$&${5.185}_{-0.058}^{+0.053}$&${116.0}_{-30.0}^{+29.0}$&${208.2}_{-8.7}^{+7.8}$&${59.29}_{-0.47}^{+0.44}$&${5667.0}_{-41.0}^{+37.0}$&${1.33}_{-0.064}^{+0.066}$\\
HD\,69123\,b&${3.09}_{-0.29}^{+0.62}$&${70.0}_{-19.0}^{+14.0}$&${110.0}_{-14.0}^{+19.0}$&${2.48}_{-0.042}^{+0.042}$&${0.224}_{-0.051}^{+0.047}$&${3.261}_{-0.019}^{+0.018}$&${87.0}_{-70.0}^{+76.0}$&${286.0}_{-16.0}^{+16.0}$&${32.98}_{-0.55}^{+0.55}$&${5703.0}_{-44.0}^{+44.0}$&${2.85}_{-0.19}^{+0.18}$\\
HD\,18015\,b&${3.3}_{-0.41}^{+0.68}$&${69.0}_{-19.0}^{+14.0}$&${111.0}_{-14.0}^{+19.0}$&${3.82}_{-0.13}^{+0.13}$&${0.092}_{-0.064}^{+0.09}$&${6.12}_{-0.24}^{+0.22}$&${96.0}_{-65.0}^{+56.0}$&${253.0}_{-103.0}^{+45.0}$&${30.6}_{-1.1}^{+1.0}$&${6888.0}_{-829.0}^{+262.0}$&${3.0}_{-0.3}^{+0.3}$\\
HD\,120066\,b&${3.31}_{-0.12}^{+0.16}$&${80.3}_{-9.4}^{+6.7}$&${99.7}_{-6.7}^{+9.4}$&${22.8}_{-5.1}^{+7.8}$&${0.875}_{-0.037}^{+0.032}$&${105.0}_{-33.0}^{+58.0}$&${44.0}_{-20.0}^{+19.0}$&${339.7}_{-1.8}^{+1.8}$&${726.0}_{-161.0}^{+248.0}$&${8121.0}_{-12.0}^{+12.0}$&${3.234}_{-0.094}^{+0.1}$\\
HD\,190984\,b&${3.58}_{-0.45}^{+1.2}$&${64.0}_{-23.0}^{+18.0}$&${116.0}_{-18.0}^{+23.0}$&${8.8}_{-1.4}^{+2.5}$&${0.745}_{-0.047}^{+0.054}$&${27.3}_{-6.1}^{+12.0}$&${108.0}_{-82.0}^{+51.0}$&${315.3}_{-3.7}^{+3.7}$&${59.0}_{-9.4}^{+17.0}$&${14428.0}_{-2245.0}^{+4507.0}$&${3.16}_{-0.26}^{+0.25}$\\
HD\,166724\,b&${3.8}_{-0.29}^{+0.65}$&${68.0}_{-16.0}^{+15.0}$&${112.0}_{-15.0}^{+16.0}$&${5.17}_{-0.49}^{+0.38}$&${0.729}_{-0.017}^{+0.018}$&${13.0}_{-1.8}^{+1.4}$&${60.0}_{-19.0}^{+26.0}$&${199.0}_{-4.8}^{+4.2}$&${114.0}_{-11.0}^{+8.4}$&${7975.0}_{-663.0}^{+530.0}$&${3.49}_{-0.12}^{+0.13}$\\
HD\,216437\,b&${3.88}_{-0.73}^{+0.73}$&${35.0}_{-6.1}^{+10.0}$&${145.0}_{-10.0}^{+6.1}$&${2.501}_{-0.037}^{+0.036}$&${0.318}_{-0.028}^{+0.028}$&${3.658}_{-0.034}^{+0.034}$&${117.0}_{-14.0}^{+14.0}$&${64.5}_{-5.4}^{+5.5}$&${93.7}_{-1.4}^{+1.4}$&${5959.0}_{-35.0}^{+38.0}$&${2.23}_{-0.083}^{+0.084}$\\
HD\,70642\,b&${3.9}_{-0.27}^{+0.29}$&${29.9}_{-2.4}^{+2.6}$&${150.1}_{-2.6}^{+2.4}$&${3.295}_{-0.021}^{+0.021}$&${0.04}_{-0.027}^{+0.034}$&${5.751}_{-0.035}^{+0.038}$&${60.7}_{-8.6}^{+8.8}$&${258.0}_{-66.0}^{+51.0}$&${112.53}_{-0.7}^{+0.71}$&${5918.0}_{-337.0}^{+358.0}$&${1.947}_{-0.073}^{+0.074}$\\
HD\,55696\,b&${4.27}_{-0.69}^{+1.0}$&${57.0}_{-13.0}^{+19.0}$&${123.0}_{-19.0}^{+13.0}$&${3.04}_{-0.17}^{+0.15}$&${0.681}_{-0.047}^{+0.048}$&${4.671}_{-0.033}^{+0.032}$&${67.0}_{-41.0}^{+71.0}$&${140.0}_{-6.8}^{+6.4}$&${39.0}_{-2.2}^{+1.9}$&${5494.4}_{-10.0}^{+8.5}$&${3.56}_{-0.41}^{+0.41}$\\
HD\,143361\,b&${4.35}_{-0.66}^{+1.2}$&${55.0}_{-15.0}^{+22.0}$&${125.0}_{-22.0}^{+15.0}$&${1.994}_{-0.018}^{+0.018}$&${0.1938}_{-0.0046}^{+0.0047}$&${2.8538}_{-0.003}^{+0.0031}$&${33.0}_{-21.0}^{+128.0}$&${240.4}_{-1.5}^{+1.4}$&${29.0}_{-0.27}^{+0.27}$&${5761.5}_{-4.1}^{+4.1}$&${3.583}_{-0.068}^{+0.069}$\\
HD\,73267\,c&${4.4}_{-1.1}^{+1.7}$&${75.0}_{-16.0}^{+10.0}$&${105.0}_{-10.0}^{+16.0}$&${11.0}_{-2.2}^{+2.5}$&${0.134}_{-0.095}^{+0.12}$&${38.0}_{-11.0}^{+14.0}$&${98.0}_{-60.0}^{+58.0}$&${56.0}_{-27.0}^{+232.0}$&${218.0}_{-43.0}^{+50.0}$&${10729.0}_{-1855.0}^{+3353.0}$&${4.01}_{-0.93}^{+1.6}$\\
HD\,142022\,Ab&${4.51}_{-0.61}^{+0.91}$&${71.0}_{-13.0}^{+13.0}$&${109.0}_{-13.0}^{+13.0}$&${2.939}_{-0.062}^{+0.062}$&${0.506}_{-0.06}^{+0.071}$&${5.297}_{-0.073}^{+0.082}$&${141.0}_{-19.0}^{+19.0}$&${168.5}_{-5.0}^{+4.1}$&${85.8}_{-1.8}^{+1.8}$&${6730.0}_{-44.0}^{+50.0}$&${4.14}_{-0.5}^{+0.82}$\\
HD\,196050\,b&${4.55}_{-0.72}^{+0.69}$&${41.0}_{-6.3}^{+10.0}$&${139.0}_{-10.0}^{+6.3}$&${2.585}_{-0.035}^{+0.032}$&${0.178}_{-0.011}^{+0.011}$&${3.813}_{-0.024}^{+0.026}$&${15.2}_{-9.2}^{+157.0}$&${165.3}_{-10.0}^{+9.4}$&${51.16}_{-0.69}^{+0.63}$&${6307.0}_{-47.0}^{+46.0}$&${2.987}_{-0.091}^{+0.084}$\\
HD\,73267\,b&${4.6}_{-1.1}^{+1.2}$&${42.0}_{-10.0}^{+19.0}$&${138.0}_{-19.0}^{+10.0}$&${2.195}_{-0.025}^{+0.024}$&${0.2625}_{-0.005}^{+0.0051}$&${3.4421}_{-0.0017}^{+0.0016}$&${76.0}_{-53.0}^{+60.0}$&${227.4}_{-1.0}^{+1.0}$&${43.77}_{-0.49}^{+0.48}$&${5594.2}_{-3.0}^{+2.9}$&${3.054}_{-0.07}^{+0.07}$\\
HIP\,106006\,b&${4.96}_{-0.45}^{+1.7}$&${64.0}_{-22.0}^{+18.0}$&${116.0}_{-18.0}^{+22.0}$&${3.208}_{-0.035}^{+0.034}$&${0.091}_{-0.0063}^{+0.0066}$&${5.612}_{-0.014}^{+0.014}$&${107.0}_{-73.0}^{+48.0}$&${289.4}_{-4.5}^{+4.6}$&${66.63}_{-0.73}^{+0.71}$&${6030.0}_{-26.0}^{+25.0}$&${4.46}_{-0.1}^{+0.1}$\\
HD\,89839\,b&${5.01}_{-0.76}^{+0.79}$&${49.5}_{-8.3}^{+14.0}$&${130.5}_{-14.0}^{+8.3}$&${4.761}_{-0.042}^{+0.042}$&${0.187}_{-0.013}^{+0.013}$&${9.421}_{-0.047}^{+0.049}$&${26.0}_{-18.0}^{+149.0}$&${161.5}_{-3.5}^{+3.5}$&${83.31}_{-0.74}^{+0.74}$&${6682.0}_{-35.0}^{+35.0}$&${3.811}_{-0.076}^{+0.077}$\\
HD\,213240\,b&${5.21}_{-0.49}^{+1.5}$&${63.0}_{-20.0}^{+17.0}$&${117.0}_{-17.0}^{+20.0}$&${1.92}_{-0.026}^{+0.026}$&${0.4201}_{-0.0093}^{+0.01}$&${2.4071}_{-0.0083}^{+0.008}$&${145.0}_{-121.0}^{+21.0}$&${201.9}_{-1.5}^{+1.4}$&${46.9}_{-0.63}^{+0.64}$&${5901.0}_{-12.0}^{+12.0}$&${4.64}_{-0.13}^{+0.14}$\\
HIP\,56640\,b&${5.6}_{-1.1}^{+1.3}$&${41.5}_{-9.4}^{+13.0}$&${138.5}_{-13.0}^{+9.4}$&${3.77}_{-0.11}^{+0.12}$&${0.101}_{-0.079}^{+0.09}$&${7.16}_{-0.21}^{+0.29}$&${73.0}_{-39.0}^{+40.0}$&${132.0}_{-31.0}^{+107.0}$&${31.0}_{-0.88}^{+1.0}$&${7141.0}_{-1137.0}^{+245.0}$&${3.7}_{-0.23}^{+0.24}$\\
HD\,80869\,b&${5.67}_{-0.84}^{+1.8}$&${69.0}_{-17.0}^{+14.0}$&${111.0}_{-14.0}^{+17.0}$&${2.877}_{-0.047}^{+0.045}$&${0.871}_{-0.019}^{+0.055}$&${4.684}_{-0.025}^{+0.024}$&${95.0}_{-62.0}^{+58.0}$&${62.4}_{-15.0}^{+4.8}$&${33.98}_{-0.56}^{+0.54}$&${5230.0}_{-19.0}^{+25.0}$&${4.91}_{-0.34}^{+1.4}$\\
HIP\,8541\,b&${5.7}_{-1.1}^{+1.2}$&${73.0}_{-16.0}^{+12.0}$&${107.0}_{-12.0}^{+16.0}$&${2.8}_{-0.25}^{+0.21}$&${0.122}_{-0.04}^{+0.078}$&${4.339}_{-0.11}^{+0.073}$&${129.0}_{-29.0}^{+24.0}$&${289.0}_{-26.0}^{+19.0}$&${18.2}_{-1.6}^{+1.4}$&${5887.0}_{-133.0}^{+80.0}$&${5.34}_{-1.0}^{+0.91}$\\
HD\,50554\,b&${5.85}_{-0.52}^{+0.9}$&${61.0}_{-12.0}^{+12.0}$&${119.0}_{-12.0}^{+12.0}$&${2.339}_{-0.029}^{+0.03}$&${0.482}_{-0.015}^{+0.015}$&${3.39}_{-0.023}^{+0.02}$&${97.0}_{-40.0}^{+50.0}$&${4.0}_{-2.1}^{+2.6}$&${75.3}_{-0.95}^{+1.0}$&${5567.0}_{-23.0}^{+21.0}$&${5.13}_{-0.19}^{+0.19}$\\
HD\,16175\,b&${5.9}_{-1.0}^{+1.8}$&${59.0}_{-19.0}^{+20.0}$&${121.0}_{-20.0}^{+19.0}$&${2.13}_{-0.08}^{+0.075}$&${0.675}_{-0.026}^{+0.026}$&${2.686}_{-0.039}^{+0.031}$&${78.0}_{-28.0}^{+27.0}$&${216.8}_{-4.7}^{+4.8}$&${35.5}_{-1.3}^{+1.3}$&${5800.4}_{-5.1}^{+4.9}$&${4.92}_{-0.56}^{+0.67}$\\
HD\,190228\,b&${6.1}_{-1.0}^{+1.2}$&${48.0}_{-10.0}^{+16.0}$&${132.0}_{-16.0}^{+10.0}$&${2.293}_{-0.031}^{+0.031}$&${0.547}_{-0.011}^{+0.01}$&${3.1391}_{-0.005}^{+0.0053}$&${48.0}_{-22.0}^{+118.0}$&${97.4}_{-1.5}^{+1.6}$&${36.46}_{-0.49}^{+0.49}$&${5815.3}_{-2.7}^{+2.8}$&${4.56}_{-0.13}^{+0.14}$\\
HD\,132406\,b&${6.2}_{-1.1}^{+2.2}$&${64.0}_{-19.0}^{+18.0}$&${116.0}_{-18.0}^{+19.0}$&${1.969}_{-0.064}^{+0.06}$&${0.303}_{-0.077}^{+0.093}$&${2.64}_{-0.11}^{+0.11}$&${74.0}_{-52.0}^{+79.0}$&${219.0}_{-19.0}^{+20.0}$&${27.91}_{-0.9}^{+0.86}$&${5414.0}_{-51.0}^{+45.0}$&${5.25}_{-0.57}^{+1.2}$\\
HD\,10697\,b&${6.77}_{-0.56}^{+0.89}$&${72.0}_{-15.0}^{+13.0}$&${108.0}_{-13.0}^{+15.0}$&${2.143}_{-0.063}^{+0.062}$&${0.1035}_{-0.0073}^{+0.0072}$&${2.9446}_{-0.0019}^{+0.0019}$&${28.0}_{-16.0}^{+18.0}$&${119.1}_{-5.2}^{+5.1}$&${64.6}_{-1.9}^{+1.9}$&${5805.0}_{-15.0}^{+15.0}$&${6.33}_{-0.37}^{+0.38}$\\
HD\,111232\,b&${7.47}_{-0.26}^{+0.6}$&${77.1}_{-12.0}^{+9.1}$&${102.9}_{-9.1}^{+12.0}$&${2.051}_{-0.024}^{+0.023}$&${0.2083}_{-0.0023}^{+0.0025}$&${3.19969}_{-0.00084}^{+0.00085}$&${100.0}_{-63.0}^{+50.0}$&${94.86}_{-0.69}^{+0.66}$&${70.98}_{-0.84}^{+0.8}$&${5870.0}_{-2.1}^{+2.0}$&${7.24}_{-0.17}^{+0.17}$\\
HD\,120084\,b&${7.6}_{-2.5}^{+3.3}$&${44.0}_{-16.0}^{+25.0}$&${136.0}_{-25.0}^{+16.0}$&${4.12}_{-0.14}^{+0.14}$&${0.7}_{-0.13}^{+0.16}$&${5.706}_{-0.091}^{+0.061}$&${97.0}_{-61.0}^{+53.0}$&${120.0}_{-11.0}^{+19.0}$&${39.7}_{-1.4}^{+1.3}$&${7007.0}_{-34.0}^{+45.0}$&${4.57}_{-0.85}^{+2.6}$\\
HD\,95544\,b&${7.74}_{-0.75}^{+1.3}$&${62.0}_{-12.0}^{+16.0}$&${118.0}_{-16.0}^{+12.0}$&${3.383}_{-0.077}^{+0.074}$&${0.037}_{-0.016}^{+0.016}$&${5.942}_{-0.05}^{+0.054}$&${146.0}_{-130.0}^{+24.0}$&${180.0}_{-26.0}^{+22.0}$&${38.55}_{-0.88}^{+0.84}$&${6703.0}_{-157.0}^{+129.0}$&${6.87}_{-0.32}^{+0.31}$\\
HD\,68402\,b&${7.9}_{-1.5}^{+1.7}$&${20.3}_{-4.1}^{+6.2}$&${159.7}_{-6.2}^{+4.1}$&${2.239}_{-0.075}^{+0.11}$&${0.225}_{-0.082}^{+0.15}$&${3.15}_{-0.14}^{+0.22}$&${87.0}_{-28.0}^{+37.0}$&${310.0}_{-300.0}^{+38.0}$&${28.5}_{-1.0}^{+1.4}$&${5481.0}_{-94.0}^{+94.0}$&${2.67}_{-0.27}^{+0.42}$\\
HD\,115954\,b&${8.45}_{-0.46}^{+0.61}$&${79.0}_{-11.0}^{+7.7}$&${101.0}_{-7.7}^{+11.0}$&${4.53}_{-0.18}^{+0.16}$&${0.457}_{-0.026}^{+0.025}$&${8.85}_{-0.49}^{+0.38}$&${41.0}_{-24.0}^{+40.0}$&${171.8}_{-7.2}^{+5.9}$&${51.8}_{-2.1}^{+1.8}$&${6072.0}_{-51.0}^{+44.0}$&${8.2}_{-0.39}^{+0.4}$\\
HD\,25015\,b&${9.42}_{-0.78}^{+0.85}$&${31.7}_{-3.5}^{+3.7}$&${148.3}_{-3.7}^{+3.5}$&${6.45}_{-0.47}^{+0.52}$&${0.341}_{-0.08}^{+0.086}$&${17.4}_{-1.9}^{+2.1}$&${119.0}_{-15.0}^{+13.0}$&${90.0}_{-15.0}^{+13.0}$&${173.0}_{-13.0}^{+14.0}$&${6016.0}_{-193.0}^{+186.0}$&${4.92}_{-0.43}^{+0.52}$\\
HD\,175167\,b&${9.8}_{-1.2}^{+1.9}$&${60.0}_{-13.0}^{+17.0}$&${120.0}_{-17.0}^{+13.0}$&${2.438}_{-0.071}^{+0.064}$&${0.539}_{-0.032}^{+0.073}$&${3.529}_{-0.078}^{+0.022}$&${87.0}_{-55.0}^{+68.0}$&${343.4}_{-4.2}^{+5.2}$&${34.24}_{-1.0}^{+0.9}$&${6171.0}_{-21.0}^{+16.0}$&${8.1}_{-0.77}^{+1.7}$\\
HD\,74156\,c&${9.84}_{-0.86}^{+0.9}$&${53.8}_{-5.9}^{+8.0}$&${126.2}_{-8.0}^{+5.9}$&${3.823}_{-0.043}^{+0.043}$&${0.3805}_{-0.0066}^{+0.0066}$&${6.697}_{-0.011}^{+0.011}$&${41.0}_{-19.0}^{+131.0}$&${270.7}_{-1.1}^{+1.2}$&${66.62}_{-0.76}^{+0.74}$&${5911.2}_{-5.6}^{+5.6}$&${7.94}_{-0.19}^{+0.19}$\\
HD\,181234\,b&${10.13}_{-0.63}^{+0.74}$&${57.3}_{-5.0}^{+5.5}$&${122.7}_{-5.5}^{+5.0}$&${7.52}_{-0.17}^{+0.16}$&${0.7254}_{-0.0094}^{+0.0094}$&${20.4}_{-0.3}^{+0.32}$&${10.4}_{-7.4}^{+163.0}$&${94.4}_{-2.7}^{+2.7}$&${155.9}_{-3.5}^{+3.4}$&${7671.1}_{-7.1}^{+7.3}$&${8.53}_{-0.38}^{+0.37}$\\
HD\,191806\,b&${10.13}_{-0.85}^{+0.91}$&${62.6}_{-5.4}^{+5.9}$&${117.4}_{-5.9}^{+5.4}$&${2.81}_{-0.1}^{+0.1}$&${0.213}_{-0.024}^{+0.022}$&${4.391}_{-0.017}^{+0.017}$&${151.0}_{-32.0}^{+16.0}$&${344.8}_{-4.8}^{+4.7}$&${42.7}_{-1.5}^{+1.5}$&${6562.0}_{-21.0}^{+20.0}$&${8.97}_{-0.65}^{+0.65}$\\
BD+631405\,b&${10.4}_{-1.1}^{+1.3}$&${23.5}_{-2.3}^{+3.3}$&${156.5}_{-3.3}^{+2.3}$&${2.115}_{-0.082}^{+0.085}$&${0.891}_{-0.016}^{+0.046}$&${3.37}_{-0.11}^{+0.12}$&${169.7}_{-16.0}^{+7.6}$&${97.9}_{-3.7}^{+17.0}$&${55.5}_{-2.2}^{+2.2}$&${5771.0}_{-83.0}^{+79.0}$&${4.07}_{-0.32}^{+0.54}$\\
HD\,219077\,b&${10.57}_{-0.17}^{+0.18}$&${82.2}_{-3.2}^{+3.2}$&${97.8}_{-3.2}^{+3.2}$&${6.13}_{-0.1}^{+0.11}$&${0.766}_{-0.003}^{+0.0033}$&${14.73}_{-0.34}^{+0.37}$&${124.0}_{-106.0}^{+39.0}$&${57.45}_{-0.48}^{+0.45}$&${209.9}_{-3.5}^{+3.7}$&${5952.5}_{-1.4}^{+1.4}$&${10.46}_{-0.15}^{+0.15}$\\
HD\,106270\,b&${11.4}_{-1.2}^{+2.3}$&${63.0}_{-15.0}^{+16.0}$&${117.0}_{-16.0}^{+15.0}$&${3.347}_{-0.085}^{+0.082}$&${0.213}_{-0.036}^{+0.035}$&${5.17}_{-0.056}^{+0.055}$&${93.0}_{-80.0}^{+52.0}$&${9.7}_{-6.4}^{+345.0}$&${35.23}_{-0.89}^{+0.86}$&${6647.0}_{-40.0}^{+37.0}$&${10.1}_{-0.58}^{+0.58}$\\
HD\,224538\,b&${11.5}_{-2.6}^{+2.7}$&${37.2}_{-7.8}^{+14.0}$&${142.8}_{-14.0}^{+7.8}$&${2.444}_{-0.031}^{+0.03}$&${0.484}_{-0.014}^{+0.014}$&${3.287}_{-0.01}^{+0.01}$&${40.6}_{-9.0}^{+10.0}$&${25.5}_{-1.5}^{+1.5}$&${30.9}_{-0.39}^{+0.38}$&${6024.4}_{-3.5}^{+3.5}$&${6.98}_{-0.22}^{+0.22}$\\
HD\,86264\,b&${12.3}_{-3.3}^{+7.4}$&${45.0}_{-14.0}^{+15.0}$&${135.0}_{-15.0}^{+14.0}$&${2.873}_{-0.06}^{+0.064}$&${0.77}_{-0.17}^{+0.15}$&${4.06}_{-0.1}^{+0.12}$&${85.0}_{-52.0}^{+54.0}$&${304.0}_{-18.0}^{+17.0}$&${42.71}_{-0.89}^{+1.0}$&${6545.0}_{-1290.0}^{+105.0}$&${7.9}_{-1.3}^{+4.4}$\\
HIP\,67537\,b&${13.7}_{-2.4}^{+5.6}$&${54.0}_{-19.0}^{+24.0}$&${126.0}_{-24.0}^{+19.0}$&${4.92}_{-0.19}^{+0.24}$&${0.589}_{-0.03}^{+0.035}$&${6.98}_{-0.29}^{+0.52}$&${55.0}_{-32.0}^{+43.0}$&${119.5}_{-4.0}^{+4.0}$&${41.5}_{-1.6}^{+2.1}$&${6290.0}_{-18.0}^{+15.0}$&${11.05}_{-0.64}^{+0.67}$\\
HD\,23596\,b&${14.6}_{-1.3}^{+1.5}$&${34.0}_{-2.9}^{+3.6}$&${146.0}_{-3.6}^{+2.9}$&${2.901}_{-0.08}^{+0.083}$&${0.292}_{-0.022}^{+0.024}$&${4.31}_{-0.055}^{+0.069}$&${31.0}_{-19.0}^{+19.0}$&${274.7}_{-4.1}^{+4.1}$&${56.1}_{-1.5}^{+1.6}$&${6328.0}_{-64.0}^{+80.0}$&${8.21}_{-0.48}^{+0.47}$\\
HD\,14067\,b&${14.9}_{-4.8}^{+6.4}$&${38.0}_{-13.0}^{+27.0}$&${142.0}_{-27.0}^{+13.0}$&${5.29}_{-0.44}^{+0.55}$&${0.694}_{-0.053}^{+0.048}$&${7.83}_{-0.91}^{+1.2}$&${122.0}_{-103.0}^{+40.0}$&${102.0}_{-5.3}^{+5.2}$&${37.6}_{-3.1}^{+3.9}$&${5794.0}_{-13.0}^{+11.0}$&${9.07}_{-0.65}^{+0.67}$\\
HD\,136118\,b&${16.5}_{-1.8}^{+1.7}$&${46.0}_{-4.7}^{+7.5}$&${134.0}_{-7.5}^{+4.7}$&${2.353}_{-0.045}^{+0.046}$&${0.35}_{-0.026}^{+0.027}$&${3.262}_{-0.051}^{+0.053}$&${87.0}_{-57.0}^{+48.0}$&${311.8}_{-3.8}^{+3.7}$&${46.62}_{-0.89}^{+0.92}$&${5367.0}_{-55.0}^{+58.0}$&${11.88}_{-0.49}^{+0.49}$\\
HD\,62364\,b&${17.46}_{-0.59}^{+0.62}$&${48.9}_{-1.7}^{+1.8}$&${131.1}_{-1.8}^{+1.7}$&${6.248}_{-0.072}^{+0.07}$&${0.6092}_{-0.0042}^{+0.0042}$&${14.156}_{-0.059}^{+0.06}$&${92.2}_{-4.4}^{+4.4}$&${358.95}_{-358.0}^{+0.69}$&${118.0}_{-1.4}^{+1.3}$&${7566.4}_{-4.3}^{+4.3}$&${13.16}_{-0.33}^{+0.33}$\\
HD\,139357\,b&${18.2}_{-5.1}^{+6.2}$&${33.4}_{-8.5}^{+15.0}$&${146.6}_{-15.0}^{+8.5}$&${2.35}_{-0.15}^{+0.13}$&${0.102}_{-0.021}^{+0.021}$&${3.088}_{-0.024}^{+0.024}$&${69.0}_{-26.0}^{+25.0}$&${238.0}_{-11.0}^{+10.0}$&${20.7}_{-1.3}^{+1.2}$&${5855.0}_{-41.0}^{+39.0}$&${10.1}_{-1.2}^{+1.2}$\\
HD\,165131\,b&${18.7}_{-1.0}^{+1.4}$&${70.0}_{-8.5}^{+12.0}$&${110.0}_{-12.0}^{+8.5}$&${3.54}_{-0.054}^{+0.054}$&${0.6708}_{-0.0019}^{+0.0019}$&${6.4138}_{-0.0036}^{+0.0035}$&${130.4}_{-3.3}^{+3.4}$&${3.61}_{-0.29}^{+0.29}$&${61.14}_{-0.94}^{+0.94}$&${5590.1}_{-1.3}^{+1.3}$&${17.56}_{-0.54}^{+0.55}$\\
HIP\,106006\,d&${19.9}_{-8.7}^{+12.0}$&${60.0}_{-29.0}^{+21.0}$&${120.0}_{-21.0}^{+29.0}$&${17.9}_{-5.0}^{+7.1}$&${0.19}_{-0.13}^{+0.23}$&${74.0}_{-28.0}^{+48.0}$&${62.0}_{-14.0}^{+13.0}$&${238.0}_{-88.0}^{+83.0}$&${373.0}_{-103.0}^{+148.0}$&${20889.0}_{-8928.0}^{+20768.0}$&${16.0}_{-10.0}^{+14.0}$\\
HD\,214823\,b&${20.3}_{-1.5}^{+1.7}$&${78.7}_{-10.0}^{+7.8}$&${101.3}_{-7.8}^{+10.0}$&${3.17}_{-0.12}^{+0.11}$&${0.1636}_{-0.0042}^{+0.0042}$&${5.0775}_{-0.0049}^{+0.0049}$&${48.2}_{-4.0}^{+4.2}$&${123.9}_{-2.1}^{+2.1}$&${31.4}_{-1.1}^{+1.1}$&${5647.0}_{-10.0}^{+10.0}$&${19.6}_{-1.4}^{+1.4}$\\
HD\,111232\,c&${20.7}_{-3.2}^{+3.4}$&${77.1}_{-12.0}^{+8.9}$&${102.9}_{-8.9}^{+12.0}$&${18.8}_{-4.1}^{+5.0}$&${0.326}_{-0.093}^{+0.1}$&${88.0}_{-27.0}^{+37.0}$&${119.0}_{-103.0}^{+46.0}$&${324.0}_{-28.0}^{+22.0}$&${649.0}_{-142.0}^{+172.0}$&${9478.0}_{-630.0}^{+656.0}$&${19.8}_{-2.8}^{+2.9}$\\
HIP\,97233\,b&${22.6}_{-3.3}^{+5.7}$&${60.0}_{-13.0}^{+14.0}$&${120.0}_{-14.0}^{+13.0}$&${2.447}_{-0.1}^{+0.09}$&${0.633}_{-0.042}^{+0.055}$&${2.886}_{-0.024}^{+0.029}$&${69.0}_{-45.0}^{+77.0}$&${248.9}_{-4.5}^{+4.2}$&${24.08}_{-0.94}^{+0.89}$&${5865.0}_{-27.0}^{+40.0}$&${19.2}_{-1.7}^{+2.5}$\\
HD\,217786\,b&${28.9}_{-3.0}^{+2.9}$&${25.2}_{-2.5}^{+3.1}$&${154.8}_{-3.1}^{+2.5}$&${2.372}_{-0.022}^{+0.022}$&${0.3158}_{-0.008}^{+0.0082}$&${3.5649}_{-0.0056}^{+0.0058}$&${164.0}_{-19.0}^{+10.0}$&${101.2}_{-1.3}^{+1.3}$&${43.17}_{-0.47}^{+0.47}$&${6038.7}_{-4.1}^{+4.3}$&${12.31}_{-0.25}^{+0.25}$\\
HIP\,84056\,b&${31.9}_{-5.3}^{+8.5}$&${5.2}_{-1.1}^{+1.1}$&${174.8}_{-1.1}^{+1.1}$&${2.065}_{-0.062}^{+0.059}$&${0.079}_{-0.053}^{+0.06}$&${2.261}_{-0.024}^{+0.027}$&${91.0}_{-17.0}^{+21.0}$&${157.0}_{-50.0}^{+44.0}$&${27.6}_{-0.82}^{+0.79}$&${5368.0}_{-92.0}^{+192.0}$&${2.92}_{-0.24}^{+0.23}$\\
HD\,125390\,b&${34.0}_{-3.3}^{+3.4}$&${41.9}_{-4.3}^{+5.8}$&${138.1}_{-5.8}^{+4.3}$&${3.184}_{-0.08}^{+0.076}$&${0.598}_{-0.017}^{+0.022}$&${4.812}_{-0.013}^{+0.013}$&${144.0}_{-6.2}^{+8.7}$&${342.1}_{-1.4}^{+1.5}$&${20.45}_{-0.51}^{+0.49}$&${5911.4}_{-8.9}^{+8.5}$&${22.6}_{-1.5}^{+1.7}$\\
HD\,122562\,b&${37.7}_{-3.5}^{+3.5}$&${42.2}_{-2.7}^{+3.1}$&${137.8}_{-3.1}^{+2.7}$&${4.23}_{-0.17}^{+0.15}$&${0.7171}_{-0.0084}^{+0.0086}$&${8.046}_{-0.091}^{+0.095}$&${27.1}_{-4.0}^{+4.1}$&${304.16}_{-0.75}^{+0.74}$&${79.6}_{-3.1}^{+2.9}$&${5929.4}_{-4.4}^{+3.8}$&${25.4}_{-1.9}^{+1.9}$\\
HD\,23965\,b&${43.4}_{-2.7}^{+3.0}$&${80.2}_{-5.0}^{+6.6}$&${99.8}_{-6.6}^{+5.0}$&${5.58}_{-0.15}^{+0.6}$&${0.805}_{-0.01}^{+0.012}$&${12.18}_{-0.42}^{+2.0}$&${75.0}_{-28.0}^{+45.0}$&${324.0}_{-2.1}^{+2.1}$&${129.5}_{-3.5}^{+14.0}$&${8953.0}_{-159.0}^{+746.0}$&${42.6}_{-2.5}^{+2.7}$\\
HD\,30246\,b&${51.8}_{-3.8}^{+8.1}$&${79.8}_{-21.0}^{+7.3}$&${100.2}_{-7.3}^{+21.0}$&${2.009}_{-0.026}^{+0.027}$&${0.761}_{-0.049}^{+0.048}$&${2.714}_{-0.015}^{+0.015}$&${31.0}_{-19.0}^{+128.0}$&${282.6}_{-5.4}^{+6.9}$&${41.19}_{-0.54}^{+0.54}$&${5323.6}_{-6.8}^{+9.2}$&${49.7}_{-3.0}^{+4.2}$\\
HD\,14348\,b&${55.5}_{-2.7}^{+2.8}$&${62.2}_{-2.4}^{+2.7}$&${117.8}_{-2.7}^{+2.4}$&${5.96}_{-0.13}^{+0.13}$&${0.4574}_{-0.0036}^{+0.0036}$&${13.01}_{-0.014}^{+0.014}$&${153.6}_{-2.7}^{+2.8}$&${64.79}_{-0.57}^{+0.58}$&${98.9}_{-2.2}^{+2.1}$&${5264.8}_{-4.5}^{+4.3}$&${49.1}_{-2.2}^{+2.2}$\\
HD\,167665\,b&${58.7}_{-2.2}^{+2.7}$&${63.4}_{-4.3}^{+3.9}$&${116.6}_{-3.9}^{+4.3}$&${5.62}_{-0.049}^{+0.05}$&${0.3383}_{-0.0036}^{+0.0036}$&${12.18}_{-0.038}^{+0.04}$&${37.9}_{-1.8}^{+1.9}$&${224.51}_{-0.64}^{+0.64}$&${182.1}_{-1.6}^{+1.6}$&${6975.0}_{-16.0}^{+16.0}$&${52.46}_{-0.92}^{+0.92}$\\
HD\,175679\,b&${59.9}_{-8.2}^{+10.0}$&${38.9}_{-5.5}^{+6.4}$&${141.1}_{-6.4}^{+5.5}$&${3.38}_{-0.13}^{+0.12}$&${0.3807}_{-0.0081}^{+0.0084}$&${3.741}_{-0.017}^{+0.018}$&${62.0}_{-14.0}^{+15.0}$&${346.5}_{-1.3}^{+1.3}$&${20.09}_{-0.78}^{+0.73}$&${5997.4}_{-6.2}^{+6.6}$&${37.7}_{-2.9}^{+2.8}$\\
HD\,74014\,b&${62.0}_{-1.6}^{+1.6}$&${55.0}_{-1.2}^{+1.2}$&${125.0}_{-1.2}^{+1.2}$&${7.144}_{-0.078}^{+0.076}$&${0.5206}_{-0.002}^{+0.0021}$&${18.56}_{-0.11}^{+0.12}$&${176.1}_{-173.0}^{+2.5}$&${325.91}_{-0.34}^{+0.34}$&${205.2}_{-2.2}^{+2.2}$&${10215.0}_{-42.0}^{+43.0}$&${50.8}_{-1.0}^{+1.0}$\\
HD\,30501\,b&${67.3}_{-1.1}^{+1.1}$&${79.6}_{-2.2}^{+2.9}$&${100.4}_{-2.9}^{+2.2}$&${3.044}_{-0.025}^{+0.024}$&${0.7453}_{-0.0024}^{+0.0024}$&${5.6797}_{-0.0073}^{+0.0074}$&${174.8}_{-2.8}^{+2.2}$&${69.65}_{-0.46}^{+0.47}$&${149.7}_{-1.2}^{+1.2}$&${5926.8}_{-5.4}^{+5.5}$&${66.2}_{-1.2}^{+1.1}$\\
HD\,48679\,b&${71.6}_{-6.6}^{+7.0}$&${31.6}_{-2.9}^{+3.5}$&${148.4}_{-3.5}^{+2.9}$&${2.168}_{-0.021}^{+0.02}$&${0.82467}_{-0.0005}^{+0.00048}$&${3.04343}_{-0.00062}^{+0.00064}$&${171.9}_{-168.0}^{+5.6}$&${155.67}_{-0.18}^{+0.17}$&${32.37}_{-0.31}^{+0.29}$&${5943.8}_{-0.28}^{+0.28}$&${37.59}_{-0.73}^{+0.7}$\\
\hline
\end{tabular}
}
\end{table}

\begin{table}
\centering
%\tabcolsep=0.25cm
%\caption{ - $continued$ Posteriors of RV Companions, Ordered by the Value of ${ M_{\rm p}}$}\label{Tabcc}
\caption{Posteriors of RV Companions, Ordered by the Value of ${ M_{\rm p}}$}\label{Tabcc}
%%Please Capitalize the First Letter of Each Notional Word in table's caption
\resizebox{\textwidth}{!}{
\begin{tabular}{lccccccccccc}
\hline \hline
 Name & $M_{\rm p}$ &$i<90^{\circ}$ &$i>90^{\circ}$ &$a$ &$e$ &$P$ &$\Omega$ &$\omega$ &$a_{rel}$ &$T_{P}-2450000$ &$M_{\rm p}\,{\rm sin}\,i $ \\
 & (${\rm M_{Jup}}$) &(\degr) &(\degr)&(AU)&&(yr)&(\degr)&(\degr)&(mas)&(day)&(${\rm M_{Jup}}$)\\
 \hline
HD\,33636\,b&${77.8}_{-6.6}^{+6.9}$&${7.07}_{-0.54}^{+0.62}$&${172.93}_{-0.62}^{+0.54}$&${3.329}_{-0.023}^{+0.022}$&${0.483}_{-0.0063}^{+0.0063}$&${5.807}_{-0.017}^{+0.016}$&${109.9}_{-5.0}^{+4.9}$&${338.2}_{-1.3}^{+1.3}$&${112.52}_{-0.77}^{+0.75}$&${5442.0}_{-13.0}^{+12.0}$&${9.57}_{-0.16}^{+0.16}$\\
HIP\,103019\,b&${83.0}_{-22.0}^{+28.0}$&${43.0}_{-11.0}^{+23.0}$&${137.0}_{-23.0}^{+11.0}$&${1.701}_{-0.016}^{+0.021}$&${0.5044}_{-0.0028}^{+0.0027}$&${2.5134}_{-0.0045}^{+0.0041}$&${112.0}_{-20.0}^{+19.0}$&${73.68}_{-0.51}^{+0.52}$&${60.8}_{-0.54}^{+0.67}$&${5599.0}_{-1.7}^{+1.6}$&${56.3}_{-1.1}^{+1.4}$\\
HD\,130396\,B&${95.1}_{-5.1}^{+5.3}$&${34.3}_{-1.8}^{+2.0}$&${145.7}_{-2.0}^{+1.8}$&${3.365}_{-0.031}^{+0.03}$&${0.4269}_{-0.004}^{+0.004}$&${5.634}_{-0.019}^{+0.019}$&${171.5}_{-3.9}^{+3.8}$&${163.32}_{-0.69}^{+0.71}$&${74.26}_{-0.67}^{+0.66}$&${5328.8}_{-9.0}^{+9.1}$&${53.6}_{-1.0}^{+1.0}$\\
HD\,203473\,B&${106.0}_{-13.0}^{+13.0}$&${36.1}_{-1.3}^{+1.4}$&${143.9}_{-1.4}^{+1.3}$&${4.32}_{-0.27}^{+0.24}$&${0.3965}_{-0.0091}^{+0.011}$&${8.1114}_{-0.0092}^{+0.0085}$&${77.5}_{-1.7}^{+1.8}$&${20.9}_{-1.8}^{+1.7}$&${59.4}_{-3.7}^{+3.4}$&${5911.0}_{-17.0}^{+16.0}$&${62.3}_{-7.8}^{+8.0}$\\
HD\,184601\,B&${117.0}_{-32.0}^{+36.0}$&${33.3}_{-7.6}^{+14.0}$&${146.7}_{-14.0}^{+7.6}$&${1.791}_{-0.045}^{+0.046}$&${0.4882}_{-0.0049}^{+0.0052}$&${2.3256}_{-0.0036}^{+0.0037}$&${110.1}_{-8.9}^{+8.7}$&${137.46}_{-0.68}^{+0.67}$&${23.23}_{-0.58}^{+0.6}$&${5931.8}_{-4.4}^{+4.3}$&${64.7}_{-3.2}^{+3.5}$\\
HD\,154697\,B&${123.3}_{-1.9}^{+1.9}$&${38.47}_{-0.31}^{+0.32}$&${141.53}_{-0.32}^{+0.31}$&${3.016}_{-0.02}^{+0.02}$&${0.1625}_{-0.002}^{+0.002}$&${5.0466}_{-0.0063}^{+0.0065}$&${64.0}_{-1.2}^{+1.2}$&${179.0}_{-1.3}^{+1.2}$&${85.17}_{-0.57}^{+0.56}$&${6258.2}_{-9.3}^{+9.4}$&${76.7}_{-1.1}^{+1.1}$\\
HD\,29461\,B&${126.9}_{-7.8}^{+8.2}$&${50.3}_{-2.8}^{+3.3}$&${129.7}_{-3.3}^{+2.8}$&${4.98}_{-0.11}^{+0.1}$&${0.6138}_{-0.0042}^{+0.0041}$&${10.22}_{-0.02}^{+0.02}$&${128.4}_{-3.2}^{+3.3}$&${52.44}_{-0.39}^{+0.37}$&${91.6}_{-2.0}^{+1.9}$&${7374.3}_{-8.0}^{+7.8}$&${97.7}_{-4.2}^{+4.2}$\\
HD\,5608\,B&${127.0}_{-10.0}^{+11.0}$&-&${148.0}_{-13.0}^{+6.2}$&${29.4}_{-4.0}^{+7.3}$&${0.65}_{-0.18}^{+0.13}$&${123.0}_{-25.0}^{+50.0}$&${84.0}_{-66.0}^{+75.0}$&${266.0}_{-34.0}^{+26.0}$&${501.0}_{-69.0}^{+125.0}$&${18047.0}_{-2238.0}^{+3972.0}$&${68.0}_{-13.0}^{+21.0}$\\
HD\,131664\,B&${131.8}_{-4.1}^{+4.1}$&${9.43}_{-0.25}^{+0.27}$&${170.57}_{-0.27}^{+0.25}$&${3.31}_{-0.03}^{+0.029}$&${0.6912}_{-0.004}^{+0.0041}$&${5.4388}_{-0.0042}^{+0.0043}$&${2.0}_{-1.3}^{+177.0}$&${151.41}_{-0.51}^{+0.52}$&${63.34}_{-0.56}^{+0.54}$&${5990.9}_{-2.5}^{+2.5}$&${21.59}_{-0.45}^{+0.46}$\\
HD\,10844\,B&${139.7}_{-7.1}^{+7.2}$&${35.7}_{-3.1}^{+3.2}$&${144.3}_{-3.2}^{+3.1}$&${9.97}_{-0.5}^{+0.57}$&${0.55}_{-0.021}^{+0.023}$&${29.8}_{-2.0}^{+2.4}$&${149.2}_{-2.2}^{+2.9}$&${264.9}_{-1.3}^{+1.2}$&${183.7}_{-9.3}^{+10.0}$&${14772.0}_{-766.0}^{+897.0}$&${81.2}_{-5.8}^{+6.3}$\\
HD\,94386\,B&${140.0}_{-23.0}^{+30.0}$&${54.0}_{-10.0}^{+19.0}$&${126.0}_{-19.0}^{+10.0}$&${2.048}_{-0.13}^{+0.095}$&${0.4213}_{-0.002}^{+0.0021}$&${2.5323}_{-0.0019}^{+0.002}$&${160.0}_{-152.0}^{+14.0}$&${37.74}_{-0.25}^{+0.25}$&${25.5}_{-1.6}^{+1.2}$&${5506.9}_{-0.63}^{+0.64}$&${115.0}_{-14.0}^{+11.0}$\\
HD 211847 B&${148.6}_{-3.6}^{+3.7}$&-&${172.32}_{-0.37}^{+0.36}$&${6.83}_{-0.063}^{+0.064}$&${0.569}_{-0.012}^{+0.011}$&${17.14}_{-0.12}^{+0.12}$&${2.2}_{-1.6}^{+3.2}$&${171.4}_{-1.8}^{+3.3}$&${140.2}_{-1.3}^{+1.3}$&${10453.0}_{-47.0}^{+48.0}$&${19.86}_{-0.79}^{+0.83}$\\
HD\,21340\,B&${150.9}_{-7.4}^{+7.8}$&${66.7}_{-3.4}^{+3.9}$&${113.3}_{-3.9}^{+3.4}$&${2.706}_{-0.057}^{+0.056}$&${0.5677}_{-0.0084}^{+0.01}$&${3.4191}_{-0.0059}^{+0.0057}$&${72.1}_{-5.8}^{+5.6}$&${129.57}_{-0.69}^{+0.73}$&${19.58}_{-0.42}^{+0.41}$&${5227.0}_{-1.4}^{+1.4}$&${138.6}_{-6.2}^{+6.4}$\\
HD\,103913\,B&${155.0}_{-11.0}^{+11.0}$&${40.7}_{-1.5}^{+1.6}$&${139.3}_{-1.6}^{+1.5}$&${3.58}_{-0.12}^{+0.11}$&${0.4036}_{-0.0075}^{+0.0073}$&${6.32}_{-0.022}^{+0.023}$&${150.6}_{-1.7}^{+1.7}$&${185.17}_{-0.62}^{+0.59}$&${41.3}_{-1.4}^{+1.3}$&${6887.4}_{-5.5}^{+5.1}$&${101.0}_{-6.6}^{+6.5}$\\
HD\,217786\,B&${167.0}_{-10.0}^{+10.0}$&-&${112.0}_{-15.0}^{+23.0}$&${222.0}_{-69.0}^{+110.0}$&${0.61}_{-0.29}^{+0.2}$&${3043.0}_{-1312.0}^{+2510.0}$&${161.0}_{-20.0}^{+10.0}$&${49.0}_{-29.0}^{+37.0}$&${4039.0}_{-1264.0}^{+1997.0}$&${987998.0}_{-504586.0}^{+901059.0}$&${152.0}_{-35.0}^{+16.0}$\\
HD\,103459\,B&${176.0}_{-20.0}^{+18.0}$&${69.7}_{-2.6}^{+2.9}$&${110.3}_{-2.9}^{+2.6}$&${3.24}_{-0.19}^{+0.16}$&${0.71124}_{-0.00031}^{+0.00031}$&${5.0139}_{-0.0039}^{+0.0037}$&${4.2}_{-3.0}^{+174.0}$&${182.589}_{-0.082}^{+0.085}$&${54.8}_{-3.2}^{+2.7}$&${5757.1}_{-0.19}^{+0.2}$&${165.0}_{-19.0}^{+17.0}$\\
BD+730275\,B&${187.0}_{-14.0}^{+15.0}$&${15.63}_{-0.88}^{+1.0}$&${164.37}_{-1.0}^{+0.88}$&${2.432}_{-0.051}^{+0.05}$&${0.8138}_{-0.001}^{+0.001}$&${3.89596}_{-0.00062}^{+0.00063}$&${62.7}_{-3.9}^{+3.8}$&${120.71}_{-0.24}^{+0.24}$&${53.1}_{-1.1}^{+1.1}$&${5903.2}_{-0.24}^{+0.24}$&${50.4}_{-2.1}^{+2.1}$\\
HD\,51813\,B&${188.0}_{-19.0}^{+20.0}$&${16.4}_{-1.5}^{+1.7}$&${163.6}_{-1.7}^{+1.5}$&${2.574}_{-0.064}^{+0.1}$&${0.749}_{-0.012}^{+0.016}$&${3.678}_{-0.022}^{+0.31}$&${88.4}_{-7.1}^{+28.0}$&${280.3}_{-3.5}^{+8.0}$&${42.7}_{-1.1}^{+1.7}$&${5925.7}_{-1.7}^{+4.4}$&${53.2}_{-2.8}^{+2.7}$\\
HD\,211681\,B&${191.9}_{-9.5}^{+9.5}$&${26.26}_{-0.47}^{+0.5}$&${153.74}_{-0.5}^{+0.47}$&${8.43}_{-0.22}^{+0.21}$&${0.4618}_{-0.0035}^{+0.0033}$&${20.59}_{-0.29}^{+0.29}$&${23.8}_{-3.1}^{+3.2}$&${307.59}_{-0.7}^{+0.71}$&${116.6}_{-3.0}^{+2.9}$&${10656.0}_{-110.0}^{+111.0}$&${84.9}_{-4.0}^{+4.0}$\\
HD\,28635\,B&${197.5}_{-9.1}^{+9.2}$&${29.87}_{-0.88}^{+0.92}$&${150.13}_{-0.92}^{+0.88}$&${4.91}_{-0.11}^{+0.1}$&${0.5583}_{-0.0057}^{+0.0062}$&${9.3244}_{-0.0038}^{+0.0036}$&${37.1}_{-2.3}^{+2.3}$&${142.6}_{-1.7}^{+1.7}$&${99.5}_{-2.2}^{+2.2}$&${7400.5}_{-6.5}^{+6.5}$&${98.4}_{-4.9}^{+4.9}$\\
BD+210055\,B&${199.0}_{-10.0}^{+10.0}$&${29.78}_{-0.8}^{+0.84}$&${150.22}_{-0.84}^{+0.8}$&${2.325}_{-0.048}^{+0.046}$&${0.4464}_{-0.001}^{+0.0011}$&${3.6193}_{-0.0022}^{+0.0017}$&${121.1}_{-1.9}^{+1.8}$&${294.39}_{-0.19}^{+0.17}$&${63.5}_{-1.3}^{+1.2}$&${5599.2}_{-0.49}^{+0.6}$&${98.8}_{-4.1}^{+3.9}$\\
HD\,53680\,B&${213.7}_{-7.0}^{+7.1}$&${17.35}_{-0.44}^{+0.46}$&${162.65}_{-0.46}^{+0.44}$&${2.77}_{-0.022}^{+0.022}$&${0.4738}_{-0.0022}^{+0.0022}$&${4.6235}_{-0.0029}^{+0.0029}$&${52.5}_{-1.6}^{+1.6}$&${226.93}_{-0.26}^{+0.26}$&${160.1}_{-1.3}^{+1.2}$&${5872.2}_{-2.0}^{+2.0}$&${63.7}_{-1.0}^{+1.0}$\\
HD\,35956\,B&${228.1}_{-8.3}^{+8.3}$&${57.8}_{-2.0}^{+2.2}$&${122.2}_{-2.2}^{+2.0}$&${2.678}_{-0.033}^{+0.033}$&${0.61631}_{-0.0006}^{+0.00061}$&${3.90602}_{-0.00039}^{+0.00036}$&${97.7}_{-2.8}^{+2.6}$&${326.504}_{-0.057}^{+0.054}$&${90.4}_{-1.1}^{+1.1}$&${6503.0}_{-0.53}^{+0.49}$&${193.1}_{-4.8}^{+4.7}$\\
HD\,87899\,B&${239.0}_{-11.0}^{+11.0}$&${89.32}_{-0.41}^{+0.39}$&${90.68}_{-0.39}^{+0.41}$&${2.645}_{-0.06}^{+0.056}$&${0.6714}_{-0.0028}^{+0.0029}$&${4.1424}_{-0.0029}^{+0.0029}$&${46.59}_{-0.31}^{+0.31}$&${181.67}_{-0.26}^{+0.25}$&${51.2}_{-1.2}^{+1.1}$&${5207.3}_{-0.87}^{+0.85}$&${239.0}_{-11.0}^{+11.0}$\\
HD\,3404\,B&${239.0}_{-22.0}^{+19.0}$&${48.8}_{-2.1}^{+2.2}$&${131.2}_{-2.2}^{+2.1}$&${2.96}_{-0.13}^{+0.11}$&${0.7403}_{-0.0029}^{+0.0058}$&${4.2183}_{-0.0031}^{+0.003}$&${100.4}_{-4.7}^{+4.8}$&${0.78}_{-0.31}^{+0.41}$&${37.4}_{-1.6}^{+1.4}$&${6539.2}_{-7.2}^{+6.5}$&${180.0}_{-20.0}^{+17.0}$\\
HD\,156728\,B&${248.5}_{-5.3}^{+4.9}$&${35.96}_{-0.16}^{+0.16}$&${144.04}_{-0.16}^{+0.16}$&${5.44}_{-0.057}^{+0.053}$&${0.348}_{-0.0018}^{+0.0018}$&${11.795}_{-0.035}^{+0.036}$&${89.71}_{-0.32}^{+0.32}$&${102.15}_{-0.49}^{+0.52}$&${130.6}_{-1.4}^{+1.3}$&${7082.2}_{-3.9}^{+4.3}$&${145.9}_{-3.0}^{+2.8}$\\
HD\,77712\,B&${256.0}_{-11.0}^{+12.0}$&${11.35}_{-0.25}^{+0.32}$&${168.65}_{-0.32}^{+0.25}$&${2.461}_{-0.034}^{+0.034}$&${0.68}_{-0.02}^{+0.026}$&${3.5919}_{-0.0052}^{+0.0052}$&${174.2}_{-2.5}^{+2.0}$&${50.6}_{-2.1}^{+1.5}$&${49.95}_{-0.69}^{+0.68}$&${5327.5}_{-5.4}^{+5.3}$&${50.2}_{-2.8}^{+3.7}$\\
HD\,101305\,B&${261.0}_{-17.0}^{+17.0}$&${28.36}_{-0.64}^{+0.65}$&${151.64}_{-0.65}^{+0.64}$&${2.977}_{-0.091}^{+0.091}$&${0.4914}_{-0.0017}^{+0.0018}$&${4.5938}_{-0.0044}^{+0.0043}$&${151.1}_{-2.2}^{+2.2}$&${241.5}_{-0.17}^{+0.17}$&${42.1}_{-1.3}^{+1.3}$&${6632.0}_{-0.88}^{+0.89}$&${124.0}_{-7.5}^{+7.7}$\\
HD\,92987\,B&${263.5}_{-9.4}^{+9.1}$&${4.316}_{-0.094}^{+0.093}$&${175.684}_{-0.093}^{+0.094}$&${11.63}_{-0.28}^{+0.29}$&${0.287}_{-0.013}^{+0.014}$&${34.36}_{-0.85}^{+1.0}$&${76.3}_{-1.6}^{+1.6}$&${177.1}_{-3.1}^{+3.1}$&${267.3}_{-6.3}^{+6.6}$&${7551.0}_{-78.0}^{+77.0}$&${19.82}_{-0.81}^{+0.81}$\\
HD\,283668\,B&${319.0}_{-19.0}^{+19.0}$&${14.2}_{-1.0}^{+1.2}$&${165.8}_{-1.2}^{+1.0}$&${3.634}_{-0.1}^{+0.091}$&${0.698}_{-0.039}^{+0.047}$&${6.984}_{-0.02}^{+0.019}$&${26.4}_{-6.2}^{+5.1}$&${277.4}_{-4.5}^{+5.9}$&${123.5}_{-3.2}^{+3.1}$&${7361.0}_{-16.0}^{+15.0}$&${78.0}_{-5.7}^{+7.0}$\\
HD\,217850\,B&${323.0}_{-23.0}^{+25.0}$&${4.67}_{-0.27}^{+0.29}$&${175.33}_{-0.29}^{+0.27}$&${5.039}_{-0.063}^{+0.062}$&${0.7587}_{-0.0014}^{+0.0015}$&${9.5987}_{-0.0072}^{+0.0075}$&${105.9}_{-3.4}^{+3.5}$&${165.61}_{-0.26}^{+0.25}$&${81.6}_{-1.4}^{+1.3}$&${7551.3}_{-1.1}^{+1.1}$&${26.3}_{-0.66}^{+0.66}$\\
HD\,5470\,B&${338.0}_{-21.0}^{+20.0}$&${41.55}_{-0.47}^{+0.48}$&${138.45}_{-0.48}^{+0.47}$&${8.48}_{-0.27}^{+0.25}$&${0.3583}_{-0.0014}^{+0.0014}$&${21.463}_{-0.062}^{+0.064}$&${111.4}_{-1.1}^{+1.1}$&${235.82}_{-0.54}^{+0.55}$&${126.4}_{-4.0}^{+3.7}$&${11000.0}_{-28.0}^{+28.0}$&${224.0}_{-14.0}^{+13.0}$\\
HD\,112988\,B&${371.0}_{-26.0}^{+26.0}$&${83.0}_{-5.4}^{+4.8}$&${97.0}_{-4.8}^{+5.4}$&${6.58}_{-0.23}^{+0.24}$&${0.7318}_{-0.0083}^{+0.0093}$&${14.32}_{-0.49}^{+0.52}$&${113.47}_{-0.5}^{+0.54}$&${160.3}_{-1.5}^{+1.6}$&${56.8}_{-2.0}^{+2.0}$&${6302.3}_{-1.6}^{+1.8}$&${367.0}_{-25.0}^{+25.0}$\\
HD\,18667\,B&${410.0}_{-27.0}^{+27.0}$&${52.0}_{-2.0}^{+2.2}$&${128.0}_{-2.2}^{+2.0}$&${5.91}_{-0.18}^{+0.16}$&${0.6659}_{-0.0011}^{+0.0011}$&${11.985}_{-0.048}^{+0.048}$&${42.6}_{-1.9}^{+2.0}$&${288.98}_{-0.12}^{+0.12}$&${33.14}_{-1.0}^{+0.91}$&${9394.0}_{-18.0}^{+18.0}$&${323.0}_{-19.0}^{+18.0}$\\
HD\,126614\,B&${416.7}_{-8.0}^{+8.4}$&${40.2}_{-1.3}^{+1.2}$&-&${211.0}_{-50.0}^{+44.0}$&${0.831}_{-0.053}^{+0.029}$&${2443.0}_{-825.0}^{+819.0}$&${70.77}_{-0.66}^{+0.73}$&${2.7}_{-1.9}^{+356.0}$&${2880.0}_{-690.0}^{+608.0}$&${7394.0}_{-221.0}^{+233.0}$&${269.2}_{-7.5}^{+7.4}$\\
HD\,108341\,B&${441.0}_{-20.0}^{+21.0}$&-&${129.0}_{-11.0}^{+18.0}$&${272.0}_{-23.0}^{+35.0}$&${0.883}_{-0.1}^{+0.048}$&${4005.0}_{-521.0}^{+778.0}$&${16.7}_{-6.8}^{+12.0}$&${34.2}_{-11.0}^{+8.5}$&${5564.0}_{-479.0}^{+706.0}$&${1152714.0}_{-175147.0}^{+269683.0}$&${342.0}_{-103.0}^{+50.0}$\\
HD\,196050\,B&${505.0}_{-20.0}^{+19.0}$&${50.4}_{-8.4}^{+8.4}$&-&${712.0}_{-127.0}^{+114.0}$&${0.885}_{-0.1}^{+0.061}$&${14740.0}_{-3791.0}^{+3644.0}$&${74.0}_{-22.0}^{+19.0}$&${253.5}_{-7.1}^{+7.5}$&${14094.0}_{-2513.0}^{+2260.0}$&${800504.0}_{-178165.0}^{+278597.0}$&${387.0}_{-51.0}^{+50.0}$\\
HIP\,84056\,B&${513.0}_{-51.0}^{+51.0}$&-&${106.5}_{-4.2}^{+11.0}$&${723.0}_{-161.0}^{+209.0}$&${0.61}_{-0.33}^{+0.24}$&${13197.0}_{-4158.0}^{+6145.0}$&${36.2}_{-6.0}^{+12.0}$&${56.0}_{-36.0}^{+285.0}$&${9668.0}_{-2153.0}^{+2795.0}$&${1873971.0}_{-699945.0}^{+1816263.0}$&${482.0}_{-64.0}^{+55.0}$\\
HD\,205521\,B&${515.0}_{-31.0}^{+31.0}$&${3.76}_{-0.17}^{+0.18}$&${176.24}_{-0.18}^{+0.17}$&${3.67}_{-0.083}^{+0.078}$&${0.171}_{-0.013}^{+0.013}$&${5.5695}_{-0.0081}^{+0.0086}$&${110.7}_{-2.6}^{+2.6}$&${222.4}_{-5.2}^{+5.0}$&${75.8}_{-1.7}^{+1.6}$&${6446.0}_{-21.0}^{+20.0}$&${33.8}_{-1.7}^{+1.7}$\\
HD\,97601\,B&${547.0}_{-22.0}^{+22.0}$&${46.79}_{-0.93}^{+1.0}$&${133.21}_{-1.0}^{+0.93}$&${6.72}_{-0.13}^{+0.12}$&${0.3315}_{-0.0045}^{+0.0046}$&${11.848}_{-0.062}^{+0.063}$&${157.7}_{-1.3}^{+1.3}$&${50.67}_{-0.76}^{+0.76}$&${56.0}_{-1.1}^{+1.0}$&${6597.3}_{-9.0}^{+9.3}$&${399.0}_{-15.0}^{+15.0}$\\
HD\,43587\,B&${584.0}_{-5.2}^{+5.2}$&${43.61}_{-0.38}^{+0.38}$&-&${12.662}_{-0.058}^{+0.058}$&${0.80928}_{-0.00091}^{+0.00093}$&${35.5}_{-0.24}^{+0.24}$&${166.8}_{-0.34}^{+0.34}$&${75.35}_{-0.24}^{+0.25}$&${654.1}_{-2.8}^{+2.8}$&${13800.0}_{-87.0}^{+89.0}$&${402.9}_{-3.6}^{+3.5}$\\
HD\,145428\,B&${605.0}_{-34.0}^{+33.0}$&${38.29}_{-0.38}^{+0.38}$&${141.71}_{-0.38}^{+0.38}$&${6.78}_{-0.2}^{+0.19}$&${0.3129}_{-0.0052}^{+0.0053}$&${13.99}_{-0.16}^{+0.16}$&${112.81}_{-0.57}^{+0.59}$&${308.82}_{-0.63}^{+0.6}$&${56.1}_{-1.6}^{+1.5}$&${5364.3}_{-4.8}^{+4.6}$&${375.0}_{-21.0}^{+20.0}$\\
HD\,142022\,B&${661.0}_{-22.0}^{+21.0}$&${70.5}_{-7.2}^{+3.9}$&-&${657.0}_{-121.0}^{+189.0}$&${0.48}_{-0.23}^{+0.22}$&${13610.0}_{-3571.0}^{+6247.0}$&${142.6}_{-5.3}^{+4.5}$&${31.0}_{-24.0}^{+322.0}$&${19172.0}_{-3534.0}^{+5514.0}$&${3619350.0}_{-864441.0}^{+1487209.0}$&${619.0}_{-36.0}^{+27.0}$\\
HD\,23596\,B&${671.0}_{-22.0}^{+22.0}$&-&${102.4}_{-7.3}^{+16.0}$&${2326.0}_{-309.0}^{+590.0}$&${0.78}_{-0.36}^{+0.2}$&${80710.0}_{-15430.0}^{+32577.0}$&${62.9}_{-5.4}^{+9.3}$&${319.0}_{-293.0}^{+21.0}$&${44959.0}_{-5969.0}^{+11412.0}$&${12958831.0}_{-2116444.0}^{+4926629.0}$&${648.0}_{-66.0}^{+30.0}$\\
\hline
\end{tabular}
}
\tablecomments{\textwidth}{\tiny{A machine-readable table will be available online as supplementary after the publication.}}
\end{table}

\bibliographystyle{raa}
\bibliography{bibtex}

\end{document}